%% file: main.tex
\documentclass[conference]{IEEEtran}
\IEEEoverridecommandlockouts
\usepackage{cite}
\usepackage{amsmath,amssymb,amsfonts}
\usepackage{algorithmic}
\usepackage{graphicx}
\usepackage{textcomp}
\usepackage{xcolor}

\usepackage{xpatch,amsthm}
\usepackage[normalem]{ulem}

\usepackage{amsmath,amssymb,amsfonts}
\usepackage{graphicx}
\usepackage{xcolor}
\usepackage{epsfig}
\usepackage{xspace}
\usepackage{graphicx}
\usepackage{balance}
\usepackage{caption}
\usepackage{hyperref}
\usepackage{multirow}
\usepackage{url}
\usepackage{enumitem}
\usepackage{threeparttable}
\usepackage{comment}

\usepackage[ruled,noend,linesnumbered]{algorithm2e}
\let\oldnl\nl
\newcommand{\nonl}{\renewcommand{\nl}{\let\nl\oldnl}}
\newcommand{\reviseOne}[1]{\noindent \textcolor{blue}{#1}}

\usepackage{subcaption}

\DeclareCaptionLabelFormat{em-noparens}{\emph{#2}}
\usepackage{todonotes}
\input{commands}

\newcommand{\tr}[1]{#1} 
\newcommand{\conf}[1]{}  

\def\BibTeX{{\rm B\kern-.05em{\sc i\kern-.025em b}\kern-.08em
    T\kern-.1667em\lower.7ex\hbox{E}\kern-.125emX}}
\begin{document}

\title{Temporal Graph Functional Dependencies \Large{[Extended Version]} \\
}

\author{\IEEEauthorblockN{Morteza Alipourlangouri}
\IEEEauthorblockA{\textit{McMaster University} \\
alipoum@mcmaster.ca}
\and
\IEEEauthorblockN{Adam Mansfield, Fei Chiang}
\IEEEauthorblockA{\textit{McMaster University} \\
\{mansfa2, fchiang\}@mcmaster.ca}
\and
\IEEEauthorblockN{Yinghui Wu}
\IEEEauthorblockA{\textit{Case Western Reserve University} \\
yxw1650@case.edu}
}

\maketitle

\begin{abstract}
Data dependencies have been extended to graphs to characterize topological and value constraints. 
Existing data dependencies are defined to capture inconsistencies in 
static graphs. Nevertheless, inconsistencies may occur over evolving graphs 
and only for certain time periods.  The need for capturing such inconsistencies 
in temporal graphs is evident in 
anomaly detection and predictive dynamic network analysis. This paper 
introduces a class of data dependencies called {\em Temporal Graph Functional Dependencies (TGFDs)}. TGFDs generalize 
functional dependencies to temporal graphs as a sequence of graph snapshots that are induced by time intervals, and enforce both topological constraints and attribute value dependencies that must be satisfied by these snapshots. (1) We establish the complexity results for the satisfiability and implication problems of TGFDs. (2) We propose a sound and complete axiomatization system for TGFDs.  (3) We also present efficient parallel algorithms to detect inconsistencies in temporal graphs as violations of TGFDs. The algorithm exploits data and temporal locality induced by time intervals, and uses incremental pattern matching and load balancing strategies to enable feasible error detection in large temporal graphs. Using real datasets, we experimentally verify that our algorithms achieve lower runtimes compared to existing baselines, while improving the accuracy over error detection using
existing graph data constraints, \eg \ \gfds and \gtars with 55\% and 74\% gain in \fscore-score, respectively. 
\end{abstract}

\begin{IEEEkeywords}
temporal graph functional dependencies, graph data quality, temporal graphs
\end{IEEEkeywords}

\input{intro}

\input{tgfd}
\input{foundations}
\input{parallel}

\input{experiments}
\input{relatedwork}

\input{conclusion}

\bibliographystyle{IEEEtran}
\bibliography{ref,main}

\end{document}

%% file: commands.tex
\makeatletter
\DeclareRobustCommand*\cal{\@fontswitch\relax\mathcal}
\makeatother

\xpatchcmd{\@thm}{\fontseries\mddefault\upshape}{}{}{}

\newcommand{\kw}[1]{{\ensuremath {\mathsf{#1}}}\xspace}

\newcommand{\stitle}[1]{\noindent{\bf #1}}

\newcommand{\fei}[1]{{\noindent \color{red}{Comment(Fei): #1}}} 

\newcommand{\commentwu}[1]{{\noindent \color{red}{[Comment(Yinghui): #1}]}} 

\newcommand{\morteza}[1]{{\noindent \color{cyan}{Comment(Morteza): #1}}}

\newcommand{\newtext}[1]{{\noindent \color{brown}{#1}}}
\newcommand{\blue}[1]{{\noindent \color{black}{#1}}}
\newcommand{\rev}[1]{{\noindent \color{black}{#1}}}

\newcommand{\wu}[1]{{\noindent \color{black}{#1}}}

\newcommand{\lit}[1]{{\sl #1}}

\newcommand{\tgfd}{\kw{TGFD}}
\newcommand{\tgfds}{\kw{TGFDs}}
\newcommand{\gfds}{\kw{GFDs}}

\newcommand{\tfds}{\kw{TFDs}}
\newcommand{\dfds}{\kw{DFDs}}
\newcommand{\gfd}{\kw{GFD}}

\newcommand{\geds}{\kw{GEDs}}
\newcommand{\ged}{\kw{GED}}
\newcommand{\gpar}{\kw{GPAR}}
\newcommand{\gtar}{\kw{GTAR}}
\newcommand{\gtars}{\kw{GTARs}}
\newcommand{\gkeys}{\kw{GKeys}}
\newcommand{\ptime}{\kw{PTIME}}
\newcommand{\npc}{\kw{NP}-complete}

\newcommand{\np}{\kw{NP}}
\newcommand{\conpc}{\kw{coNP}-complete}
\newcommand{\conp}{\kw{coNP}}

\newcommand{\tedincremental}{\kw{IncTED}}
\newcommand{\seqted}{\kw{SeqTED}}
\newcommand{\isounit}{\kw{IsoUnit}}
\newcommand{\tedbatchnaive}{\kw{NaiveTED}}
\newcommand{\tedbatchopt}{\kw{BatchTED}}
\newcommand{\tedparallel}{\kw{ParallelTED}}
\newcommand{\tedparallelReb}{\kw{ParallelTED_{reb}}}
\newcommand{\tednobalance}{\kw{ParallelTED_{NoBal}}}
\newcommand{\gfdbasic}{\kw{GFD}-\kw{Parallel}}
\newcommand{\gtarSubIso}{\kw{GTAR}-\kw{SubIso}}
\newcommand{\gfdmultigraph}{\kw{GFD}-\kw{MultiGraph}}

\newcommand{\gmark}{\kw{gMark}}
\newcommand{\changerate}{\kw{chg}}
\newcommand{\errorrate}{\kw{err}($\mathcal{G}_T$)}
\newcommand{\snapshot}{$G_{t}$}
\newcommand{\snapshoti}{$G_{i}$}

\newcommand{\matches}{$\mathcal{M}(Q,G_{t})$}
\newcommand{\matchesJ}[1]{$\mathcal{M}_{r}(Q,G_{{#1}})$}

\newcommand{\fragments}{$\{F_{tr}\}$\xspace}
\newcommand{\fragment}{$F_{tr}$}
 
\newcommand{\querypath}{$Q_k$} 
\newcommand{\vecmatchfull}{${\bar{\beta}}$(\matchone)}
\newcommand{\vecmatch}{${\bar{\beta}}$}
\newcommand{\unsatfull}{$\kw{unSat}_k$(\matchone)}
\newcommand{\unsat}{$\kw{unSat}_k$}

\newcommand{\dbpedia}{\kw{DBpedia}}
\newcommand{\pdd}{\kw{PDD}}
\newcommand{\imdb}{\kw{IMDB}}
\newcommand{\yago}{\kw{Yago}}
\newcommand{\wikidata}{\kw{Wikidata}}

\newcommand{\synthetic}{\kw{Synthetic}}

\newcommand{\val}{\kw{val}}
\newcommand{\tempgraph}{$\mathcal{G}_T$}

\newcommand{\component}{$\mathcal{C}_{S}$}
\newcommand{\errorSigma}{$\mathcal{E}($\tempgraph$, \Sigma)$}
\newcommand{\errorSigmaj}{$\mathcal{E}_r(F_{r}, \Sigma)$}
\newcommand{\errorSigmaC}{$\mathcal{E}_c($\tempgraph$, \Sigma)$}
\newcommand{\errorsigma}{$\mathcal{E}($\tempgraph$, \sigma)$}
\newcommand{\errorsigmaj}{$\mathcal{E}_{r}(F_{r}, \sigma)$}
\newcommand{\matchpair}{$\{h_{i}$, $h_{j}\}$}
\newcommand{\matchone}{$h_{i}$}
\newcommand{\matchtwo}{$h_{j}$}

\newcommand{\intone}{$I_{h_{i}}$}
\newcommand{\inttwo}{$I_{h_{j}}$}
\newcommand{\vio}{\kw{vio}}
\newcommand{\permissrange}{$\rho(i)$}
\newcommand{\hashmapX}{$\Pi_{X}$}
\newcommand{\hashmapXY}{$\pi_{XY}$}
\newcommand{\timehashmapX}{$\Gamma(\Pi_{X})$}
\newcommand{\timehashmapXY}{$\Gamma(\pi_{XY})$}
\newcommand{\mapfunc}{$\mu_{X}(h_{i})$}

\newcommand{\simplejob}{$\mathcal{J}_{tr}$}
\newcommand{\job}{$\mathcal{J}_{tr}(\sigma, $ \fragment$, $ \tempgraph$)$}
\newcommand{\jobSigma}{$\mathcal{J}_{tr}(\Sigma, $ \fragment$,$ \tempgraph$)$}
\newcommand{\joblet}{$\omega_{trk}(Q_{k},$\fragment$, \mathcal{G}(v',d))$}
\newcommand{\timebuff}{$\zeta$}

\newcommand{\tempgraphk}{$\mathcal{G}(v', d)$}

\newcommand{\candQk}{$Candid(Q_k,$\fragment$)$}

\newcommand{\snapshotdiff}[3]{$\delta_{#1,#2}(#3)$}

\newcommand{\eat}[1]{}

\newcommand{\thesis}[1]{}

\makeatletter
\DeclareRobustCommand*\cal{\@fontswitch\relax\mathcal}
\makeatother

\newcommand{\sstab}{\rule{0pt}{8pt}\\ [-3ex]}

\newcommand{\bi}{\begin{itemize}}
	\newcommand{\ei}{\end{itemize}}
	{\end{itemize}} 

\newcommand{\be}{\begin{enumerate}}
	\newcommand{\ee}{\end{enumerate}}
\newcommand{\beqn}{\begin{eqnarray*}}
	\newcommand{\eeqn}{\end{eqnarray*}}

\newcommand{\eetitle}[1]{\vspace{0.8ex}\noindent{\underline{\em #1}}}

\newcommand{\ie}{\emph{i.e.,}\xspace}
\newcommand{\eg}{\emph{e.g.,}\xspace}
\newcommand{\wrt}{\emph{w.r.t.}\xspace}

\newcommand{\kwlog}{\emph{w.l.o.g.}\xspace}

\newcommand{\eop}{\hspace*{\fill}\mbox{$\Box$}}

 \newcounter{example}
 \renewcommand{\theexample}{\arabic{example}}
 \newenvironment{example}{
       \refstepcounter{example}
       {\noindent\bf Example \theexample:}}{
         \eop
         }

\newcommand{\nthesection}{\arabic{section}}

 \newcounter{lemma}
 \renewcommand{\thelemma}{\arabic{lemma}}
 \newenvironment{lemma}{\begin{em}
 		\refstepcounter{lemma}
 		{
 		\noindent\bf Lemma \thelemma:}}{
 	\end{em}\eop
 	}

 \newcounter{axiom}
 \renewcommand{\theaxiom}{\arabic{axiom}}
 \newenvironment{axiom}{\begin{em}
 		\refstepcounter{axiom}
 		{
 		\noindent\bf Axiom \theaxiom:}}{
 	\end{em}\eop
 	}

 \newcounter{theorem}
 \renewcommand{\thetheorem}{\arabic{theorem}}
 \newenvironment{theorem}{\begin{em}
         \refstepcounter{theorem}
         {
         \noindent\bf  Theorem  \thetheorem:}}{
         \end{em}\eop
         }

\newenvironment{ctheorem}[1]{\begin{em}
        \refstepcounter{theorem}
        {\vspace{1ex}{\noindent \bf Theorem  \thetheorem~[#1]}: }}{
        \end{em}\eop\vspace{1.5ex}}

\newcounter{cor}
\renewcommand{\thecor}{\arabic{cor}}

\newcounter{prop}
\renewcommand{\theprop}{\arabic{theorem}}

 \newcounter{definition}[section]
 \renewcommand{\thedefinition}{\nthesection.\arabic{definition}}
 \newenvironment{definition}{
         \refstepcounter{definition}
         {\noindent\bf Definition {\bf \thedefinition}:}}{\eop
 }
 
\newcounter{alg}[section]
\renewcommand{\thealg}{\nthesection.\arabic{alg}}

\newcounter{arule}
\renewcommand{\thearule}{\arabic{arule}}

\newenvironment{proofS}{
	{\noindent\bf Proof sketch:\ }}{\eop
	}

 \newcommand{\tbf}{\textbf{\textcolor{red}{X}}\xspace}

\definecolor{gray}{rgb}{0.5,0.5,0.5}

\newcounter{ccc}

\renewcommand{\thetheorem}{\arabic{theorem}}

\newenvironment{ab}
{\mathactivatecomma
	\mathcode`\,=\string"8000
	\ignorespaces}
{\ignorespacesafterend}

\newcommand{\mathactivatecomma}{%
	\begingroup\lccode`~=`\,
	\lowercase{\endgroup\edef~}{\mathchar\the\mathcode`\,\penalty0 }}
	
	\renewenvironment{proofS}{
        \vspace{1ex}
        {\noindent\bf Proof sketch:\ }}{\eop\vspace{1ex}}

\newcommand{\bab}{\begin{ab}}
	\newcommand{\eab}{\end{ab}\xspace}

\newcommand{\GFD}{\kw{GFD}}

\newcommand{\GQRs}{\kw{GQRs}}

\newcommand{\NGDs}{\kw{NGDs}}

\newcommand{\PTIME}{\kw{PTIME}}

\renewcommand{\prec}{\kw{precision}}
\newcommand{\rec}{\kw{recall}}
\newcommand{\fpr}{\kw{fpr}}
\newcommand{\fscore}{\kw{F_1}}



%% file: intro.tex
\section{Introduction}
\label{sec:intro}
\eat{
Graphs are increasingly being used to model information about entities, their properties, and relationships between entities.  
Examples include relationships between customers, their product purchases, and inter-relationships between products. Large-scale knowledge bases such as \dbpedia~\cite{lehmann2015dbpedia}, \yago~\cite{mahdisoltani2013yago3}, \wikidata~\cite{vrandevcic2014wikidata}, and Amazon product graphs operate in dynamic data environments, where having accurate and consistent information is critical for downstream decision making and fact checking.}

\eat{Graphs have been widely used to 
such as knowledge graphs~\cite{lehmann2015dbpedia,mahdisoltani2013yago3,vrandevcic2014wikidata}, e-commerce~\cite{Dong2020AutoKnowSK} and social networks~\cite{wasserman1994social}. }

Data constraints 
have been extended for graphs to 
capture inconsistencies and errors in 
graph data~\cite{fan2016functional,fan2015keys,ma2019ontology}. Given a graph $G$, a graph data dependency 
is often \blue{of} the form of $(Q, X\rightarrow Y)$, where 
$Q$ is a graph pattern that specifies a set of  
subgraphs in $G$ via graph pattern matching 
(\eg subgraph isomorphism) such that each 
subgraph should satisfy the value
constraints enforced by $X\rightarrow Y$ 
(where $X$ and $Y$ are literals \blue{from the graph pattern}). 
Notable examples include 
\wu{
graph functional dependencies (\gfds)~\cite{fan2016functional}, 
graph keys (\gkeys)~\cite{fan2015keys} and 
graph association rules (\gtars)~\cite{fan2015association,wang2020extending}.}

Existing graph dependencies are often designed \wu{to capture inconsistencies in} static graphs. \wu{Nevertheless, data errors also occur in {\em evolving} 
real-world graphs. In such scenarios, node attribute values and edges in graphs experience constant changes over time, and data consistency may persist only over a fraction of graphs 
 specified by certain time periods.}
 \wu{
 The need for modeling data consistency in temporal graphs 
 that are time-dependent is evident in 
 time sensitive applications such as policy-making~\cite{cdngov} 
 and anomaly detection 
 in health care~\cite{fdacovid,PG18}. 
 Data constraints for static graphs are often 
 insufficient to satisfy such needs, as illustrated 
 next. }

\begin{figure*}
    \centering
    \includegraphics[width=7in,keepaspectratio]{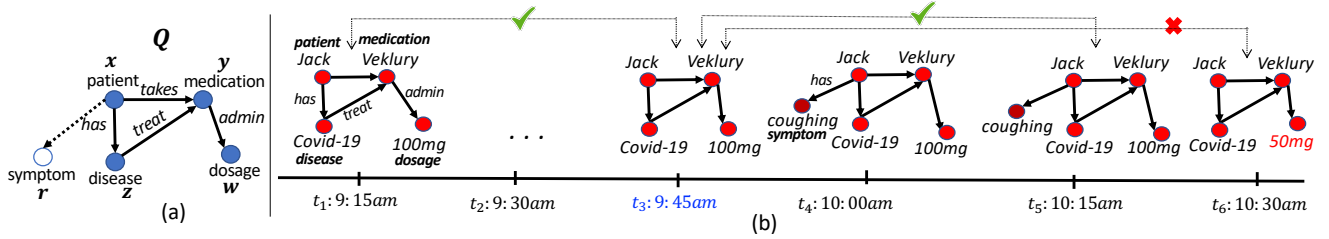}
    \caption{\rev{Temporal data constraint to validate drug dosage at $t_{3}$.}}
        \vspace{-4ex}
    \label{fig:mainExample}
\end{figure*}

\vspace{-.5ex}

\begin{example} \label{exm:motivating}
\wu{
\rev{Consider a graph pattern $Q$ (consisting of the solid, blue vertices) shown in Figure~\ref{fig:mainExample}(a) describing patients with a  specific disease are treated with an administered medication and dosage within a time interval.  Figure~\ref{fig:mainExample}(b)
shows a temporal graph with matches of $Q$ that monitors medicare activities~\cite{PG18}.} A snapshot describes patient diagnosis, medication, and dosage taken at a timestamp.} \blue{Federal drug regulating agencies, such as the Food and Drug Administration, receive numerous medication error reports due to missed doses, incorrect preparation and administration of drug formulations. To counter this, safe practice recommends that administered dosages be verified against patient characteristics, and require past and subsequent doses be correctly timed~\cite{ismp22}.} 

\rev{Figure~\ref{fig:mainExample}(b) shows patients treated with medication \emph{Veklury},} which requires if the intravenous (IV) infusion is between 30 to 120 mins, then the dosage must be 100mg~\cite{fdacovid}.  At the current time $t_c = t_3$, continual validation of the dosage compares to past drug administrations (at least 30 but no more than 120 mins away) that must be 100mg \rev{(correct dosages indicated via green check marks).}  An incorrect dosage of 50mg is captured at $t_6$.  
This time-sensitive rule can be expressed by 
posing a value constraint on the dosage to 
 patients who match $Q$, {\em conditioned} 
by a time infusion period. 
To capture violations 
requires  comparing a 
fraction of ``snapshots'' 
induced by a time period. 
Existing data constraints 
cannot express such 
consistency criteria for temporal graphs. 
\eat{
Existing data constraints 
cannot express such criteria  
which requires to 
capture possible violations by the comparison over ``snapshots'' induced by a time period. }
\end{example}

\blue{The above example calls for the need to model data consistency with respect to (w.r.t.) topological constraints and attribute values that are 
contextualized over a time interval duration. 
For each ``current time'' $t_c$, 
there is a need to compare matches at $t_c$ against historical and future matches that are at least $p$ but no more than $q$ time units away. These ``minimum'' and ``maximum'' time bounds pose additional requirements where structural and literal conditions specified via graph pattern matching should hold.}

\eat{
Using a \tgfd mining algorithm that we developed together with this work~\cite{NoronhaC21}, we profiled \dbpedia and found over 140 temporal data constraints with graph patterns involving variable and constant literals.  Of these constraints, 28\% contained at least one inconsistency, with an average of 7 distinct (error) values per constraint, spanning a duration of 5 time units.}

Such schema and temporal constraints cannot be readily captured by prior graph dependencies and rules~\cite{fan2016functional,fan2019dependencies,namaki2017discovering}. \blue{ For example, \gfds cannot express the temporal constraints that perform necessary pairwise comparisons of single snapshots induced by minimum and maximum time ranges.  Even if time periods were captured via graph attributes, modeling such semantics would involve an excessive number of \gfds, making them infeasible to 
\wu{be verified} in practice.  \gtars detects delayed co-occurrences of events via graph pattern matching, 
and do not model value dependencies~\cite{namaki2017discovering}.  
}We address the above challenges, and study several questions. (1) How to formalize such temporal data constraints in dynamic graphs? (2) What is the hardness of the fundamental problems (satisfiability, implication,  validation) for these temporal data constraints? 
 (3) How to efficiently detect violations with these dependencies in evolving graphs?

\stitle{Contributions.}  
\blue{
\wu{We introduce Temporal Graph Functional Dependencies (\tgfds), a class of data dependencies to model the time-dependent consistency of temporal graph data, and study its fundamental problems and inconsistency detection. }
}

\wu{
\sstab(1) We introduce a formal model of 
\tgfds. \tgfds model time-dependent data consistency of 
temporal graphs by enforcing value dependencies that are  
conditioned by topological and temporal constraints in terms of temporal graph pattern matching. 
\tgfds apply conditionally to temporal graphs, 
and hold on graph snapshots that are induced by 
time intervals with lower and upper bounds. 
A special case of \tgfds with size-bounded 
graph patterns and their benefit in capturing 
data errors has been justified by  
our pilot study~\cite{NoronhaC21}. 
Providing a formal model 
for general \tgfds is our first 
contribution. 
}

\sstab (2) We study the satisfiablity, implication, and validation problems for \tgfds. 
\blue{We introduce an implication algorithm for \tgfds, and the notion of \emph{temporal closure} that considers the time intervals of inferred values.}  We also develop a sound and complete \tgfd axiom system for the 
implication of \tgfds (Section~\ref{sec:foundations}).  

\sstab (3)  We introduce two \tgfd-based inconsistency detection algorithms for temporal graphs:  (i) an incremental detection algorithm (\tedincremental) that re-uses the matches and inconsistencies captured from earlier graph snapshots \blue{to compute updated matches and errors given graph changes}; and (ii) a parallel algorithm (\tedparallel) that \blue{performs fine-grained estimations of the workload by computing the fraction of data induced by the literal conditions and the time interval duration in a \tgfd.} We show that these algorithms have time costs that are independent of the size of temporal graphs, and are feasible for large graphs  (Section~\ref{sec:parallel}).


\sstab (4) We conduct an extensive evaluation over real data collections. We verify the efficiency of \tedparallel over a wide range of parameters achieving 120\% and 29\% speedup over sequential and \gfd-based baselines, respectively, demonstrating \tgfd error detection is feasible over real graphs.  We show the effectiveness of \tedparallel \ over error detection using \gfds and \gtars \ with up to 55\% and 74\% gain in \fscore-score, respectively. Lastly, we conduct a case study with real examples of \tgfds, and 
\wu{the detected inconsistencies} to demonstrate the utility of \tgfds in practice (Section~\ref{sec:exp}).

%% file: tgfd.tex
\vspace{-1ex}
\section{Temporal Graph Functional Dependencies}
\label{sec:tgfd}

\vspace{-1ex}
We start with a notion of temporal graph 
pattern matching, 
and then introduce \tgfd syntax and semantics.

\vspace{-1ex}
\subsection{Temporal Graph Pattern Matching}
\label{sec:prelim}

\vspace{-1ex}
\stitle{Temporal Graphs.}
\wu{
We consider a temporal graph 
\tempgraph \ = $\{G_1, \ldots ,G_T\}$ 
as a sequence of graph snapshots. Each snapshot 
$G_t$ ($t\in[1,T]$) is a graph 
$(V, E_t, L_t, F_{A_t})$ with a fixed node set $V$ 
and edge set $E_t\subseteq V \times V$. 
Each node $v \in V$ (resp. edge $e \in E_t$) has a label $L_t(v)$ (resp. $L_t(e))$ at time $t$. 
For each node $v$, $F_{A_t}(v)$ 
is a tuple $(A_{1,t} = a_1, \ldots ,A_{n,t} = a_n)$
specifying a value of each attribute of $v$ at time $t$. 
}

\eat{
We consider a temporal graph as a sequence of graph snapshots, with fixed vertices $\mathcal{V}$ and possibly varying edges, and attribute values. Let \tempgraph \ be a temporal graph defined over a period of time, consisting of a sequence of graph snapshots $\{G_1, \ldots ,G_T\}.$  Each snapshot 
$G_t$ = $(V_t, E_t, L_t, F_{A_t})$ at a timestamp $t \in [1, T]$ is a directed graph with: (i) a vertex set $V_t$; (ii) an edge set $E_t = V_t \times V_t$ \commentwu{This looks 
incorrect. $\subseteq$?}; (iii) each node $v \in V_t$ (edge $e \in E_t$) has a label $L_t(v) \ (L_t(e))$ at time $t$; and (iv) for each vertex $v$, $F_{A_t}(v)$ is a tuple specifying the value for each attribute of $v$ at time $t$, i.e., a tuple of the form $(A_{1,t} = a_1, A_{2,t} = a_2, \ldots ,A_{n,t} = a_n)$, for attribute $A_{i,t}$ with constant $a_i$, representing the properties of vertex $v$ at time $t$.
}


\stitle{Temporal graph pattern matching}. \wu{A graph pattern is a directed, connected graph $Q[\bar{x}] = (V_Q, E_Q, L_Q, \mu)$ with a set of nodes $V_Q$ and edges $E_Q \subseteq V_Q\times V_Q$. For each node $u \in V_Q$ and each edge $e \in E_Q$, the function $L_Q$ assigns a label $L_Q(u)$ and $L_Q(e)$ to $u$ and $e$, respectively. $\bar{x}$ is a set 
of variables, and   
$\mu$ is a function that maps each $u \in V_Q$ to a 
distinct variable in $\bar{x}$. 
We shall refer to $\mu(u)$ as 
$u$ for simplicity.
}

\wu{
\eetitle{Matches}. Given a temporal graph \tempgraph \ and pattern $Q$, a \emph{match} $h_t(\bar{x})$ 
between a snapshot \snapshot \ of \tempgraph \ and pattern $Q$ is a subgraph $G'_t = (V'_t, E'_t, L'_t, F'_{A_t})$ induced by $V'_t$ of \snapshot \
that is isomorphic to $Q$. 
That is, there exists a bijective function 
(a matching) $h_t$ from $V_Q$ to $V'_t$ 
such that: (1) for each node $u \in V_Q$, $L_Q(u)=L'_{t}(h_t(u))$; and (2) for each edge $e = (u, u') \in Q$, there is an edge $e' = (h_t(u), h_t(u'))$ in $G'_t$ such that $L_Q(e)=L'_t(e')$.  If $L_Q(u)$ is ‘\_’, then 
it matches with any label for any 
timestamp $t$. 

As a matching $h_t$ uniquely 
determines a match (subgraph) for $\bar{x}$ 
at any time $t$, 
we refer to a match as $h_t$ 
for simplicity. Table~\ref{tab:symbols} summarizes our notation and symbols.
}

\begin{table}
  \centering
  \caption{Notations and symbols.}
  \vspace{-2mm}
  \label{tab:symbols}
  \setlength\tabcolsep{2 pt}
  \small \begin{tabular}{l l}
    \hline
    Symbol              & Description \\
    \hline
   \tempgraph, $G_t$, $G'_t$,  & temporal graph,
   snapshot, and subgraph \\
    $Q[\bar{x}]$ & graph pattern \\
    $\sigma, \Sigma$ & \blue{a single \tgfd, and a set of \tgfds} \\
    $h_{t}$ & a match of $Q[\bar{x}]$ in \snapshot \ at time $t$ \\
    \errorsigma, \errorSigma & violations of $\sigma, \Sigma$ in \tempgraph \\
   \matches & match set of $Q$ in \snapshot \\
    \intone, \inttwo & time domain of \matchone, \matchtwo \\
    \permissrange & permissable range of \matchone \\
    \hashmapX, \hashmapXY & matches with shared values in $X$, $XY$ \\
    \timehashmapX & timestamps of matches in \hashmapX  \\
    \mapfunc & returns \hashmapXY \ where $h_{{i}}$ belongs \\
    \hline
    \end{tabular}
    \vspace{-3ex}
\end{table}

\vspace{-1ex}
\subsection{Temporal Graph Functional Dependencies}
\stitle{Syntax.}
A \emph{Temporal Graph Functional Dependency} (\tgfd) $\sigma$ is a triple $(Q[\bar{x}], \Delta, X \rightarrow Y)$, where:

\begin{itemize}
\makeatletter\@topsep0pt\makeatother 
 \item $Q[\bar{x}]$ is a graph pattern;
 \item \wu{$\Delta$ = $(p, q)$ is a time interval, specified by a lower bound $p$, and an upper bound $q$ \wu{($p,q$ are integers and} $q \geq p \geq 0)$; }
 \item \wu{ a 
 value dependency
 $X \rightarrow Y$, where 
 $X$ and $Y$ are two (possibly empty) sets of literals defined on $\bar{x}$. 
 }
 
\end{itemize}

\eat{
\stitle{Syntax2 (Morteza).}
\newtext{A \emph{Temporal Graph Functional Dependency} (\tgfd) $\sigma$ is a triple (Q[\bar{x}],\bar{a}, \Delta, X \rightarrow Y), where
\begin{itemize}[topsep=0pt]
\item $Q[\bar{x}]$ is a graph pattern;
 \item $\bar{a}$ contains set of anchor vertices from $\bar{x}$; 
 \item $\Delta$ is a time interval, defined as a pair $(p, q),  q \geq p \geq 0$, 
 \item $X$ and $Y$ are two (possibly empty) sets of literals of $\bar{x}$.
\end{itemize}
}
}


\wu{Literals are of the form $u.A = c$ (constant literal) or $u.A$ = $u'.A'$ (variable literal), where $u \in \bar{x}$, $A$ and $A'$ are attributes, and $c$ is a constant.}  
A \tgfd $\sigma$ enforces three constraints: 
\begin{enumerate}[noitemsep,nolistsep,align=left]
\makeatletter\@topsep0pt\makeatother 
    \item topological and label constraint of pattern $Q$, 
    \item a value dependency specified by $X \rightarrow Y$, and    
    \item \wu{a time interval constraint $\Delta$, which specifies, for 
    a time $t$, two time windows 
    $[t-q, t-p]$ and $[t+p, t+q]$. 
    The time windows induce the set of snapshots 
    over which the constraint 
    should hold (see ``Semantics'').}
\end{enumerate}

For simplicity, we consider \kwlog~\tgfds in a normal form, \ie with a dependency 
$X\rightarrow Y$ where $Y$ contains a single literal. 
We justify the normal form in Section~\ref{sec:axioms}.
\eat{
without loss of generality, we decompose the consequent $Y$ into single literals, i.e., $\sigma$ can be equivalently represented as $\sigma' = (Q[\bar{x}],\Delta,X\rightarrow y_1)$, and $\sigma'' = (Q[\bar{x}],\Delta,X\rightarrow y_2)$, for $Y = \{y_1, y_2\}$; we refer to this as the Decomposition Axiom in Section~\ref{sec:axioms}. Similar to \gfds, we call this the \emph{normal form} representation, and assume henceforth, all \tgfds $\sigma= (Q[\bar{x}],\Delta,X\rightarrow y)$ are in normal form~\cite{fan2016functional}.  
}

\begin{example} \label{exm:tgfdDefinition}
\blue{
We define a \tgfd $\sigma=(Q[\bar{x}]$, $\Delta=(30, 120)$, $[x.name, z.name= \lit{Covid19}, y.name= \lit{Veklury}] \rightarrow [w.\val = \lit{100mg}])$. If a patient has \lit{Covid-19} and is treated with \lit{Veklury} over IV infused between 30 to 120 mins, then the dosage is \lit{100mg}.
}
\end{example}

\stitle{Semantics.}
\wu{Given a temporal graph 
\tempgraph, a \tgfd 
enforces value dependencies between 
any pair of snapshots that are (1) matches via graph pattern matching, and (2) having timestamps within time ranges in $\Delta$, for  
any timestamp in $[1,T]$}. 
Consider a \tgfd $\sigma = (Q[\bar{x}], \Delta, X \rightarrow Y)$, and a pair of matches 
$(h_i, h_j)$ of $Q$ in 
$G_i$ and $G_j$, respectively ($i,j\in [1,T]$). 
We say $(h_i, h_j)$  {\em matches} $\sigma$ 
if 
\begin{itemize}
   \makeatletter\@topsep0pt\makeatother
   \item $(h_i, h_j)$ satisfies $X$ (denoted as 
        $(h_i, h_j)\models X$), i.e., for each constant literal 
        $l$ = $(u.A = c)$ (resp. variable literal 
        $u.A = u'.A'$) in $X$, $h_i(u)=h_j(u)=c$ 
        (resp. $h_i(u)$ = $h_j(u')$); and 
    \item $|j-i|\in \Delta$. 
\end{itemize}

 \wu{
The term $|j-i|\in \Delta$ conveniently express 
temporal semantics including
``future'' and ``past'' from their 
conventional counterparts in temporal integrity constraints~\cite{Wijsen2009TemporalD} 
to temporal graphs. 
For any specific timestamp $i\in [1,T]$, 
(a) if $i\leq j$, then $|j-i|\in \Delta$ 
specifies any timestamp $j$ from a ``future'' time window $[i+p, i+q]$; 
(b) if $i>j$, then $|j-i|\in \Delta$ 
specifies timestamps $j$ from a ``past'' time window 
$[i-q, i-p]$. 
Given a time interval $\Delta$, a \tgfd $\sigma$ enforces data consistency 
over all pairs $(h_i,h_j)\in [1,T]$ 
that matches $\sigma$ in terms of $\Delta$. 
}

\wu{For any pair $(h_i, h_j)$ that does not match 
$\sigma$, $(h_i, h_j)$ ``trivially'' satisfies $\sigma$.
We say a temporal graph \tempgraph \ {\em nontrivially} \ {\em satisfies} a \tgfd $\sigma$, denoted as \tempgraph $\models \sigma$, if 
(a) there exists at least a pair of matches $(h_i, h_j)$ that also matches $\sigma$, 
and (b) $(h_i,h_j)\models Y$. 
 \tempgraph \ satisfies a set of \tgfds $\Sigma$ if for every $\sigma \in \Sigma$, \tempgraph \  $\models \sigma$. 
}

\blue{ 
\noindent \begin{example} \label{exm:tgfdSemantic}
Figure~\ref{fig:mainExample}(b) shows matches $(h_1,h_3)$ and $(h_3,h_5)$ satisfy $\sigma$ (denoted with a green check mark), but $(h_3,h_6) \not \models \sigma$, having the wrong dosage of \lit{50mg} at $t_6$ (denoted with a red x) given the required infusion time of 30 to 120 minutes.  
\end{example}
}

\stitle{Remarks}. \wu{
As justified in our pilot study~\cite{NoronhaC21} on a special case of the general \tgfds with 
bounded pattern size over a
real knowledge graph \dbpedia, we found 
over 140 temporal data constraints in 
real knowledge base \dbpedia.  Of these constraints, 28\% 
can capture at least one inconsistency; and 
each constraint can capture 
on average 7 distinct erroneous attribute 
values that span over $5$ timestamps. 
}

\stitle{Relationship to Other Dependencies.}
\gfds define topological and attribute dependence over \emph{static} graphs. \eat{ without temporal semantics, involving a single match ~\cite{fan2016functional}.  As expected,} \tgfds subsume \gfds as a special case when $\Delta$ $=(0,0)$.  \eat{Graph keys (\gkeys), and their ontological variant~\cite{fan2015keys,ma2019ontology} 
specify topological and value constraints to 
enforce node equivalence but do not consider attribute dependencies.  Graph Entity Dependencies (\geds) extend \gfds to support equality of entity ids, and subsume \gfds, but again, are defined over static graphs.}  \blue{\gtars are soft rules that use an approximate subgraph isomorphism matching, \eat{in contrast to \tgfds which are hard constraints enforcing strict subgraph isomorphism matching based on $Q$.  In addition, \gtars}only consider matching with time intervals $p = 0$, and do not include historical matches as part of their semantics.}

%% file: foundations.tex
\vspace{-2ex}
\section{Foundations}
\label{sec:foundations}

\vspace{-1ex}
We next study the satisfiability, implication and validation problems for \tgfds, and present an axiomatization.

\vspace{-2ex}
\subsection{Satisfiability}
\label{sec-sat}

\wu{
A set $\Sigma$ of \tgfds is {\em satisfiable}, if there exists a temporal graph 
\tempgraph, such that \tempgraph \  $\models \Sigma$. 
The satisfiability problem is to determine whether $\Sigma$ is satisfiable. Satisfiability checking helps to decide whether a set of \tgfds $\Sigma$ are ``inconsistent'' before being applied for error detection in temporal graphs.  
}

\begin{example} \label{exmp:Satisfiability} 
\rev{Define pattern $Q'$ by augmenting $Q$ with an edge from \lit{patient} ($x$) to \lit{symptom} ($r$) (shown as a dotted edge in Figure~\ref{fig:mainExample}(a).}  Consider $\sigma'$ defined over $Q'$  as $\sigma'=(Q'[\bar{x}], \Delta=(30, 120), [x.name, r.name, z.name = \lit{Covid19}, y.name=\lit{Veklury}] \rightarrow [w.\val = \lit{20mL}])$, with $Q \subseteq Q'$.  The consequent literal requires the value of $w$ to be equal to \lit{100mg} and \lit{20mL} simultaneously, leading to ``conflicting'' value constraints. Since any match of $\sigma$ will also match $\sigma'$, there is no temporal graph \tempgraph \ that satisfies both. Thus, $\{\sigma, \sigma'\}$ are not satisfiable.
\end{example}

The ``conflicting'' value constraints do not necessarily lead to unsatisfiable \tgfds.  \blue{In the above example, suppose the $\Delta$ time interval durations for $\sigma$, $\sigma'$ were $(30,120)$ and $(20,25)$, respectively, then one can verify that they do become satisfiable as the time intervals are not overlapping.  This requires computing the time intervals of when derived values and literals are expected to hold. }
This illustrates that \tgfd satisfiability analysis is more involved than \gfds: the ``conflicting'' value constraints are conditioned upon both \emph{pattern matching and the temporal constraints}.  To characterize this, we introduce a notion of \tgfds~{\em embeddings}. The notion has its foundation in~\cite{fan2016functional}, and is extended for \tgfds with temporal constraints.

\stitle{Pattern embedding}~\cite{fan2016functional}. We say a graph pattern $Q'[\bar{x}'] = (V_Q', E_Q', L_Q', \mu')$ is {\em embeddable} in another pattern $Q[\bar{x}] = (V_Q, E_Q, L_Q, \mu)$, if there exists a 
mapping $f$ from $V_Q'$ to a subset of nodes in $V_Q$ that 
preserves node labels of $V_Q'$, all the edges induced by $V_Q'$ and corresponding edge labels. 
\wu{Moreover, $f$ induces a ``renaming'' 
of each variable $x\in \bar{x}$ to a distinct variable $x'$ in $\bar{x}'$, 
\ie for each variable $x \in \bar{x}$, 
$\mu'(f(\mu^{-1}(x)))$ = $x'\in \bar{x}'$.  
}

\stitle{TGFDs Embedding}. 
Given two \tgfds $\sigma$ = $(Q[\bar{x}]$, $\Delta,$ $f(X')$ $\rightarrow f(Y'))$ and $\sigma'=(Q'[\bar{x}],\Delta', X' \rightarrow Y')$, 
We say that $\sigma$ is a (temporally) \emph{overlapped} \tgfd of $\sigma'$ \wrt graph pattern $Q$,
if (1) $Q'$ is embeddable in $Q$, and 
(2) $(\Delta \cap \Delta') \neq \emptyset$.  
Moreover, $\sigma$ is a (temporally) \emph{embedded} \tgfd of $\sigma'$ \wrt $Q$ if  $\Delta' \subseteq \Delta$. 
Given a set of \tgfds $\Sigma$ and a graph 
pattern $Q$, 
we denote as $\Sigma_{Q}$ (resp. $\Sigma_{Q,\{\Delta,\Delta'\}}$) the set of embedded (resp.  overlapped \tgfds) \wrt $Q$. 

\begin{example} \label{emp: embedding}
\blue{
In \rev{Figure~\ref{fig:mainExample}(a)}, $Q$ is an embedded pattern in $Q'$ as there exists a subgraph isomorphism mapping from $Q$ to a subgraph of $Q'$. Moreover, consider a \tgfd $\sigma''$ with $\Delta'' = (20, 60)$, then $\sigma''$ is a (temporally) overlapped \tgfd of $\sigma$.  Lastly, $\sigma$ is a (temporally) embedded \tgfd of $\sigma'$.}
\end{example}

\blue{
In contrast to \gfds, \tgfd satisfiability requires checking whether conflicting literal values occur during overlapping time intervals, leading to  non-satisfiability.  \eat{If differing attribute values occur during non-overlapping intervals for a set of \tgfds, then each unique literal value is expected to hold over a distinct time interval, independent of time intervals from other \tgfds.}  This requires us to consider the notion of {\em temporal closure} for \tgfds.
}

\begin{definition} (\emph{Temporal Closure})
Given a set of overlapped \tgfds $\Sigma_{Q,\{\Delta,\Delta'\}}$ \wrt
a graph pattern $Q$, the {\em temporal closure} of a set $\Sigma_{Q,\{\Delta,\Delta'\}}$, denoted as 
\emph{closure}($X, \Sigma_{Q,\{\Delta,\Delta'\}}$), 
refers to the set of literals that are derivable  
via transitivity of equality atoms in $X$ 
over $\Sigma_{Q,\{\Delta,\Delta'\}}$. 
The temporal closure 
of a set of \tgfds $\Sigma$ (denoted as 
{\em closure($\Sigma)$}), refers to 
all the literals \emph{closure}($X, \Sigma_{Q,\{\Delta,\Delta'\}}$) 
with $Q$ ranging over the patterns 
from the \tgfds in $\Sigma$.
\end{definition}

\vspace{-1ex}
We outline an algorithm below to compute 
closure($\Sigma)$.  First, given a set of \tgfds $\Sigma$, 
it first adds all $Y$ seen in a \tgfd $\sigma\in\Sigma$, 
if $(Q[\bar x], \delta, \emptyset\rightarrow Y)\in\Sigma$.  
Second, for each pattern $Q$ seen in 
$\Sigma$, it then computes the overlapped \tgfds 
 $\Sigma_{Q,\{\Delta,\Delta'\}}$. 
 For each literal $Y$ seen in 
 $X\rightarrow Y$ in a \tgfd $\sigma\in$ 
 $\Sigma_{Q,\{\Delta,\Delta'\}}$, 
 if $X\subseteq$ closure$(\Sigma)$ or can be 
 derived via transitivity of equality atoms in 
 closure$(\Sigma)$, 
 then \blue{it} adds $Y$ to closure($\Sigma)$. 
 For example, if $x.A = u$ and $y.B = u$ 
 are in closure($\Sigma)$, then $x.A = y.B$ 
 can be derived and added to closure($\Sigma)$.   
This will give us the set of literals that are to be enforced in $\Sigma_{Q,\{\Delta,\Delta'\}}$.

\blue{For each derived value, we compute the time intervals over which each literal value is expected to hold.  
} We say \emph{closure($X, \Sigma_{Q,\{\Delta,\Delta'\}}$)} is {\em in conflict}, 
if \emph{closure($X, \Sigma_{Q,\{\Delta,\Delta'\}}$)} contains literals $x.A = a$ and $y.B = b$, and $a \neq b$. Given $\Sigma$, the temporal closure $closure(\Sigma)$ is in conflict, if 
there exists a pattern $Q$ from a $\sigma \in \Sigma$ such that \emph{closure}($X, \Sigma_{Q,\{\Delta,\Delta'\}}$) is in conflict. We show the following result. 


\begin{lemma}
\label{lm-tgfdsat}
A set $\Sigma$ of \tgfds is not satisfiable 
if and only if closure($\Sigma$) is in conflict. 
\end{lemma}


\vspace{-0.5em}
The result below verifies that checking \tgfd  satisfiability is not harder than their \gfd counterparts.  \blue{For brevity, all proofs in this section are in the extended version~\cite{AMCW21}.}

\begin{theorem}
\label{thm-tgfdsat}
The satisfiability problem for \tgfds is \conpc.
\end{theorem}

\tr{
\begin{proofS}
We introduce an \np algorithm for checking the 
complement of \tgfd satisfiability, 
which is to decide if a given 
set of \tgfds $\Sigma$ is not satisfiable. 
Based on Lemma~\ref{lm-tgfdsat}, the 
algorithm guesses (1) a graph pattern 
$Q$ seen in $\Sigma$, (2) a subset 
$\Sigma'\subseteq \Sigma$,  
(3) a subgraph isomorphism 
mapping from each pattern 
$Q'$ seen in $\Sigma'$ to $Q$. 
It then verifies if 
$\Sigma'$ is a set of 
overlapped \tgfds \wrt $Q$, 
by verifying, 
for each pair of \tgfds 
$\sigma_{1}, \sigma_{2} \in \Sigma'$ with 
time intervals $\Delta_1$ and $\Delta_2$, 
(a) whether their respective patterns $Q_1, Q_2$ are isomorphic to a subgraph in $Q$, and (b)
whether the time intervals 
$(\Delta_1 \cap \Delta_2) \neq \emptyset$. 
If so, 
It then invokes the aforementioned algorithm 
to check if the temporal closure \emph{closure}($X, \Sigma_{Q,\{\Delta,\Delta'\}}$) is in conflict. 
If so, it returns ``yes'' ($\Sigma$ 
is not satisfiable). The above process is in \np 
given bounded number of pairwise verification 
and conflict checking in \PTIME. 

The lower bound is verified by a reduction from 
\gfd satisfiability with constant literals~\cite{fan2016functional}. 
A \tgfd counterpart can be constructed 
by introducing, for each \gfd, a time interval 
$[0,0]$, and a temporal graph with a single 
snapshot. As \gfd satisfiability with 
constant literals is already \conp-hard, 
the lower bound follows. 
\eat{
We extend the algorithm for checking the complement of \gfd satisfiability to consider the $\Delta$ time intervals as \tgfds may overlap in time with conflicting literals~\cite{fan2016functional}.  We randomly select: (i) a subset $\Sigma' \subseteq \Sigma$; (ii) the largest $Q$ in $\Sigma$ (w.r.t. the number of edges) with literals in $\Sigma$.  A mapping $f$ is induced between each pattern $Q'$ in $\Sigma'$ and $Q$.  For each pair of $\sigma_{1}, \sigma_{2} \in \Sigma'$, check whether their respective patterns $Q_1, Q_2$ are isomorphic to a subgraph in $Q$, and whether their time intervals $(\Delta_1 \cap \Delta_2) \neq \emptyset$.  If so, compute the set $\Sigma_{Q,\{\Delta,\Delta'\}}$ from $\Sigma'$, $Q$.  Check whether \emph{closure}($X, \Sigma_{Q,\{\Delta,\Delta'\}}$) is in conflict.  If so, then we return yes, and $\Sigma$ is not satisfiable.  
In the worst case, the upper bound to check \tgfd satisfiability in \conp, similar to \gfds~\cite{fan2016functional}.  The lower bound requires a reduction from subgraph isomorphism, which is known to be \npc.  This result naturally extends from the \gfd case, as there are a bounded number of pairwise matches in the closure to evaluate for conflict.  
}
\end{proofS}
} 

\vspace{-2ex}
\subsection{Implication}
\label{sec:implication}

\vspace{-1.2ex}
Given a set of \tgfds $\Sigma$ and a $\sigma \not \in \Sigma$, we say $\Sigma$ implies $\sigma$, denoted $\Sigma \models \sigma$, if for any \tempgraph, if  \tempgraph \  $\models \Sigma$, then \tempgraph \  $\models \sigma$. If $\Sigma \models \sigma$, then $\sigma$ is a logical consequence of $\Sigma$.  Given $\Sigma$ and $\sigma$, the implication problem is to determine whether $\Sigma \models \sigma$. Implication analysis helps us to remove redundant \tgfds and perform error detection with a smaller number of rules.
To check whether $\Sigma \models \sigma$, we 
extend the temporal closure in Section~\ref{sec-sat} to a counterpart for embedded \tgfd set $\Sigma_Q$ (denoted as \emph{closure($X, \Sigma_Q$)}). 
The temporal closure \emph{closure($X, \Sigma_Q$)} is a set of pairs $(Y, \Delta)$,  
where $Y$ is a literal, and $\Delta$ is an 
associated time interval called the {\em validity period} in 
which $Y$ should hold. 
We outline a new algorithm that 
computes \emph{closure($X, \Sigma_Q$)}.  \blue{In contrast to \gfd \ implication,  we must compute the sub-intervals on which derived literals are expected to hold after applying each \tgfd.  The increased space of literals is managed by indexing literals, and searching for overlapping time intervals at each apply step.}
\eat{
Next, we need to compute the subsumed set $\Sigma_{Q}$ of $\Sigma$, where $Q$ is the graph pattern of $\sigma$. $\Sigma_{Q}$ is different from $\Sigma_{Q,\Delta}$ from the satisfiability section, where $\Sigma_{Q}$ is only based on the graph pattern $Q$ of a given $\sigma$ (\emph{i.e.,} only the \tgfds that are embedded in $Q$) and also, $\Sigma_{Q}$ does not partition the set based on the time interval $\Delta$. After finding the $\Sigma_Q$ from the given set $\Sigma$ and the \tgfd $\sigma$, \fei{Why does $\Sigma_Q$ need to be dependent on $\sigma$ in general?} 
}
\vspace{-0.5em}
\begin{itemize}[nolistsep,align=left]
\item[(i)] \uline{Initialize}: for each literal $x \in X$, add $(x,\Delta)$ to \emph{closure}($X,\Sigma_Q$). 
\item[(ii)] \uline{Apply}: for each $\sigma'$ $\in \Sigma_{Q}$, where $\sigma'$ = $(Q'[\bar{x}]$, $\Delta'$, $X' \rightarrow Y')$, if all literals $x' \in X'$ can be derived via transitivity of equality of values 
from the literals 
in  \emph{closure}($X, \Sigma_Q$), 
then add $(Y', \Delta' \cap \Delta'')$ to \emph{closure}($X, \Sigma_Q)$.
\item[(iii)] \uline{Merge}: for a literal $Y''$ and all $\{(Y''$, $\Delta_1'')$, $\dots ,(Y'',\Delta_m'')\}$ in the \emph{closure}($X, \Sigma_Q$), merge the time intervals and replace the pairs with a single pair $(Y'',(\Delta_1'' \cup ... \cup \Delta_m'')$).
\end{itemize}

The above process is in \PTIME. 
We say a literal $Y$ is \emph{deducible} from $\Sigma$ and $X$, if there exists a $\Sigma_Q$ derived from $\Sigma$, and a pair $(Y, \Delta'')$ $\in$ \emph{closure($X, \Sigma_Q$)} such that $\Delta \subseteq \Delta''$. 

\begin{lemma}
\label{lm-imp}
Given a set of \tgfds $\Sigma$ and a \tgfd 
$\sigma$ = $(Q(\bar x), \Delta, X\rightarrow Y)$, 
$\Sigma \models \sigma$ if and only if $Y$ is deducible from $\Sigma$ and $X$. 
\end{lemma}







\begin{theorem} \label{thrm:implication}
The implication problem for \tgfds is \npc.
\end{theorem}

\tr{
\begin{proofS}
We provide an 
\np algorithm that, given  $\Sigma$ and $\sigma$ = 
$(Q(\bar x), \Delta, X\rightarrow Y)$,  
guesses a subset $\Sigma'$ of $\Sigma$, 
and a mapping from the patterns of each \tgfd in $\Sigma'$ 
to the pattern of $\sigma$, to verify 
if $\Sigma'$ is the embeddable set of \tgfds 
in $\Sigma_Q$. If so, it invokes the 
aforementioned procedure to compute 
the temporal closure of $\Sigma_Q$, 
and verify if $Y$ is deducible. 
Specifically, it checks the validity period of the enforced literals during the Apply step, i.e., checking whether $(\Delta' \cap \Delta'') \neq \emptyset$ for time intervals $\Delta', \Delta''$ from $\sigma'\in\Sigma_Q$, and literal $(Y'', \Delta'')$, respectively.  The verification is in \ptime for a finite number of literals in the closure and $|\Sigma_Q|$. 
For the lower bound, the implication of \tgfd is 
\np-hard by reduction from the implication of \gfds, 
which is known to be \npc~\cite{fan2016functional}. 
\eat{
The proof follows by extension of its \gfd counterpart, where implication has been shown to be \npc~\cite{fan2016functional}.  For \tgfds, we check the validity period of the enforced literals during the Apply step, i.e., checking whether $(\Delta' \cap \Delta'') \neq \emptyset$ for time intervals $\Delta', \Delta''$ from $\sigma'$, and literal $(y'', \Delta'')$, respectively.  This verification can be computed in \ptime for a finite number of literals in the closure and $|\Sigma_Q|$, and does not increase the complexity bound. 
}
\end{proofS}
}


\input{axioms}

\vspace{-2ex}
\subsection{Validation}
\eat{
Given $\Sigma$, and a temporal graph \tempgraph, the validation problem is to decide whether \tempgraph \ $\models \Sigma$. For each $\sigma=(Q[\bar{x}], \Delta, X \rightarrow y) \in \Sigma$, we compute the pairs of matches $(h_{t_i}(\bar{x}),h_{t_j}(\bar{x}))$ in \tempgraph \, such that if $(h_{t_i}(\bar{x}),h_{t_j}(\bar{x})) \models X$, then $(h_{t_i}(\bar{x}),h_{t_j}(\bar{x})) \models y$, for $|t_j - t_i| \in \Delta$.   If this is not true, i.e.,  $(h_{t_i}(\bar{x}),h_{t_j}(\bar{x})) \not\models y$, \eat{or  $|t_j - t_i| \not \in \Delta$,} then $(h_{t_i}(\bar{x}),h_{t_j}(\bar{x}))$ is a violation of $\sigma$. 
We add such violations $(h_{t_i}(\bar{x}),h_{t_j}(\bar{x}))$  to an error set $\mathcal{E}($\tempgraph$, \Sigma)$.  
The validation problem is to answer the decision problem of whether any violations exist, i.e., is $\mathcal{E}($\tempgraph$, \Sigma)$  empty.  
}

\vspace{-1ex}
Given $\Sigma$, and a temporal graph \tempgraph, the validation problem is to decide whether \tempgraph \ $\models \Sigma$. A practical application of validation 
is to detect {\em violations} of $\Sigma$ in \tempgraph. 
We say a pair of matches 
\wu{$(h_i, h_j)$ is a violation (``error'')}
of a \tgfd $\sigma$ = $(Q(\bar x), \Delta, X\rightarrow Y)\in\Sigma$, if 
\wu{$(h_i, h_j)$ matches $\sigma$ and 
$(h_i, h_j)\not\models Y$.}
We denote the set of violations of $\Sigma$ 
in \tempgraph \ as $\mathcal{E}($\tempgraph$, \Sigma)$. 
The validation problem is to decide whether  $\mathcal{E}($\tempgraph$, \Sigma)$ is empty. 


\begin{theorem}
The validation of \tgfds is \conpc.
\end{theorem}

\tr{
\begin{proofS}
The 
lower bound can be verified from 
the GFDs counterpart, where the validation problem is \conpc~\cite{fan2016functional}.
For \tgfds, an \np algorithm returns ``yes'' if \tempgraph$\not\models\Sigma$ as follows. 
It (1) guesses a \tgfd $\sigma$, and 
guesses and verifies a pair of mappings $(h_{t_i}(\bar{x}),h_{t_j}(\bar{x}))$ 
over a finite number of sub-intervals $\{[t_i, t_j] \mid 1 \leq i, j \leq T,  |t_j - t_i| \in \Delta\}$, and 
  (2) checks  whether $(h_{t_i}(\bar{x}),h_{t_j}(\bar{x})) \models X$, but $(h_{t_i}(\bar{x}),h_{t_j}(\bar{x})) \not \models y$, if so, return ``yes''.  Checking all matches over the intervals is in \ptime.  Since the complement to the decision problem still remains in \np, the validation problem for \tgfds is \conpc, no harder than its \gfds counterpart.
\end{proofS}
}

\vspace{-0.2cm}

%% file: axioms.tex
\subsection{Axiomatization}
\label{sec:axioms}
\vspace{-4pt}
\blue{We present an axiomatization for \tgfds.  The first five axioms also apply to \gfds (no axioms were defined for \gfds~\cite{fan2016functional}).  Our axiomatization is sound and complete~\cite{AMCW21}.} 
\vspace{-2pt}

\begin{axiom} {(Literal Reflexivity)} \label{axiom:reflex} For a given $Q[\bar{x}]$, $X,Y$ are sets of literals, if $Y \subseteq X$, then  $X \rightarrow Y$.
\end{axiom} 
\vspace{-0.5em}

In Axiom~\ref{axiom:reflex}, a set of literals $Y$ that is a subset of literals $X$, will induce a trivial dependency $X \rightarrow Y$ for any $\Delta$.

\begin{axiom} {(Literal Augmentation)} \label{axiom:subset_X} If  $\sigma'$ = $(Q[\bar{x}], \Delta, X' \rightarrow Y)$, $\sigma$ = $(Q[\bar{x}], \Delta, X \rightarrow Y)$, and $X' \subseteq X$, then $\sigma' \models \sigma$.
\end{axiom}

If $X' \rightarrow Y$ holds, then literals in $X', Y$ are in \emph{closure($X, \Sigma_Q$)}.  Since $X' \subseteq X$, we can derive $Y$, and $\sigma$ holds.

\begin{axiom} {(Pattern Augmentation)} \label{axiom:subgraph}
If  $\sigma'$ = $(Q'[\bar{x}'], \Delta, X \rightarrow Y)$, and $\sigma$ = $(Q[\bar{x}], \Delta, X \rightarrow Y)$,   $Q' \subseteq Q$, then $\sigma' \models \sigma$.
\end{axiom} 

In Axiom~\ref{axiom:subgraph}, if \eat{we extend $Q'$ with new nodes and edges such that} $Q'$ is isomorphic to a subgraph of $Q$, and \blue{if $\sigma'$ (with pattern $Q'$) holds, it will continue to hold under $Q$.}

\begin{axiom} {(Transitivity)} \label{axiom:trans}
If $\sigma'=(Q'[\bar{x}'], \Delta, X \rightarrow W)$, $\sigma =(Q[\bar{x}], \Delta, W \rightarrow Y)$, where $Q' \subseteq Q$, 
then for any $\sigma'' =(Q[\bar{x}], \Delta, X \rightarrow Y)$, it follows that $\{\sigma, \sigma'\} \models \sigma''$.
\end{axiom}

In Axiom~\ref{axiom:trans}, all matches that satisfy $\sigma'$ will also be contained within matches satisfying $\sigma$ since $Q' \subseteq Q$.  By transitivity of equality \wrt the literals in $W$, these  matches of $Q$ will satisfy $X \rightarrow Y$, thereby showing $\{\sigma, \sigma'\} \models \sigma''$.

\begin{axiom} {(Decomposition)} \label{axiom:decomposition}
If  $\sigma=(Q[\bar{x}],\Delta,X \rightarrow Y)$ with $Y=\{l_1,l_2\}$, then for $\sigma'=(Q[\bar{x}],\Delta,X \rightarrow l_1)$ and $\sigma''=(Q[\bar{x}],\Delta,X \rightarrow l_2)$, it follows that $\sigma \models \{\sigma', \sigma''\}$.
\end{axiom}

Verifying the literals in $Y$ can be done simultaneously in one verification (via $\sigma$), or in conjunction (via $\sigma', \sigma''$). 

\begin{axiom} {(Interval Intersection)} \label{axiom:intoverlap}
If $\sigma=(Q[\bar{x}],(p, q), X \rightarrow Y)$, $\sigma'$ = $(Q[\bar{x}]$, $(p', q')$, $X \rightarrow Y)$,
then for $\sigma''$ = $(Q[\bar{x}]$, $(p'', q'')$, $X \rightarrow Y)$, where $(p'', q'') = (p,q) \cap (p', q')$, it follows that $\{\sigma, \sigma'\} \models \sigma''$.
\end{axiom}

For $\sigma, \sigma'$ with matches satisfying $X \rightarrow Y$ over $\Delta, \Delta'$, respectively, requires verifying all pairwise matches over all sub-intervals $\Delta'' \subseteq (\Delta \cap \Delta')$, thereby showing  $\{\sigma, \sigma'\} \models \sigma''$.

\begin{axiom} {(Interval Containment)} \label{axiom:intsubsume}
If  $\sigma$ = $(Q[\bar{x}], \Delta, X \rightarrow Y)$, and  $\sigma'$ = $(Q[\bar{x}], \Delta', X \rightarrow Y)$, $\Delta' \subseteq \Delta$, then $\sigma \models \sigma'$.  
\end{axiom}

If $\sigma$ holds over $\Delta$, then it will hold over any subsumed $\Delta'$ that is more restrictive.  Pairwise matches over all sub-intervals $\Delta' \subseteq \Delta$ must also satisfy the dependency, thus $\sigma'$ holds.  

\begin{theorem}
The axiomatization is sound and complete.
\end{theorem}

\tr{
\begin{proof}
 Axioms~\ref{axiom:reflex}-\ref{axiom:decomposition} are sound as described above.  To show completeness of Axioms~\ref{axiom:reflex}-\ref{axiom:trans}, recall that the \emph{closure($X, \Sigma_Q$)} is defined over all embedded \tgfds for a pattern $Q$, i.e., by Axiom~\ref{axiom:subgraph}, if $\sigma'$ holds over $Q' \subseteq Q$, then $\sigma$ holds for pattern $Q$.  In computing \emph{closure($X, \Sigma_Q$)}, inference of valid time intervals is done by checking Axiom~\ref{axiom:intoverlap}, where the interval overlap is non-empty.  For $\sigma= (Q[\bar{x}],\Delta,X\rightarrow y)$ can be inferred from $\Sigma$, if and only if there exists a literal $(y, \Delta'')$ $\in$ \emph{closure($X, \Sigma_Q$)} such that $\Delta \subseteq \Delta''$.  
 
To show completeness, it is sufficient to show that for any $\sigma$, if $\Sigma \not \models \sigma$, then there exists a \tempgraph \ such that \tempgraph $\models \Sigma$, but \tempgraph $\not \models \sigma$, i.e, $\Sigma$ does not logically imply $\sigma$.  First, we show that \tempgraph $\models \sigma'$, for all $\sigma' \in \Sigma$, and then show $\sigma$ is not satisfied by \tempgraph.  Suppose $\sigma'= (Q[\bar{x}],\Delta, V \rightarrow W)$ is in $\Sigma$ but not satisfied by \tempgraph.  Then $(V, \Delta) \in$ \emph{closure($X, \Sigma_Q$)}, and $(W, \Delta)$ nor any subset of literals of $W$ can be in \emph{closure($X, \Sigma_Q$)}, otherwise, $\sigma'$ would be satisfied by \tempgraph.  Let $(w, \Delta)$ be a literal of $W$ not in \emph{closure($X, \Sigma_Q$)}.  We have $\sigma''= (Q'[\bar{x}],\Delta, X \rightarrow V)$, since $V$ is in the closure, and by Axiom~\ref{axiom:subgraph} and Axiom~\ref{axiom:trans}, we have $\sigma'''= (Q[\bar{x}],\Delta, X \rightarrow W)$.  By Axiom~\ref{axiom:reflex}, we can infer $W \rightarrow w$ over the interval $\Delta$, and by transitivity w.r.t. $W$ over pattern $Q$, we have $X \rightarrow w$, which implies that $(w, \Delta) \in$ \emph{closure($X, \Sigma_Q$)}, which is a contradiction.  Hence, \tempgraph $\models \Sigma$, for all $\sigma' \in \Sigma$. 
Second, we show that for a $\sigma$, if $\Sigma \not \models \sigma$, then \tempgraph $\not \models \sigma$.  Let's suppose \tempgraph $\models \sigma$, then $(y, \Delta)$ $\in$ \emph{closure($X, \Sigma_Q$)}.  However, if this is true, then $\Sigma \models \sigma$, which is a contradiction.  Hence, \tempgraph $\not \models \sigma$, and the axiomatization is complete. 
\end{proof}
}

%% file: parallel.tex
\section{Parallel \tgfd Error Detection}
\label{sec:parallel}

We introduce a parallel \tgfd error detection algorithm, \tedparallel, that includes fine-grained workload estimation, temporal pattern matching with match maintenance over evolving graph fragments, and incremental error detection that avoids all pairwise match enumerations.  


\eat{
For an increasing number of \tgfds, there will be increasing similarity in data accesses among the graph pattern queries.  Past work have shown that queries over social network graphs share over 82\% spatial similarity~\cite{graphm}.  In addition, the same subgraphs may be accessed by different queries over the same time interval.  To reduce this data access overhead among graph pattern queries, we exploit this spatial and temporal locality, and propose a parallel algorithm.  
}

\stitle{A Sequential Algorithm}. 
Given a \tgfd $\sigma$ = 
$(Q(\bar x), \Delta, X\rightarrow Y)$, a  
sequential algorithm computes \errorSigma \ as follows. 
(1) For each snapshot \snapshot $\in$ \tempgraph, 
it computes from scratch all matches of $Q$ in \snapshot, and those in the subsequent snapshots $G_j$ \wu{where $|j - t| \in \Delta$ (if not computed yet). 
(2) It then verifies, for each pair of 
matches of $Q$ $\{(h_i, h_j)\}$  with 
$|j - i| \in \Delta$,  
if $(h_i, h_j)\models X$ 
(\ie $(h_i, h_j)$ matches $\sigma$), 
and $(h_i, h_j)\not\models Y$. 
\eat{
\ie the set of pairs that 
match $\sigma$, 
to verify if $(h_{i}(\bar{x}),h_{j}(\bar{x}))\models X$ and $(h_{i}(\bar{x}),h_{j}(\bar{x}))\not\models Y$.
}
If so, a violation $(h_i, h_j)$ is identified 
and added 
to \errorSigma. 
This approach separately 
performs pattern matching and 
\tgfd validation, and is 
infeasible 
for large graphs due to 
an excessive number of 
subgraph isomorphism tests 
(which is known to be \npc~\cite{GJ90}), 
and comparisons 
over each pairs of snapshots.}
As most changes in 
real-world temporal graphs 
affect a small portion (\eg 5-10\% 
of nodes weekly~\cite{FJJ11})
and 80-99\% of the changes are highly localized~\cite{BBT18}, this naturally motivates incremental and parallel solutions 
for \tgfd-based error detection.

We next present \tedparallel, a \emph{\uline{Parallel} \uline{T}GFD \uline{E}rror \uline{D}etection} algorithm, for 
error detection in large \tempgraph. 
\wu{
The algorithm fully exploits 
the temporal interval constraints to 
interleave parallel 
 pattern matching 
and incremental error detection, 
in at most $T$ rounds of 
parallel computation.}

\stitle{Overview}. 
\wu{
The algorithm \tedparallel (Algorithm~\ref{alg:tedparallel}) 
works with  
a set of $n$ worker machines $M_1 \ldots M_n$ and a coordinator 
$M_c$. 
It executes in total 
$T$ supersteps. 
Each superstep $t$ 
processes a 
 {\em fragmentation} of a snapshot 
$G_t$ as a set of 
subgraphs \fragments of 
$G_t$ ($t \in [1, T]$ and $r \in [1, n]$), 
Intuitively, we  
perform parallel computation to 
detect and maintain the set of 
violations 
($\mathcal{E}$($\{G_1, \ldots, G_t\}, \Sigma)$. )
over ``currently observed'' snapshots 
$\{G_1, \ldots, G_t\}$, at each superstep $t$. 
The algorithm terminates 
after $T$ supersteps, and ensures 
the correct computation of 
$\mathcal{E}($\tempgraph$, \Sigma)$ 
given the correct update of 
violation set at each superstep. 
 
\stitle{Algorithm}. 
\tedparallel maintains the following 
structures that are dynamically updated. 
(1) Each worker $M_r$ maintains 
(a) a set of local 
matches of $Q$ of snapshot $G_t$ 
$M_r(Q, G_t)$, 
(b) a set of local violations 
$\mathcal{E}_r($\tempgraph, $\Sigma)$, 
by only accessing local graphs 
(initialized as fragment \fragments), 
and (c) a set of 
``cross-worker'' 
violations $\mathcal{E}_r($\tempgraph, $\Sigma)$, 
which requires 
the comparison of two matches 
from $M_r$ and a different worker. 
The coordinator maintains 
a set of global 
violations $\mathcal{E}($\tempgraph,$\Sigma)$ 
to be assembled from local violations. 

\tedparallel
first invokes a procedure~\kw{GenAssign} to initialize and assigns a set of joblets (a set of lightweighted validation tasks; see Section~\ref{sec-workload}) to all the workers, and initializes 
global structures for 
incremental maintenance of 
the pattern matches (see Section~\ref{sec:querymatch}) (lines 1-2).  
This cold-starts the parallel error detection 
upon the processing of the first fragmented snapshot 
in \tempgraph. 
It then runs in $T$ supersteps 
in parallel (lines~3-10), following a bulk 
synchronous model, and 
processes one fragmented snapshot a time. 

 In each superstep $t$, each $M_r$ 
 executes the following two steps in parallel (lines 4-7):  
 (1) incrementally updates a set of local matches $M_r(Q, G_t)$ and 
violations $\mathcal{E}_r($\tempgraph,$\Sigma)$  over fragmented snapshot \fragments 
by invoking \kw{LocalVio}
(line 4); and
(2) 
 requests a small amount of edges from 
other workers, and invokes~\kw{IncTED} to incrementally 
detect the violations \errorSigmaC \ across 
two workers (line 5; see Section~\ref{sec:querymatch}). 
$M_r$ then returns the updated 
local violations (augmented with 
cross-worker violations) to $M_c$ (lines~6-7).  The coordinator $M_c$
incrementally maintains a set of global violations upon receiving 
the updated local 
violations (line 8), and  
invokes~\kw{GenAssign} to rebalance the workload for the next superstep (lines~9-10) 
upon a triggering condition. 
This validates 
$\Sigma$ over the snapshots 
$\{G_1, \ldots, G_t\}$ 
``seen'' thus far.}

\stitle{Optimizations.} 
\wu{
To 
cope with skewed localized updates that lead to stragglers and reduce communication overhead, 
\tedparallel adopts two strategies 
below.  

\sstab
(1) An adaptive time-interval aware workload balancing strategy (procedure~\kw{GenAssign}; line~6) to  
automatically (re-)partition 
the \tgfd validation workload 
to maximize parallelism. 
The idea is to 
decompose validation tasks to 
a set of highly parallizable, 
small subtasks that are 
induced by 
path patterns, 
time-intervals, and 
common literals 
(``Joblets''), and 
dynamically 
estimates a bounded subgraph 
induced by the joblets 
to be assigned and executed 
in parallel. 

\sstab
(2) \tedparallel also adopts  
an incremental pattern matching 
scheme (Algorithm~\kw{IncTED}; line~5) to maintain 
the local violations. The 
strategy performs case analysis of 
edge updates and only 
processes a necessary amount of 
validation, to reduce the 
communication and local 
detection cost. 
} We next introduce the details of \kw{GenAssign} and 
\kw{IncTED}. 

\eat{
We illustrate \tedparallel in Algorithm~\ref{alg:tedparallel}, and 
}


\subsection{Time interval-Aware Workload 
Balancing}
\label{sec-workload}

\input{Algorithms/parallelTED}

\vspace{-2ex}
\tr{Large, evolving graphs pose additional efficiency challenges over static graphs, where delta changes between snapshots need to be incorporated.} 

\noindent Given a set of \tgfds $\Sigma$ and 
\tempgraph, procedure~\kw{GenAssign} 
creates a set of ``joblets'' and 
estimates their processing cost for 
balanced workload assignment (as illustrated 
in Algorithm 1). 

\stitle{Joblets and Jobs}. 
A \emph{joblet} characterizes a small \tgfd validation task that can be conducted by a worker 
in parallel. A joblet at 
superstep $t$ is a triple \joblet, 
where 
\begin{itemize}
  \makeatletter\@topsep0pt\makeatother
    \item  $Q_k(v_k, d)$ is a sub-pattern of a pattern $Q$ with a designated center node $v_k$ and a radius $d$ \wrt $v_k$, for a \tgfd $\sigma$ = $(Q[\bar x], \Delta, X\rightarrow Y)$ in $\Sigma$; 
    \item \fragment~is a fragment of snapshot $G_t$ on worker $M_r$; and  
    \item \tempgraphk = $\bigcup_{s,s'\leq t}\{ \{G_s(v',d), G_s'(v',d)\}||s' - s| \in \Delta\}$, 
    where $v$ and $v'$ have the same label, and 
    $G_s'(v',r)$ is a subgraph of $G_s$ 
    induced by $v'$ and its $d$-hop neighbors. 
\end{itemize}

Intuitively, a joblet encodes a fraction of a validation task to 
identify violations of a \tgfd $\sigma\in\Sigma$.  
It is dynamically induced by incorporating a small fragment of $Q$ and only 
the relevant fraction of snapshots in 
\tempgraph\ that should be checked to detect  violations 
of $\sigma$ with pattern $Q$ 
and time interval $\Delta$.

\eat{
``time-interval aware'' task that processes a 
single pattern path $Q_k$ on fragment \fragment over \tempgraphk = $\{G_1(v',d), \ldots , G_{i}(v',d)\}$, $v' \in $ \candQk, such that $\forall s, s' \leq i, |s' - s| \in \Delta$.  The fragment \fragment \ contains node $v'$, and \tempgraphk \ are the induced subgraphs $G_t(v',d)$ $d$-hops around $v'$ for diameter $d$ of $Q_k$ for $t \leq i$. 
}


\wu{
A {\em job} \job \ refers to a set of 
joblets that encode the workload 
to validate a \tgfd $\sigma$ with pattern $Q$.  A job contains all joblets for all subqueries $Q_k$ 
from $Q$.  
We denote as \jobSigma~the jobs for validating a set of \tgfds $\Sigma$, \ie \jobSigma = $\cup_{\sigma \in \Sigma}$ \job.
}

We next introduce a path decomposition strategy 
adopted by \kw{GenAssign}. The joblets are 
created accordingly for a given 
decomposition of $Q$ = $\{Q_1, \ldots Q_k\}$.

\begin{figure}
    \centering
    \includegraphics[width=3.49in,keepaspectratio]{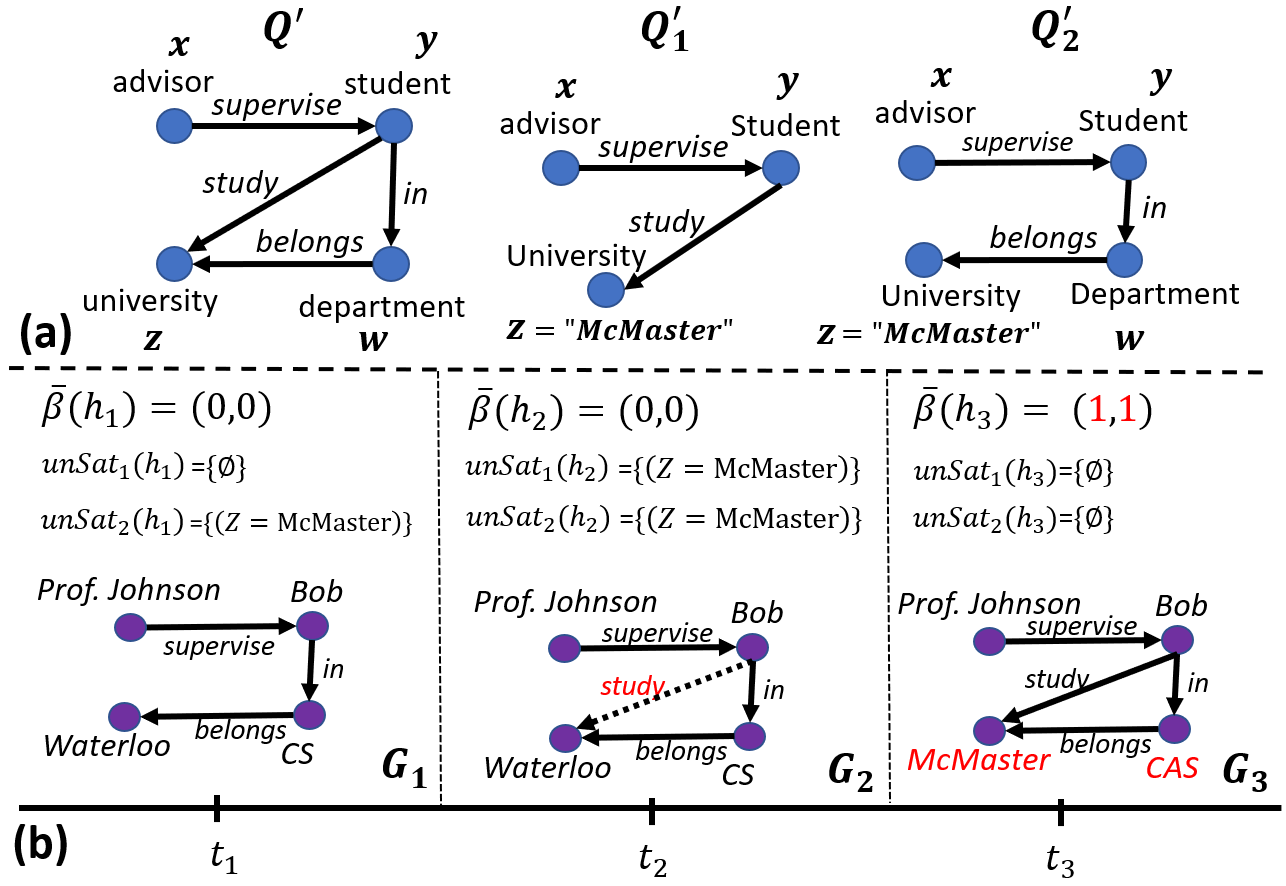}
    \vspace{-0.5cm}
    \caption{Pattern paths and partial matches.}
    \vspace{-4ex}
    \label{fig:matchingExample}
\end{figure}

\stitle{$K$-Path decomposition} (not shown in 
Algorithm 1).  
\wu{It has been shown that the cost of 
graph query processing can be effectively 
estimated using path queries~\cite{SRK07}.
\kw{GenAssign} adopts a $K$-path pattern decomposition strategy for 
joblet creation and cost estimation. 
For each $\sigma \in \Sigma$, \kw{GenAssign} decompose a pattern $Q$ into a set of paths $\{Q_1, \ldots Q_K\}$, such that 
$Q$ is the union of $Q_k$ $(k\in[1,K])$. 
Each $Q_k$ is a maximal path from a 
 node $v_{k_1}$ to a destination node $v_{k_m}$   that cannot be further 
extended by pattern edges. 
Each $Q_k$ is augmented with value constraints posed by the literals from $X\cup Y$. 

Let $v_k$ be a center node with minimum radius in each path pattern $Q_k$ ($k\in[1,K]$), where the radius $d$ is the longest shortest path between $v_k$ and any node in $Q_k$. The joblets are 
then created with $K$-path decomposition 
and induced subgraphs \wrt $v_k$ and $Q_k$ $(k\in[1,K])$ accordingly. 
\reviseOne{Figure~\ref{fig:matchingExample}(a)} shows two (maximal) pattern paths $Q_1'$ and $Q_2'$ of the pattern $Q'$ with the literal set $\{z=$\lit{McMaster}$\}$.  
}

\stitle{Workload Estimation} (line~4 of Procedure~\kw{GenAssign}).
\label{sec:estimation}
The workload is estimated as the size of joblets/jobs, and the expected communication cost among the workers.  
\eat{
Given these estimates, we derive an initial assignment of jobs to machines that balances the workload across the machines.  
To compute \errorSigma, we compute the matches of $Q$ for all $\sigma \in \Sigma$.  Lastly, we adjust our job estimations and workload assignments to adapt to changes in \tempgraph.}
To estimate the joblet size, %
we 
adapt the cardinality estimation for graph queries~\cite{SRK07}. We compute a probability distribution of the expected number of matches for each edge $(v_l,v_{l+1})$ in $Q_k$ with label 
$a_l$.  The distribution provides a probabilistic estimation of the number of matches of $v_{l+1}$, given the matches of $v_l$ that are connected via an edge 
with label $a_l$ and satisfy the literals over $v_l$ and $v_{l+1}$. 
We estimate using the mean and standard deviation of the number of matches of $v_{l+1}$ \wrt $v_l$.  We compute the distribution function of $Q_k$ as the product of the distribution functions of each consecutive edge in $Q_k$. We estimate the size of a job $|$\job$|$ \ as the cardinality of the pattern $Q$, i.e.,  the number of matches of $Q$,  computed as the upper-bound of the minimum cardinality of all $Q_k$. 


\eat{
\begin{equation} \label{eq:jobletEstimation_mu}   \mu(v_l,v_{l+1})=\dfrac{1}{|v_l|}\sum_{a_l}  |(a_l, v_{l+1})|
\end{equation}
\begin{equation} \label{eq:jobletEstimation_VA}   \sigma^2(v_l,v_{l+1})=abs((\dfrac{1}{|v_l|}\sum_{a_l}  |(a_l, v_{l+1})|^2)-\mu^2(v_l,v_{l+1}))
\end{equation}
\begin{equation} \label{eq:jobletEstimation_VA}   D(v_l,v_{l+1})=\mu(v_l,v_{l+1})\pm k\sigma(v_l,v_{l+1})
\end{equation}
}

\eetitle{Communication cost estimation} (
line 10 of Algorithm~\tedparallel; line 4 of Procedure\kw{GenAssign}). 
For a joblet \joblet, fragment \fragment \ may contain a partial match of a path $Q_k$, \eat{over the induced subgraphs $G_t(v',d), t \leq i$,} i.e.,\eat{with the query path $Q_k$ that resides in $M_j$ at timestamp $i$, the fragment $F_{ij}$ may only contain a partial match, i.e.,} some nodes and edges that match $Q_k$ reside in machine $M_{r'}, r' \neq r$.  We must exchange these small number of edges between workers that we estimate as the communication cost of a joblet, defined as \eat{as the total number of such edges that must be transferred from all $M_j'$ to $M_j$,} $CCost(\omega_{trk})=\sum_{\substack{(v_1, v_2) \in G_t(v', d), \\ \neg((v_1, v_2) \in F_{tr})}}  |(v_1, v_2)|$.  
We aggregate over all the joblets containing pattern paths $Q_k$ of $Q$ to define the communication cost of a job, $CCost(\mathcal{J}_{tr}) = \sum_{\substack{k}} CCost(\omega_{trk})$.



\eat{
\begin{equation} \label{eq:datashipment_Joblet}
   CCost(\omega_{ijk})=\sum_{\substack{(v_1, v_2) \in G_t(v', d), \\ \neg((v_1, v_2) \in F_{ij})}}  |(v_1, v_2)|  
\end{equation}
}




\stitle{Workload Assignment and Rebalancing} (lines~9-10 of \tedparallel; line~5 of \kw{GenAssign}). \label{sec:InitialWorkloadAssignment} We distribute jobs to workers such that the communication cost $C$, and the makespan $\tau$ is minimized.  We solve a \emph{general assignment (GA) problem}~\cite{shmoys1993approximation}. Given $C$ and $\tau$, the GA problem is to find an assignment of each job 
such that the total parallel cost is bounded by $C$, and minimizes the makespan. Following~\cite{shmoys1993approximation}, 
we develop a pseudo-polynomial-time, 2-approximation algorithm.  It performs a bisection search on the range of estimated makespan and keeps assignments with an objective function value less than half of the current best 
solution. It solves a linear program at most $log(wP)$ times ($w$ is the number of jobs and $P$ is the job with maximum estimated workload) and selects the one with the smallest makespan.  We introduce the details in~\cite{AMCW21}. 

\tr{
Given lower and upper bounds $(t_l,t_u)$ for job runtimes, Shmoys and Tardos present a 2-approximation algorithm that computes a schedule with communication cost at most $C$, and makespan at most $2\tau$; as long as $t_l \leq t_{\mathcal{J}_{ij}} \leq t_u$.
We assume these time bounds are given as input.  To find an assignment that is at most $2\tau$, the algorithm performs a bisection search on the range of possible makespan values by keeping only those assignments with an objective function value less than half of the objective function value of the best assignment found thus far.
The algorithm solves the linear program of the GA problem at most $log(wP)$ times, where $w$ is the total number of jobs and $P$ is the maximum workload among all jobs~\cite{shmoys1993approximation}. Among all the current assignments, we select the assignment with minimum makespan.
}

The initial workload distribution guarantees the makespan is at most $2\tau$ provided $t_l \leq t_{\mathcal{J}_{tr}} \leq t_u, \forall t_{\mathcal{J}_{tr}}$, for given $(t_l, t_u)$.  Changes among the fragments can imbalance the workload, and increase the makespan $\tau$.  To avoid excessive workload re-distributions, we define a \emph{burstiness time buffer}, \timebuff, representing the allowable percentage change in job runtime due to workload bursts. We add \timebuff \ to $(t_l, t_u)$ to define $t'_{l} = ((1 -$ \timebuff) \ $\cdot \ t_{l})$, and $t'_{u} = ((1 +$ \timebuff) \ $\cdot \ t_{u})$.  If a job runtime, $t_{\mathcal{J}_{tr}}$, lies beyond  $(t'_{l}, t'_{u})$, $M_c$ triggers a workload redistribution.  In Section~\ref{sec:exp-adaptiveLoadBalance}, we study the overhead and frequency of re-distributions over real workloads for \timebuff \ $= 0.1$ under varying rates of change, and found the overhead is about  6\% of the total runtime.

\subsection{Parallelized Incremental Temporal Matching 
}
\label{sec:querymatch}

Another bottleneck is to compute the pattern matches over evolving and distributed fragments. 
We next introduce the procedures 
\kw{LocalVio} and \kw{IncTED} that efficiently  
maintain the matches in \tedparallel. 

\stitle{Incremental Local Matching}.
\wu{
The local violation computation (Procedure~\kw{LocalVio}, line~4 of Algorithm~\tedparallel) relies on 
fast computation of local pattern matches. 
At each superstep $t$, jobs are executed at each worker to compute matches of each query path $Q_k$ corresponding to $Q$.  For each $Q_k$ in 
a joblet, it performs subgraph isomorphism matching on \tempgraphk \ to find local matches. 
Matches that require edges from multiple workers are shipped to a single worker. 
The matches of $Q$ are obtained by 
(1) taking the intersection of the nodes identified 
by the matches across all paths $Q_k$ of $Q$, 
and (2) a verification to 
find true matches.  
}

\eetitle{Coping with edge updates.}\label{sec:queryMatchWithChanges}
\blue{New matches of $Q_k$ can appear and existing matches disappear (or be updated) as changes occur to fragment \fragment.  To adapt to these changes, we present an incremental matching strategy that avoids redundant computations, and tracks partial matches for each $Q_k$ with their missing topological and literal conditions.  
}
We introduce a boolean vector \vecmatchfull \ = $(b_1,\dots, b_{\kappa})$ over a candidate match \matchone \ of $Q$ in \fragment.  \vecmatchfull \ contains $\kappa$ components, where the $k$-th component is the matching status of \joblet.  That is, if there exists a match \matchone \ of $Q_k$ in \fragment, then \vecmatchfull \ = 1 (\kw{true}), otherwise, \vecmatchfull \ = 0 (\kw{false}).  In the latter case, a partial match \matchone \ may be isomorphic to  $Q_k$ but not satisfy its literals.  We define \unsatfull \ as the set of unsatisfied literals $x' \in \bar{x}$ in \matchone \ \wrt $Q_k$, \eg \unsatfull \ = $\{x'.A = c \mid h_{i}({x'}).A \neq c\}$.  We use  \vecmatch \ and \unsat \ when \matchone \ is clear from the context. 

Each worker initializes and populates \vecmatch \ and \unsat \ as matches are found.  As changes occur over the fragments, \tedparallel \ checks \vecmatch, and \unsat \ to determine if new matches arise, and updates existing matches to minimize the need for (expensive) subgraph isomorphism operations.  If the $k$-th component of \vecmatch \ equals \kw{false}, there are two cases to consider: (i) a topological match of $Q_k$ exists, but there are unsatisfying literals in \unsat; or (ii) an empty \unsat \ represents no topological match of $Q_k$ in \fragment. For a  job $\mathcal{J}_{tr}$, we evaluate the following cases for an input change $c$: 

\begin{enumerate}
   \makeatletter\@topsep0pt\makeatother
\item[(a)] \uline{attribute insertion/deletion/update:} \eat{since there is no change in the topology in \fragment\ (no need for subgraph isomorphism operations),} we check whether $c$ adds/removes literals from \unsat, and update \vecmatch \ to denote the insertion/deletion of a match \matchone \ \wrt $Q_k$.
\item[(b)] \uline{edge insertion:} we perform subgraph isomorphism matching to determine whether a partial match is upgraded to a complete match \wrt $Q_k$. 
\item[(c)] \uline{edge deletion:} \eat{an edge deletion in \fragment \ does not lead to new matches,} if $c$ causes a prior match \matchone \ \wrt $Q_k$ to be removed, we then set \vecmatch(\matchone)[$k$] \ to \kw{false}, and add any unsatisfying literals to \unsat(\matchone). 
\end{enumerate}

\conf{We present the details of the procedure LocalVio in~\cite{AMCW21}.} 
\tr{Algorithm~\ref{alg:LocalMatch} describes the details.  For attribute updates, we update \vecmatch \ and \unsat without performing any subgraph isomorphism operations (Lines 4-12). \eat{Removing an attribute cannot make a partial match to be a match, so it might only remove a match (Lines 9-11).} For edge insertions, we perform subgraph isomorphism along $Q_k$ to check for a new match (Lines 13-16). For edge deletions, which can only remove matches, we update $b_k$ for $Q_k$ \eat{However, removing an edge might only make a match to be disappeared and it is being handled by setting $b_k=0$, if necessary}(Lines 17-19). A match \matchone \ of $Q$ exists if $b_k=1$ for all $k$, and we add  \matchone \ to   \matchesJ{i} \ (Lines 20-22).  
}

\tr{
\input{Algorithms/QueryMatching}
}



\begin{example}\label{example:DefineVectors}
\reviseOne{Figure~\ref{fig:matchingExample}(b)} shows (partial) matches of patterns $Q'_{1}, Q'_2$.  At $t_1$, \vecmatch($h_{1}$) denotes no match of either $Q'_1$ nor $Q'_2$, while  \vecmatch($h_{1}$)[2] indicates a topological match of $Q'_2$ with non-empty $\kw{unSat}_2$ containing literal $z=$ \lit{McMaster}.  At $t_2$, with an edge insertion labeled \lit{study} from \lit{Bob} to \lit{Waterloo}, we perform a subgraph isomorphism matching and find a topological match of $Q'_1$.  However, \vecmatch($h_{2}(\bar{x})$) remains  \kw{false} due to the unsatisfying \lit{McMaster} literal in $\kw{unSat}_{1}(h_{2}$).  Lastly, the update at $t_3$ to \lit{University} from \lit{Waterloo} to \lit{McMaster} clears all unsatisfying literals, and updates \vecmatch($h_{3}$) \ to \kw{true} for $Q'_1$ and $Q'_2$, without requiring any additional matching.
 \end{example}

\input{Algorithms/IncTED}
\noindent Validating each $\sigma \in \Sigma$ 
may require us to store and pairwise compare 
all the local and cross-worker matches. 
We next introduce procedure \tedincremental, an \uline{Inc}remental \uline{T}GFD \uline{E}rror \uline{D}etection algorithm. 
The algorithm avoids the need to store 
the matches, and reduces the cost of the exhaustive pairwise match comparisons by 
performing efficient set difference operations. 

\stitle{Permissible Ranges}. Rather than 
exhausting the comparison of 
any pair of matches, \tedincremental 
verifies
\blue{match pairs with respect to a ``current'' time, $c = i$, and induces an allowable time range between \matchone \ and \matchtwo\  as defined by $\Delta$.}  We define \intone and \inttwo as follows: 

\vspace{-0.6cm}
\begin{equation} \label{eqn:intervalbounds}
\begin{split}
I_{h_{i}} & = [1, |T|-p] \\
I_{h_{j}} & = [(i + p), j'], \text{where } j' = 
\left\{
    \begin{array}{l}
      T,  (i + q) > T \\
      (i + q), o.w.
    \end{array}
\right.
\end{split}
\end{equation}
\vspace{-0.6cm}

\blue{
If matches \matchone, \matchtwo \ have the same values in $X$, we want to identify matches \matchtwo \ to compare against \matchone \ by defining the allowable time range.  We call this \permissrange,  the \emph{permissable range} of \matchone, as \permissrange = $\{j \mid |j - i| \in \Delta\}$. 

\stitle{Auxiliary Structures}. 
\wu{\tedincremental
uses the following notations and auxiliary structures. 
To avoid enumeration of all pairwise matches in \permissrange, we define a hash map, \hashmapX$(i)$, that partitions all matches of $Q$ according to their values in $X$ up to and including $i$.  \blue{Specifically, let \hashmapX $= \{$\matchtwo $\mid$ \matchone$.A =$ \matchtwo.$A, \forall$ (\matchone$.A$, \matchtwo.$A) \in X, i \leq j\}$}. We simply use \hashmapX \ when $i$ is clear from the context.   We sub-partition the matches in \hashmapX \ according to their distinct values in $Y$ to identify error matches with different consequent values within the permissable range.  We record the timestamps of such matches \hashmapX \ (resp. \hashmapXY) in \timehashmapX \ (resp. \timehashmapXY), \ie \timehashmapX = $\{j \mid $\matchtwo \ $ \in $ \hashmapX $\}$.  We define a mapping function that returns the \hashmapXY \ partition in which \matchone \ belongs as \mapfunc = \hashmapXY \ such that \matchone\  $\in$ \hashmapXY.}  For example, \rev{Figure~\ref{fig:mainExample}(b)} shows matches of $\sigma$ with \hashmapX $=\{h_{1}, \dots , h_{6}\}$, \timehashmapX$=\{t_1, \dots, t_6\}$, and \hashmapXY$=\{\{h_{1}, \dots, h_{5}\} \{h_{6}\}\}$, \timehashmapXY$=\{\{t_1, \dots, t_5\} \{t_6\}\}$.  These global structures and variables are maintained by the coordinator $M_c$ and shared among all workers. 

}

\stitle{Algorithm}. 
The algorithm \tedincremental is illustrated in Algorithm~\ref{alg:IncTED}.  
After initializing the local error set (line 1),
it iterates over the local matches and records the timestamps of each match in the auxiliary structures (lines 3-4). For each match $h_i$, the \permissrange \ is bookkept based on the timestamp of $h_i$ (line 5). For variable \tgfds, if there exists another match $h_j$ within the permissable range of $h_i$ such that $h_i$ and $h_j$ have different set of timestamps for \hashmapXY, then \tedincremental adds the pair to the violation set (lines 6-7). For constant \tgfds, if the match $h_i$ violates a constant literal in $Y$, then it adds $h_i$ to the violation set (lines 8-10).  \conf{More details (subroutines) are provided in~\cite{AMCW21}.}

\eat{
 For each $\sigma \in \Sigma$, \tedincremental \ incrementally adds each match \matchone \ to the hashmap of matched values so far \wrt $X$ and $Y$ for each \fragment.  If \matchone \ and \matchtwo \ share equal values in $X$ but different values in $Y$, where $|j - i| \in \Delta$, then \matchpair \ is in error, and is added with $(i, j)$  to the error set   \errorsigmaj \ for worker $M_r$.  We define a violation as a pair, \vio $=\{$\matchpair, $(i, j)\}$, such that \matchpair $\not \models \sigma$.  Let \intone and \inttwo represent the time intervals where matches \matchone \ and \matchtwo \ are found.  
 }



Using \tedincremental algorithm, each worker computes its local set of errors, \errorsigmaj.  For matches that span two workers, the coordinator \eat{ Matches from each $M_j$ are then sent to $M_c$, who annotates each \matchone \ with the originating machine $M_j$.} verifies matches $h_{i}, h_{{i'}}$ from $M_r$, $M_{r'}, r \neq r'$, respectively.  This is done by selecting matches from $\mu_{X}(h_{i}) =$ \hashmapXY$(h_{i}$) and $\mu_{X}(h_{{i'}}) =$ \hashmapXY$(h_{{i'}}$) such that $|i - {i'}| \in \Delta$, and computing $\{\mu_{X}(h_{{i'}})\} \setminus \{\mu_{X}(h_{i})\}$, \ie checking whether the set difference is non-empty.  If so, we add $\{(h_{i}, i), (h_{{i'}}, {i'}) \}$ to the cross-machine error set \errorSigmaC.  At each timestamp, we update \errorSigma \ with  \errorSigmaC \ $\bigcup_r$ \errorSigmaj. The incremental strategy improves the efficiency of a batch counterpart for \tgfd-based error detection by 3.3 times (see Section~\ref{sec:exp}). 
\tr{
\input{Algorithms/LocalVio}

Algorithm~\ref{alg:localVio} provides details of the matching and local error detection at each $M_j$.  We compute matches and errors over the first graph snapshot $G_1$ (Lines 4-5), and then incrementally compute matches (Line 9), and errors (Line 10) via \kw{LMatch} and  \kw{LError}, respectively.  The set of local matches and errors for each $M_j$ are sent to $M_c$ (Line 11).  
Algorithm~\ref{alg:tedparallel} shows \tedparallel \ which first estimates the jobs and communication cost before assigning jobs to the machines (Lines 4-7).  At each $t_i$, $M_c$ sends the changes (computed from log files) to all machine workers.  Each $M_j$ runs Algorithm~\ref{alg:localVio} to compute matches and the local error set, which is sent to the coordinator node after each timestamp to minimize idle time at $M_c$ (Lines 8-11).  Lastly, $M_c$  computes errors between $M_j, M_j'$ by comparing matches from $M_j'$ (Line 13) with those from $M_j$ (Lines 14-15), and returns the total error set (Line 16).
}

\begin{figure}
    \centering
    \includegraphics[width=2.7in,keepaspectratio]{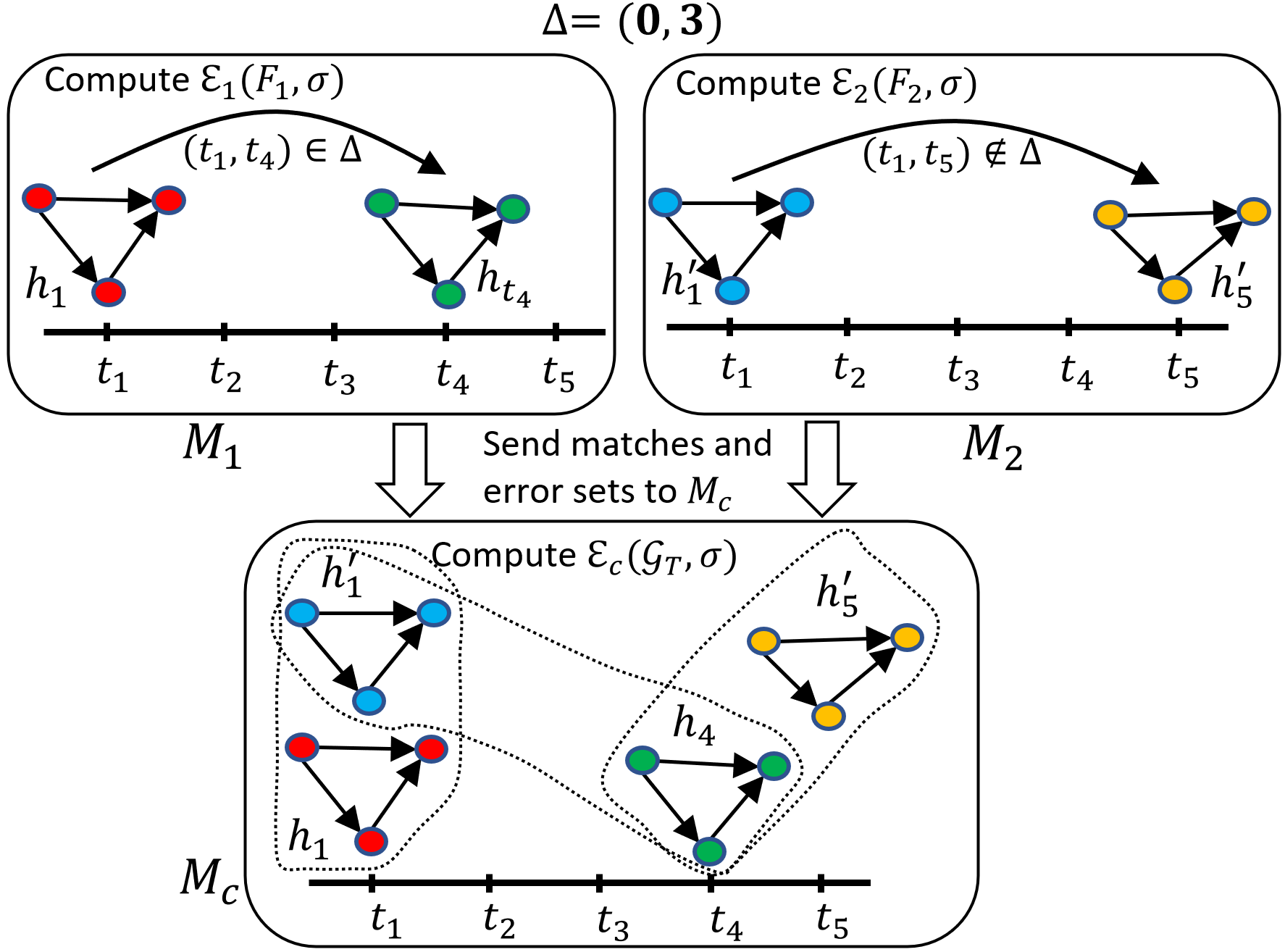}
    \vspace{-0.1cm}
    \caption{Error detection in \tedparallel}
    \label{fig:TEDParallel}
    \vspace{-3ex}
\end{figure}

\begin{example}\label{example:TEDParallel}
Figure~\ref{fig:TEDParallel} shows matches \{$h_{1}$, $h_{4}$\} and \{$h'_{1}, h'_{5}$\} for a $\sigma$ with $\Delta = (0,3)$ computed locally (via \kw{LMatch}~\cite{AMCW21}) at $M_1$ and $M_2$, respectively.  $M_1$ computes local error set $\mathcal{E}_1$ via \tedincremental, and $\mathcal{E}_{2} = \emptyset$ since $|t_5 - t_1| \not \in \Delta$.  $M_c$ receives $\{h_{1}, h'_{1}\}, \{h_{4}, h'_{1}\}, \{h_{4}, h'_{5}\}$, validates the pairs (shown by dotted lines) via \tedincremental, and adds the violations to \errorsigma += $\mathcal{E}_c($\tempgraph, $\sigma$) $\bigcup \mathcal{E}_1$.
\end{example}

\rev{\textbf{Runtime Analysis.} In Algorithm~\ref{alg:tedparallel}, \kw{GenAssign} runs in polynomial time taking at most $2\tau$ to assign joblets~\cite{shmoys1993approximation}.  The initial subgraph matching dominates the runtime (\kw{LocalVio}) as  an exponential number of subgraphs may match $Q$ for each $\sigma$.  Fortunately, subsequent, incremental matching takes linear time proportional to the number of changes at each worker due to our $K$-Path decomposition.  Since $|Q|$ is small in practice, and we localize the matching to small graphs making it less likely to find exponentially many isomorphic subgraphs.  At each superstep $t$, incremental violation detection (\kw{IncTED}) runs in linear time w.r.t. the number of matches in \snapshot.  \\
}




%% file: Algorithms/parallelTED.tex
\begin{algorithm}[tb]
\kw{Initialize} \hashmapX, \hashmapXY, \timehashmapX, \timehashmapXY;\\
\kw{GenAssign}(\tempgraph, $\Sigma$);\\
\ForEach{$t \in T$}
{ 

        \nonl/* at worker $r$ in parallel */\\
         \wu{\errorSigmaj := \errorSigmaj $\cup$ \ $\kw{LocalVio}($\fragment$,$\simplejob$, t)$; }\\
         \wu{\errorSigmaC := \errorSigmaC $\cup$ \kw{IncTED}($G_{t}$, $\Sigma$, \matchesJ{t}$)$;}\\
         \wu{\errorSigmaj := \errorSigmaj $\cup$ \errorSigmaC; \\
         \KwRet{\errorSigmaj}; }\\
    \nonl/* at the coordinator side */\\
    \wu{\errorSigma \ := \errorSigmaC $\cup$~ $\bigcup_r$ \errorSigmaj;} \\
    \If{($t_{\mathcal{J}_{tr}}<((1-\zeta) \cdot t_l$))$\lor$($t_{\mathcal{J}_{tr}} > ((1+\zeta) \cdot t_u$)) }
    { \kw{Assign}(\simplejob, $CCost$); /* rebalance workload */ \\   
    }
}
\KwRet{\errorSigma};\\

\vspace{1ex}
\nonl \textbf{Procedure} \kw{GenAssign}(\tempgraph, $\Sigma$)\\
\nonl/* at the coordinator side */\\
\setcounter{AlgoLine}{0}
\errorSigma := $\emptyset$; \errorSigmaC := $\emptyset$; $CCost:= \emptyset$;\\ 
\ForEach{$\sigma \in \Sigma$} 
{
    Define \job \ and estimate  $|$\simplejob$|$; \\ 
    Estimate communication cost  $CCost($\simplejob$)$ \\ 
}
\kw{Assign}(\simplejob, $CCost$);\\
Send \simplejob \ across workers $M_r$;\\

\caption{\tedparallel(\tempgraph, $\Sigma$)}
\label{alg:tedparallel}
\end{algorithm}

%% file: Algorithms/QueryMatching.tex
\begin{algorithm}[tb]
\matchesJ{i} := $\emptyset$;\\
Initialize \vecmatch \ and \unsat \ from \matchesJ{i-1};\\
$\mathcal{M}_{chg}$ = $\{h_{t_{i-1}}(\bar{x}) \in$ \matchesJ{i-1} $|$ c applied to $G_{t-1} \}$ \\
\If{$c$ is \kw{insert/update}$(x.A=a)$} 
{
    \If{$b_k=0$ and $(x.A=a) \in $\unsat} 
    {
        Remove $(x.A=a)$ from \unsat;\\
        \textbf{if} \unsat$=\emptyset$ \textbf{then}
            $b_k=1$;
    }
    \ElseIf{$b_k=1$ and $(x.A=b)\in Q_k$} 
    {
        Add $(x.A=b)$ to \unsat; $b_k=0$;
    }
}
\ElseIf{$c$ is \kw{delete}$(x.A=a)$} 
{
    \If{$b_k=1$ and $(x.A=b)\in Q_k$} 
    {
        Add $(x.A=b)$ to \unsat; $b_k=0$;
    }
}
\ElseIf{$c$ is \kw{insert}$(e)$} 
{
        Add $e$, do subgraph isomorphism to evaluate $b_k$;\\
        \If{$c$ creates a new match for $Q_k$} 
        {
            Duplicate all vectors \vecmatch \ and  
            set $b_k=1$;
        }
}
\ElseIf{$c$ is \kw{delete}$(e)$} 
{
    \If{\vecmatchfull[$k$] = 1} 
    {
        $b_k=0$; \unsatfull$=\emptyset$;
    }
}
\ForEach{ \matchone \ $\in$ $\mathcal{M}_{chg}$}  
{
   \If{(\vecmatchfull[$k$] = 1, $\forall k$)}
    {
     \matchesJ{i} += \matchone; 
    }
}
\KwRet{\matchesJ{i}};
\caption{\kw{LMatch}($\mathcal{J}_{ij}$, \matchesJ{i-1}, $F_{ij}$, $c$)} 
\label{alg:LocalMatch}
\end{algorithm}

%% file: Algorithms/IncTED.tex
\begin{algorithm}[tb]
\errorsigma \ := $\emptyset$; \\
\ForEach{\matchone $\in$ \matches}
{
    $\{$\hashmapX, \hashmapXY$\}$ := $\{$\hashmapX, \hashmapXY$\}$ $\cup$ \ \matchone \\
    $\{$\timehashmapX, \timehashmapXY$\}$ := $\{$\timehashmapX, \timehashmapXY$\}$ $\cup$ \ $i$; \\
    \permissrange \ := $[\kw{max}(1,{i}-p), \kw{min}({i}+q,T)]$; \\ 
    \If{$\{\mu_{X}(h_{j)}\} \setminus $ \mapfunc$\} \neq \emptyset, \forall j \in$ \permissrange$\}$} 
    {
        \errorsigma \ := \errorsigma \ $\cup$ \ $\{$(\matchone, $i$), (\matchtwo, $j$)$\}$; \\   
    }
    \ForEach{$(x.A = a) \in Y$}
    {
        \If{\matchone$.A \neq a$} 
        {
            \errorsigma \ := \errorsigma \ $\cup$ \ $\{$(\matchone, $i$)$\}$; \\   
        }
    }
}
\KwRet{\errorsigma};\\
\caption{\kw{IncTED}$($\tempgraph$, \sigma, $\matches$)$
}
\label{alg:IncTED}
\end{algorithm}

%% file: Algorithms/LocalVio.tex
\begin{algorithm}[tb]
\errorSigmaj := $\emptyset$;\\
\kw{Initialize} \hashmapX, \hashmapXY, \timehashmapX, \timehashmapXY;\\
        Compute matches and errors over $G_1$; \\
        (\matchesJ{1}, $h_{t_i}(\bar{x})$) := \isounit($Q$, $G_{t_1}$) \\
        \errorSigmaj +=\kw{IncTED}($G_1$, $\sigma$, \matchesJ{1}, \hashmapX, \hashmapXY, \timehashmapX, \timehashmapXY); \\
    Compute matches and errors over subsequent $G_i, i \geq 2$ \\
    \matchesJ{i}$=\emptyset$;\\
    \ForEach{$c \in changes$} 
    {
        \matchesJ{i} += \kw{LMatch}$(\mathcal{J}_{ij},$\matchesJ{i-1}$,c)$;\\
        \errorSigmaj
        +=\kw{IncTED}$(F_{ij}, \Sigma$, \matchesJ{i}, \hashmapX, \hashmapXY, \timehashmapX, \timehashmapXY);
    }
Send \matchesJ{i}, \errorSigmaj \ to $M_c$\\
\caption{\kw{LocalVio}$(F_{ij}, \mathcal{J}_{ij}, t_i)$}
\label{alg:localVio}
\end{algorithm}

%% file: experiments.tex
\vspace{-2ex}
\section{Experiments}
\label{sec:exp}


\vspace{-1ex}
We evaluate our algorithms to test: (1) its scalability and impact of varying parameters; (2) the comparative effectiveness of \tgfd-based error detection; and (3) case study that verifies real-world \tgfds and errors that can be captured.

\vspace{-2ex}
\subsection{Experimental Setup}
\label{sec:expsetup}
\vspace{-1ex}

\stitle{Datasets.} 
We use two real, and one synthetic graph.  All datasets and source code are publicly available at \cite{datasite}. \\
(1) \uline{\dbpedia}\cite{lehmann2015dbpedia}:  The graph contains in total 2.2M entities with 73 distinct entity types, and 7.4M edges with 584 distinct labels from 2015 to 2016, with snapshots 
every 6 months.\\ 
(2) \uline{\imdb}~\cite{imdb}: The data graph contains 4.8M entities with 8 types and 16.7M edges. IMDB provides diff files, where we extract 38 monthly updates from Oct. 2014 to Nov. 2017.  \\
(3) \rev{\uline{\pdd}}~\cite{wang2017pdd}: \rev{The graph contains over 4.2M entities, 10.2M edges describing patient, drug, disease, and  prescriptions. We extracted 31 consecutive daily snapshots, and defined (with domain expertise) 12 \tgfds, as reported in~\cite{datasite}.}  \\
(4) \uline{\synthetic:} 
\blue{
We use the \gmark benchmark~\cite{bagan2016generating}, \eat{that generates static, domain independent synthetic graphs to support user-defined schemas and queries.  We} and generate data graphs with up to 21M vertices, 40M edges, and 11 attributes per node. We transform the static graph into a temporal graph of $T$ timestamps by randomly generating updates 4\% the size of the graph (\wrt the number of edges) to create subsequent graph snapshots.}


\noindent \textbf{TGFDs Generation}.  
\wu{We generated 40 \tgfds (\dbpedia, \imdb), and 20 \tgfds (\synthetic) \eat{I moved the next sentence from case study to here - make sense? If not, we may better clarify why two different ways of generating \tgfds are used for testing and for case study.} by  
using a discovery 
algorithm in our pilot study~\cite{NoronhaC21}. 
\rev{The discovery algorithm mines minimal and frequent $k$-bounded \tgfds.  We set the support level to find strict, and approximate \tgfds that allow for exceptions.  We consider matches with the most frequent consequent values as the ground truth.  Lower support values lead to more exceptions and error values, and larger $k$ lead to a larger number of expressive \tgfds with more literals. We rank the \tgfds according to decreasing support, and select the top-40 (20 for \synthetic).}  The pattern size, time intervals and size of \tgfds are reported in Table~\ref{tbl:defaults}. 
}

\thesis{We generate \tgfds from the data by defining 
patterns $Q$ composed of frequent edges, where the size of a 
pattern $|Q|$, which is the number of edges, ranges from 2 to 10 (by default 6 edges). We 
set the 
value constraints $X \rightarrow Y$ based on the attributes from the frequent 
matches of $Q$.  We define $\Delta$ based on finding a time interval that captures at least 50\% of satisfying matches.  In addition, we manually define a set of \tgfds based on domain expertise over \dbpedia \ and \imdb.}     

\stitle{Comparative Baselines.} 
\wu{We implemented the following.} 

\sstab
(1) \uline{\tedbatchnaive:} We compute matches at each snapshot using the VF2 matching algorithm ~\cite{foggia2001performance}.  We verify matches between two snapshots if their time intervals lie within $\Delta$.  

\thesis{\uline{\tedbatchopt:} This algorithm extends \tedbatchnaive to include incremental error detection (Section~\ref{sec:incdetect}) but computes matches over each single snapshot. We compare \tedbatchopt \ against \tedbatchnaive \ to evaluate the benefit of incremental error detection. }
\sstab
(2) 
\uline{\seqted:} We implement a sequential \tgfd error detection algorithm by running \tedincremental on a single machine. We compute matches by using an incremental subgraph isomorphism matching algorithm, \isounit~\cite{FJJ11}.  We run \seqted \ on a Linux machine with AMD 2.7 GHz, 256GB RAM. 

\sstab
(3) 
\uline{\gfdbasic:}  We implement the parallel \GFD error detection algorithm 
~\cite{fan2016functional}. For a set of \gfds $\Sigma_{\gfd}$ (defined in~\cite{datasite}), we compute matches for each $\varphi \in \Sigma_{\gfd}$ over each \snapshoti, and check whether \snapshoti $\models \Sigma_{\gfd}$ (pairwise matches between snapshots are not compared). 

\sstab
(4) 
\uline{\gtarSubIso:}  
\blue{Given a set of \tgfds $\Sigma$, we transform each $\sigma \in \Sigma$ to a \gtar.  We remove the $X \rightarrow Y$ dependency, and define a $\Delta$t time interval $(0, q)$, for a pattern $Q$ (serving as both the antecedent and consequent) containing the same constants for each literal in $\bar{x}$.  We evaluate this class of of subgraph isomorphism-based \gtars for its error detection accuracy and performance. 
}
 \\

\vspace{-2ex}
\stitle{Error Injection, Parameters, and Metrics.}
We inject positive and negative errors according to varying error rates.  For instance, for positive errors, we randomly select \kw{err\%} of pairwise matches with equal values in $X$, and update their values in $Y$ to 
\wu{create violations}, and add the pair to the set of positive errors $\Gamma^+$. For negative errors, we similarly pick \kw{err\%} of pairwise matches \wrt a \tgfd $\sigma$, and update a value in $Y$ to a value in the domain of $X'$ \wrt another \tgfd $\sigma'$, where $\{Y \cap X' \neq \emptyset\}$, and add to $\Gamma^-$.  

We use the following commonly adopted 
measures:  \prec$=\dfrac{|\mathcal{E}(\mathcal{G}_T, \Sigma) \land \Gamma^+|}{|\mathcal{E}(\mathcal{G}_T, \Sigma)|}$ and \rec$=\dfrac{|\mathcal{E}(\mathcal{G}_T, \Sigma) \land \Gamma^+|}{|\Gamma^+|}$. The false positive rate is computed as \fpr$=\dfrac{|\mathcal{E}(\mathcal{G}_T, \Sigma) \land \Gamma^-|}{|\Gamma^-|}$, and \fscore$=2\times \dfrac{\prec\times \rec}{\prec + \rec}$. 

Table~\ref{tbl:defaults} summarizes the parameters and their default values. 

\begin{table}
\begin{small}
\begin{center}
\caption{Parameter values (defaults in bold) \label{tbl:defaults}}
\vspace{-2mm}
\begin{tabular}{|l|l|l|}
  \hline
  \textbf{Symbol} & \textbf{Description} & \textbf{Values} \\
  \hline
  $|\Sigma|$  & \#\tgfds  & \textbf{10}, 20, 30, 40 (\imdb, \\
  & & (\dbpedia),  \textbf{12} (\pdd) \\
  & & \textbf{20} (\synthetic) \\
  \hline
  $|Q|$  & graph pattern size & 2, 4, \textbf{6}, 8, 10 \\
  \hline
   $\Delta$ & time interval & 5, \textbf{10}, 15, 20, 25 \\
  & & month (\imdb) \\
  \hline
  $T$  & total timestamps & 5, \textbf{10}, 15, 20 (\synthetic)  \\ 
  \hline
  $|G| = (|V|,|E|)$  & graph size (in M) & \textbf{(5, 10)}, (10, 20), \\
  & & (15, 30),  (20, 40) \\   
  \hline  
 \changerate  & change rate & 2\%, \textbf{4\%}, 6\%, 8\%, 10\% \\ 
  \hline
 \errorrate & error rate & 1\%, \textbf{3\%}, 5\%, 7\%, 9\% \\  
  \hline  
  $n$  & \#processors & 2, \textbf{4}, 8, 16 \\
  \hline
\end{tabular}
\end{center}
  \vspace{-4ex}
\end{small}
\end{table}


\wu{
\stitle{Implementation}. 
We implement all our algorithms using Java v.13 and Scala.  We run 
the tests on a cluster of 16 Amazon EC2 Linux machines, each with 32GB RAM, 8 cores at 2.5 GHz.
The full set of data constraints including \tgfds, source code, and datasets are   available at~\cite{datasite}.
}

\subsection{Exp-1: Scalability} 


\begin{figure*}[tb!]
    \centering
	{\includegraphics[width=13cm,height=0.55cm]{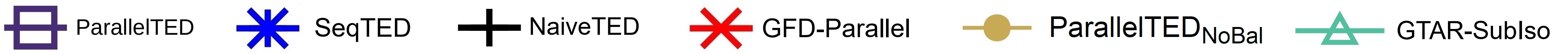}}
	\label{fig:legend}
	\vspace{-2ex}
\end{figure*}

\begin{figure*}[tb!]
	\captionsetup[subfloat]{format=plain, justification=centering}
	\centering
		\subfloat[\small{Vary $|\Sigma|$ (\dbpedia)}] {\label{fig:sigma(dbpedia)}
			{\includegraphics[width=4.2cm,height=3.2cm]{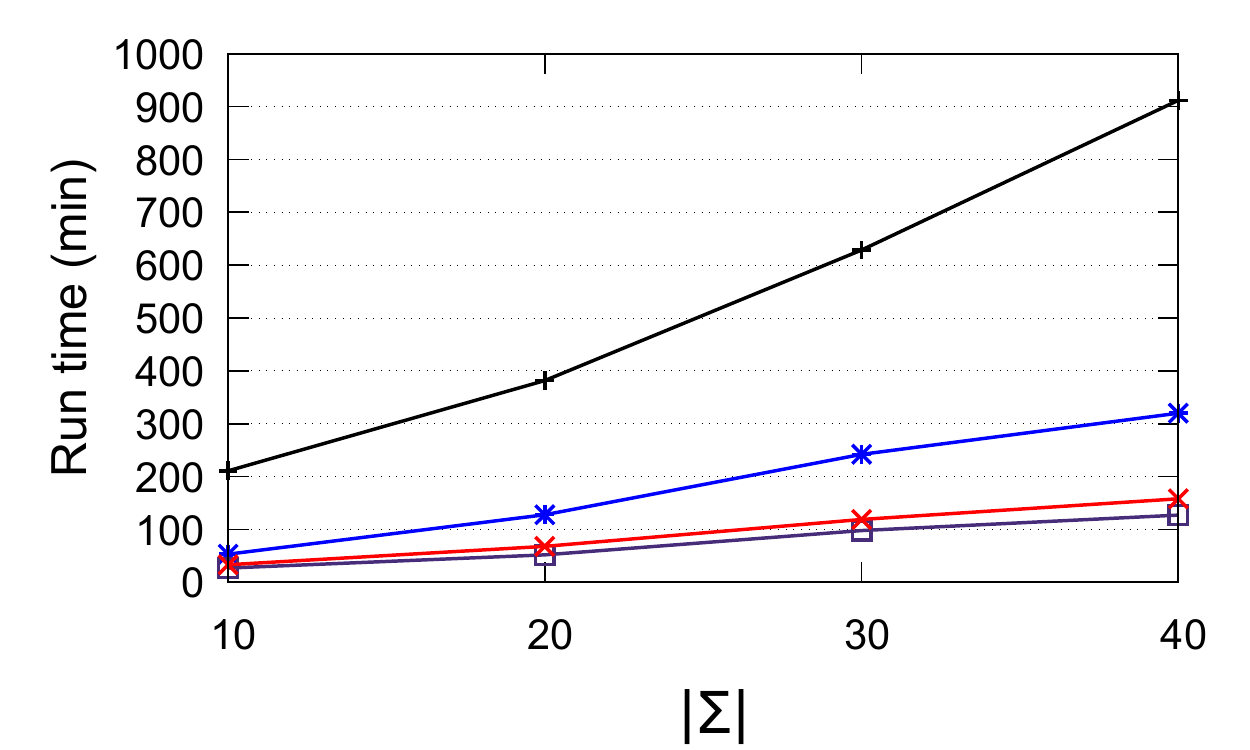}}}
		\hfill\subfloat[
		\small{Vary $|\Sigma|$ (\imdb)} ]
		{\label{fig:sigma(imdb)}
			{\includegraphics[width=4.2cm,height=3.2cm]{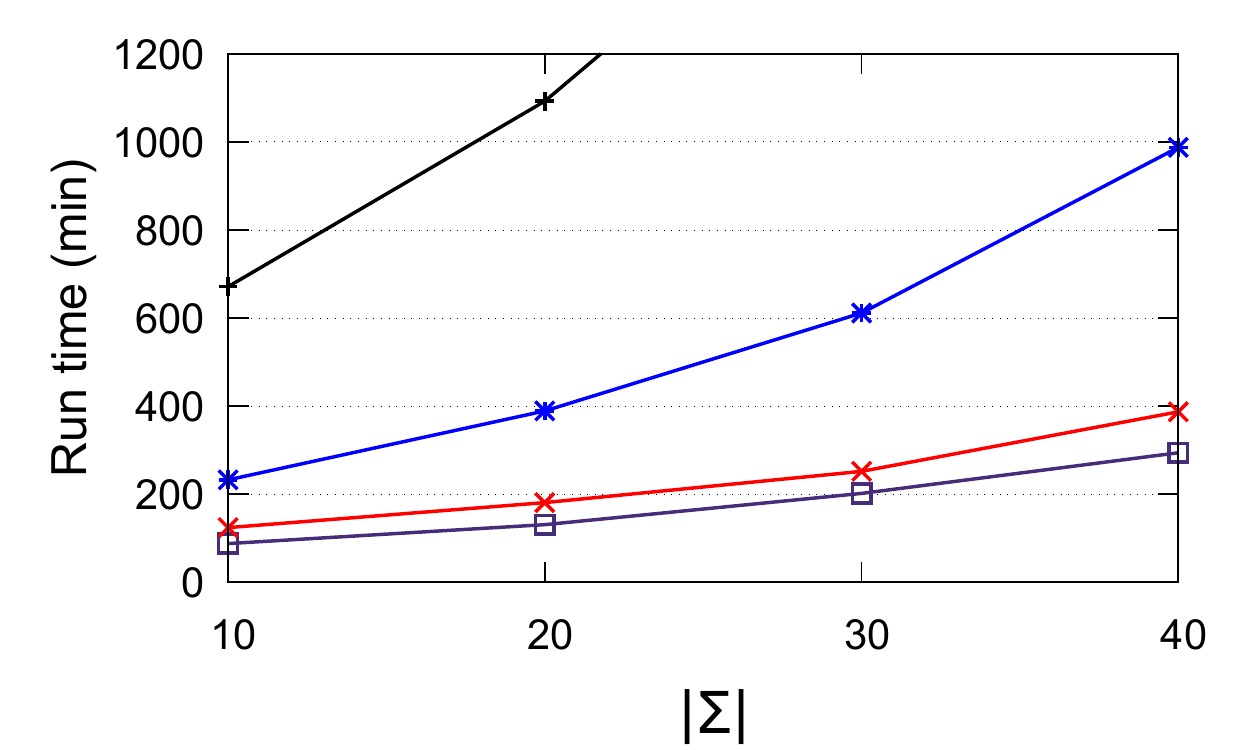}}}
\thesis{\hfill\subfloat[
		\small{Run time breakdown (\imdb) \morteza{Do we need to keep this? We can just say it in text in my opinion.} } 
		]{\label{fig:time_breakdown(imdb)}
			{\includegraphics[width=4.2cm,height=3.2cm]{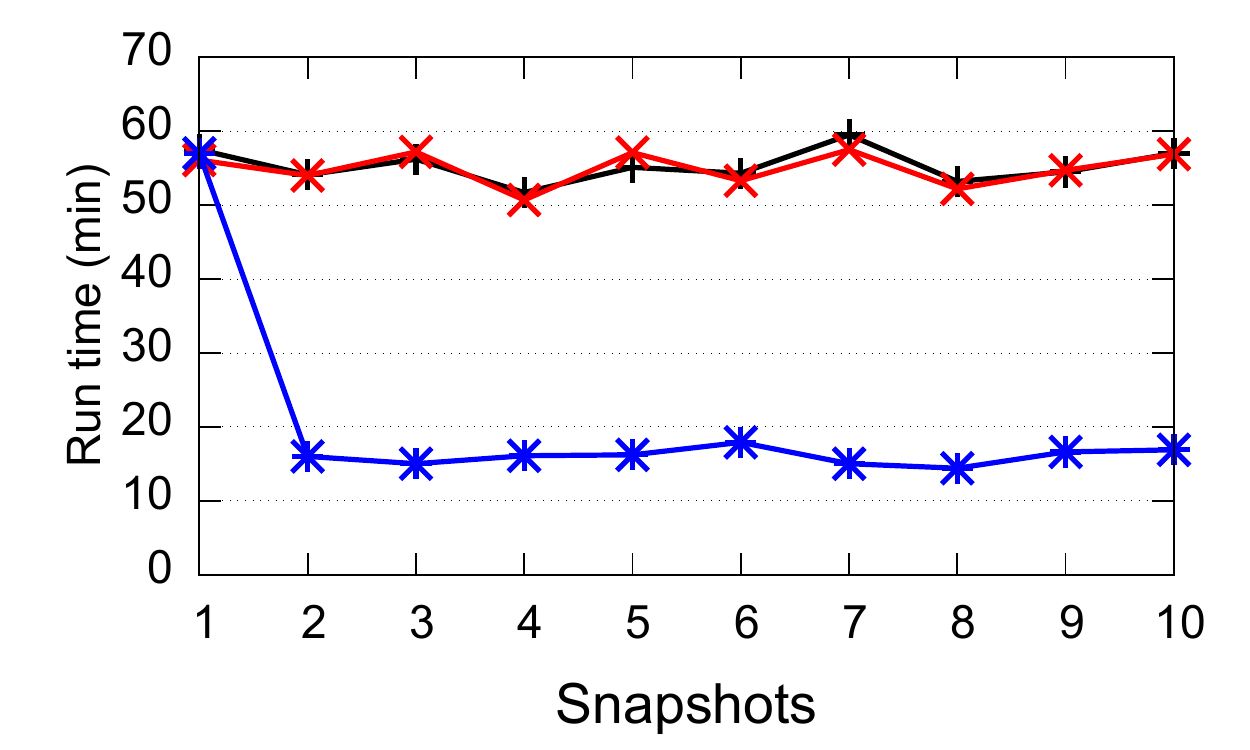}}}}
		\hfill\subfloat[
		\small{Vary $|Q|$ (\dbpedia)}]
		{\label{fig:patternSize(dbpedia)}
			{\includegraphics[width=4.2cm,height=3.2cm]{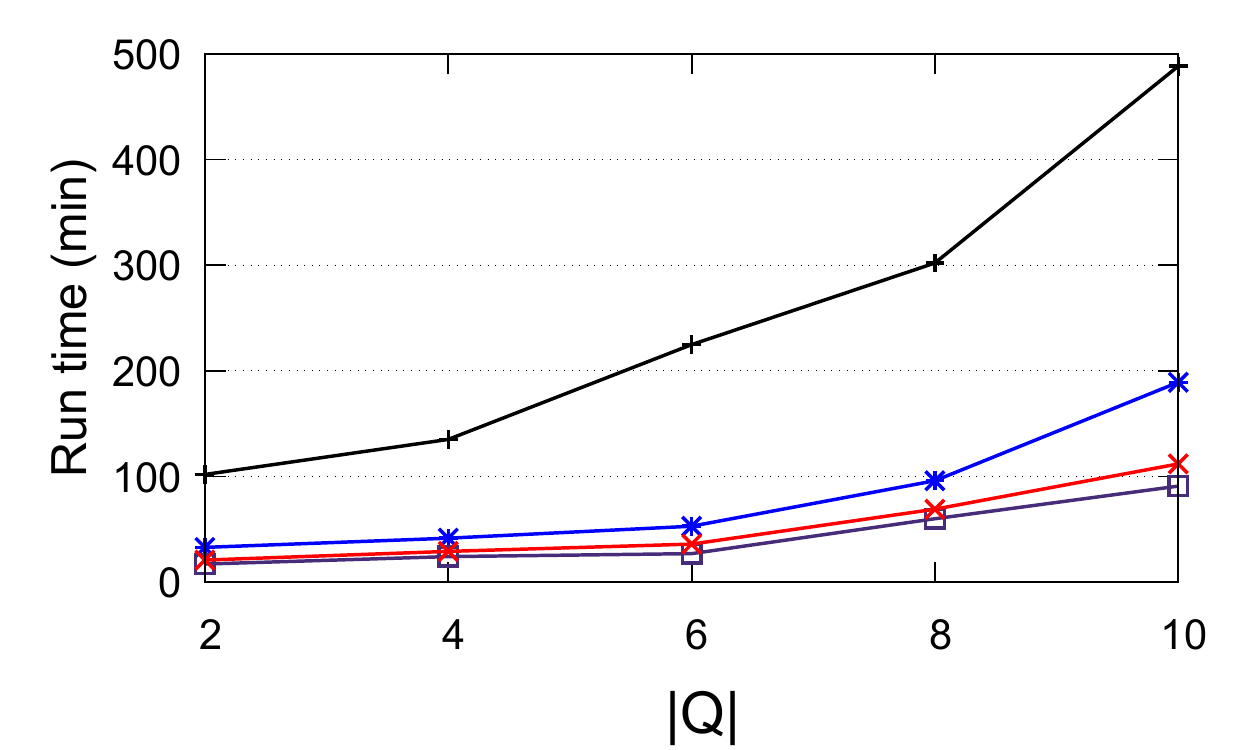}}}
		\hfill \subfloat[\small{Vary $|Q|$ (\imdb)}]
		{\label{fig:patternSize(imdb)}
			{\includegraphics[width=4.2cm,height=3.2cm]{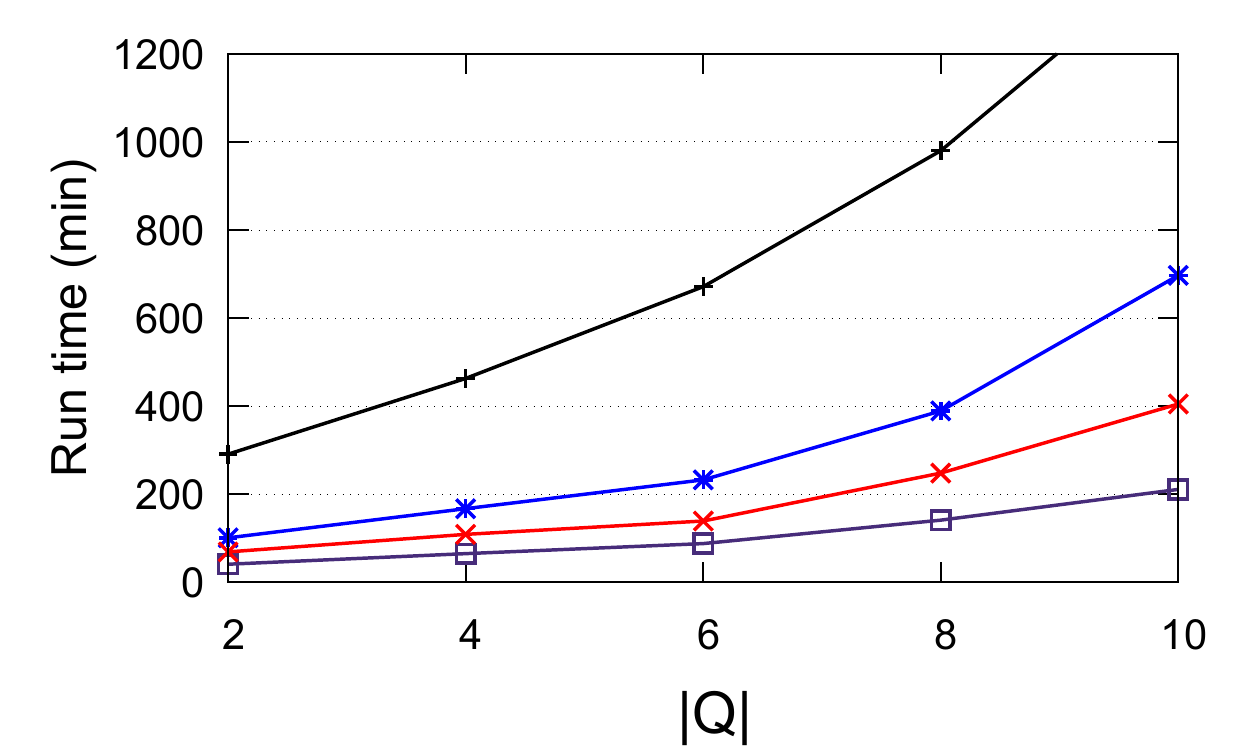}}}		
\tr{		\hfill\subfloat[\small{Vary $\Delta$ (\imdb)}]{\label{fig:delta(imdb)}
			{\includegraphics[width=4cm,height=3cm]{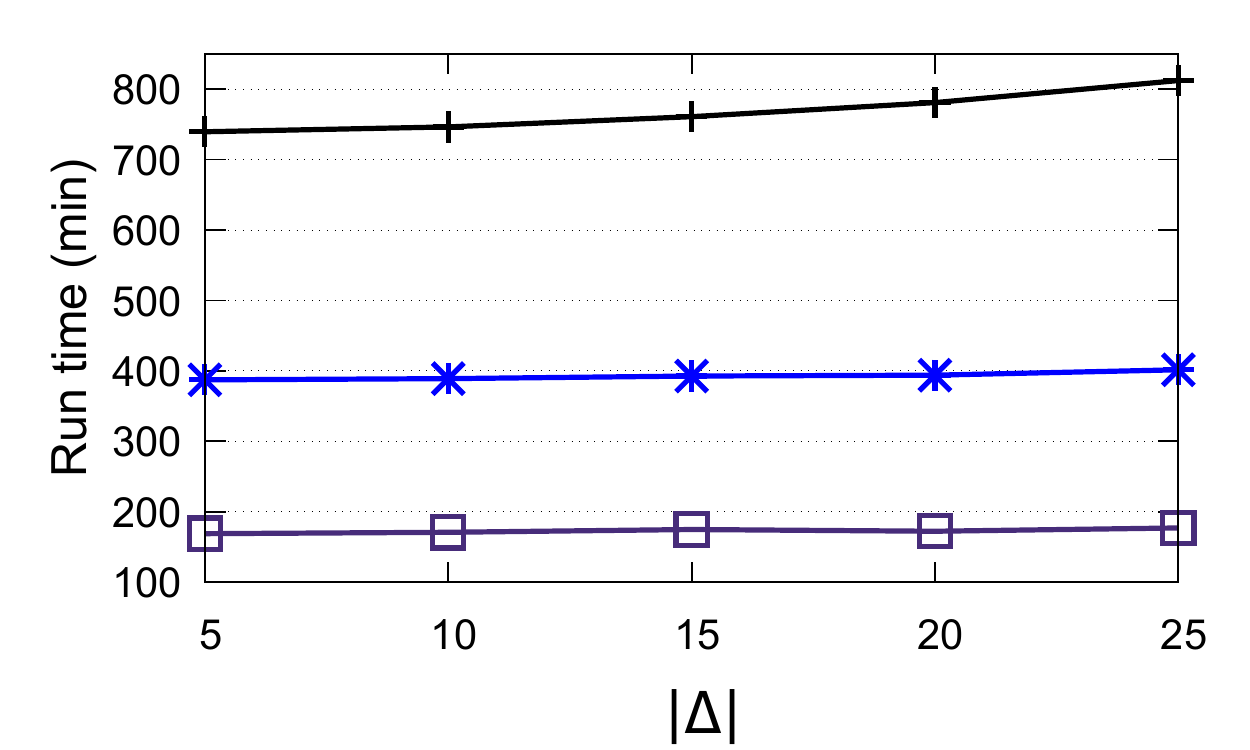}}}
	}
		\hfill\subfloat[\small{Vary \changerate (\dbpedia) }]
	    {\label{fig:change(dbpedia)}
			{\includegraphics[width=4cm,height=3cm]{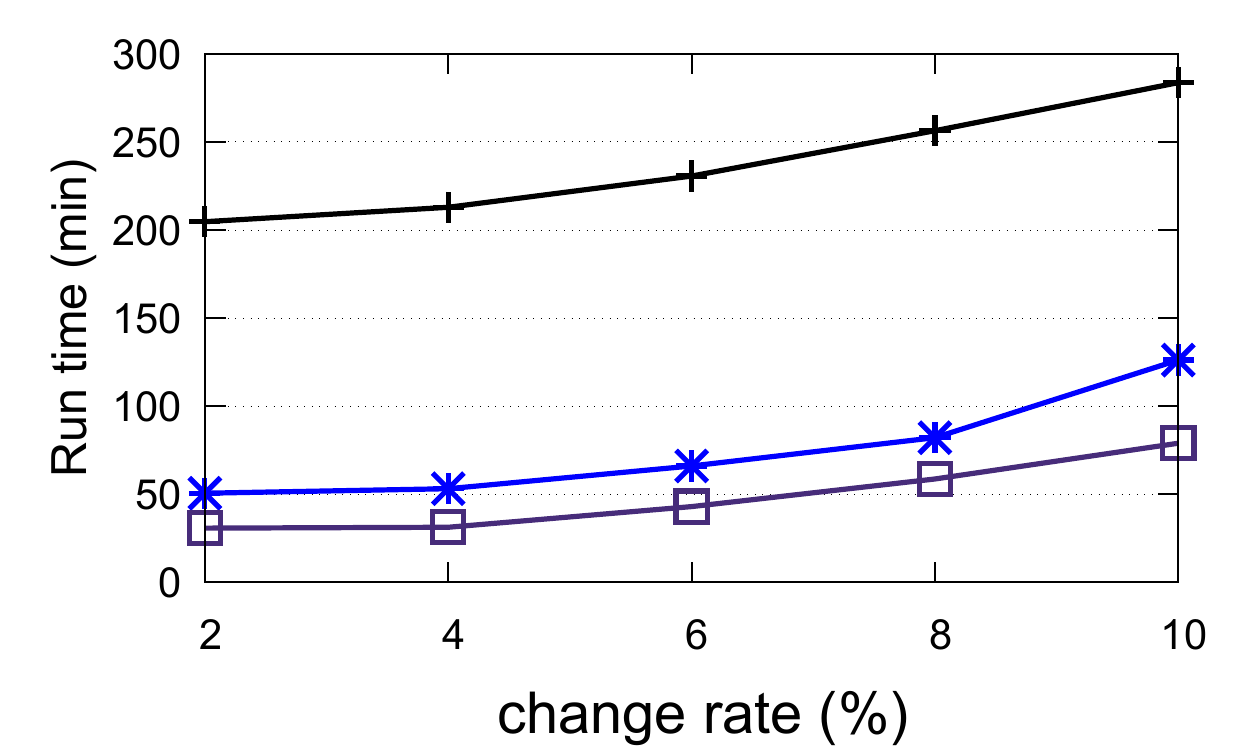}}}
		\hfill\subfloat[\small{Vary $|G|$ (\synthetic)}]
		{\label{fig:graphSize(synthetic)}
			{\includegraphics[width=4cm,height=3cm]{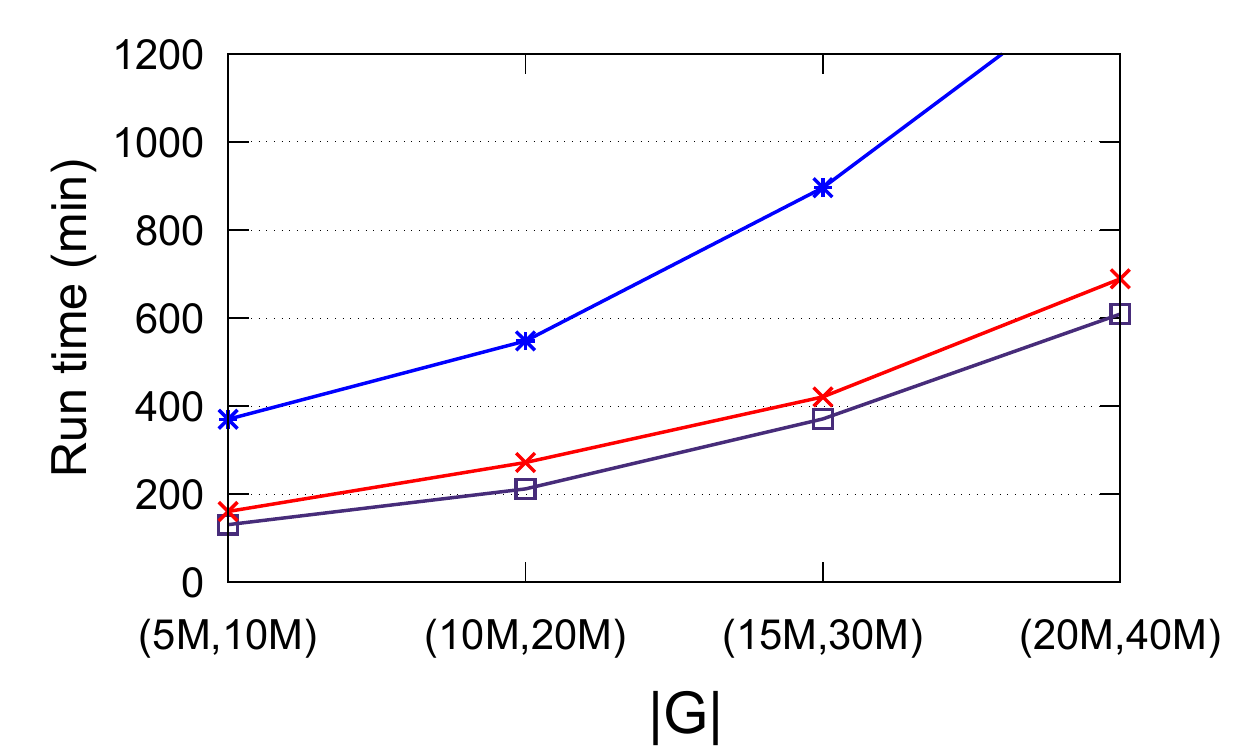}}}
		\hfill\subfloat[\small{Vary $|T|$ (\synthetic)}]
		{\label{fig:snapshots(synthetic)}
			{\includegraphics[width=4cm,height=3cm]{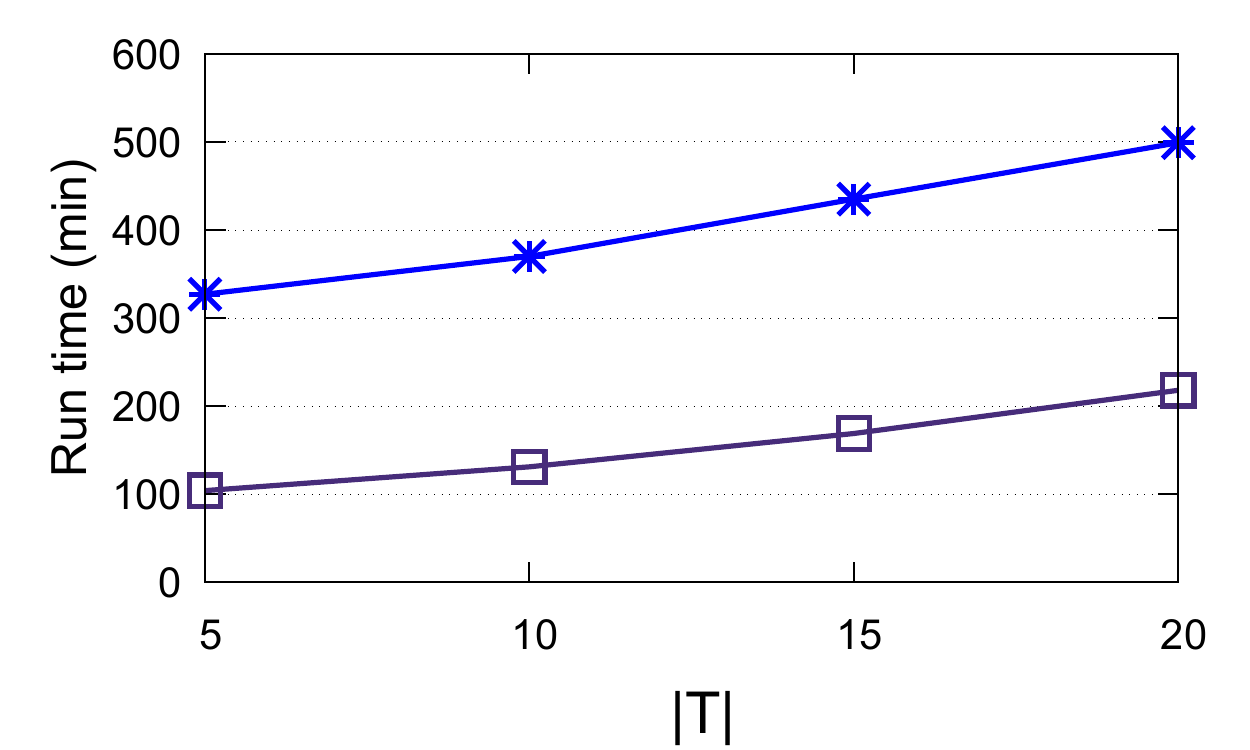}}}
    	\hfill\subfloat[\small{Vary $n$ (\dbpedia)}]
		{\label{fig:cores(dbpedia)}
			{\includegraphics[width=4cm,height=3cm]{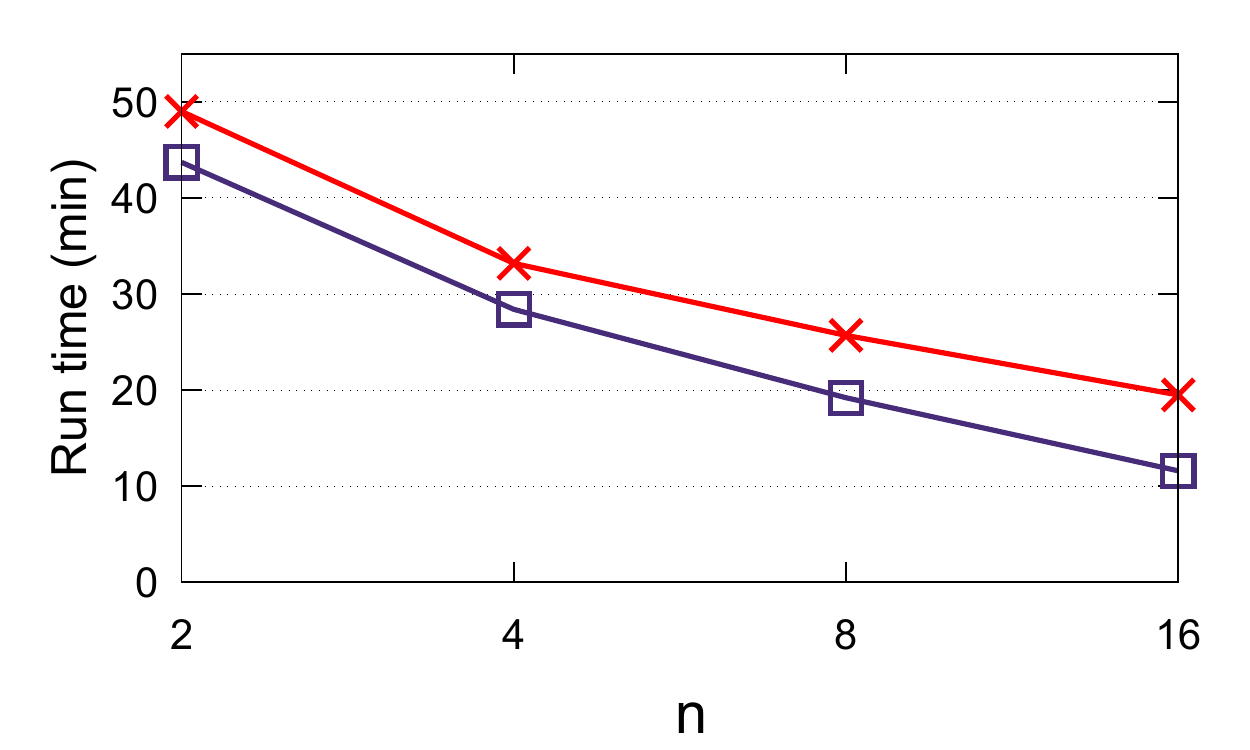}}}
\thesis{		\hfill\subfloat[\small{Vary $n$ (\imdb) \morteza{Run \gfdmultigraph}}]{\label{fig:cores(imdb)}
			{\includegraphics[width=4cm,height=3cm]{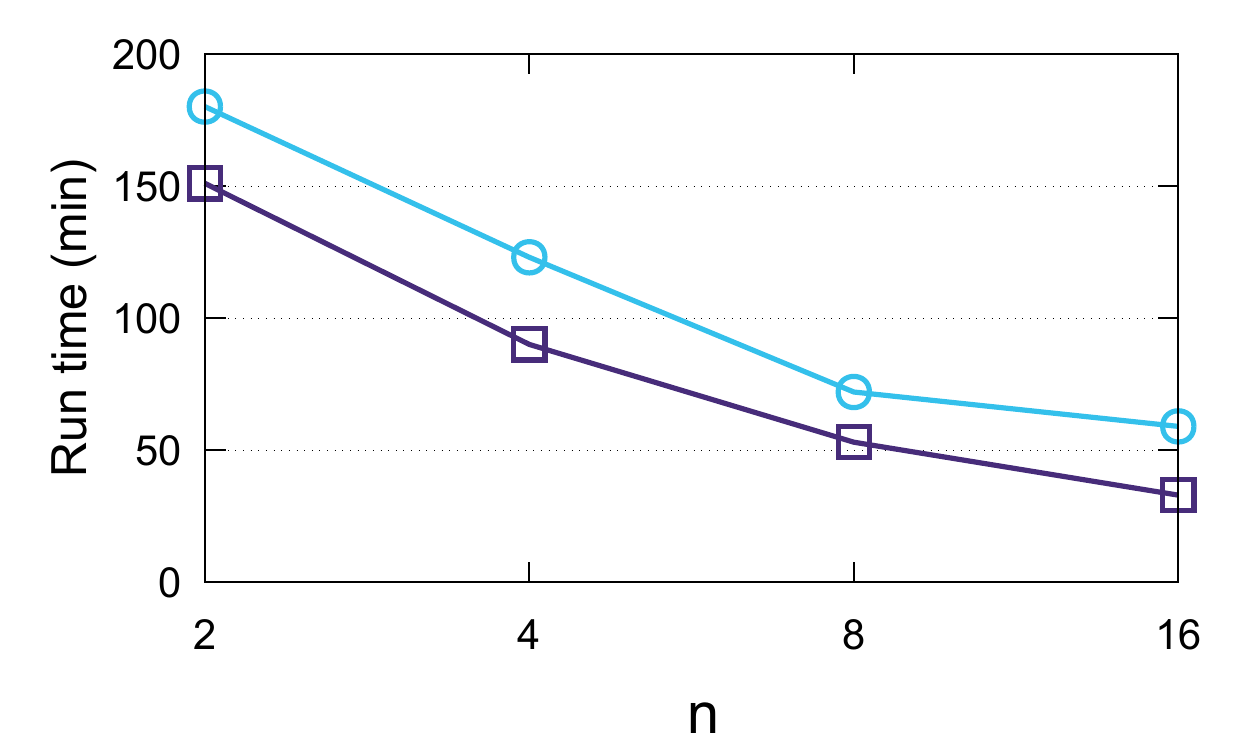}}}
}	
		\hfill \subfloat[\small{Vary $n$ (\dbpedia)}]
		{\label{fig:ccost(dbpedia)}
			{\includegraphics[width=4cm,height=3cm]{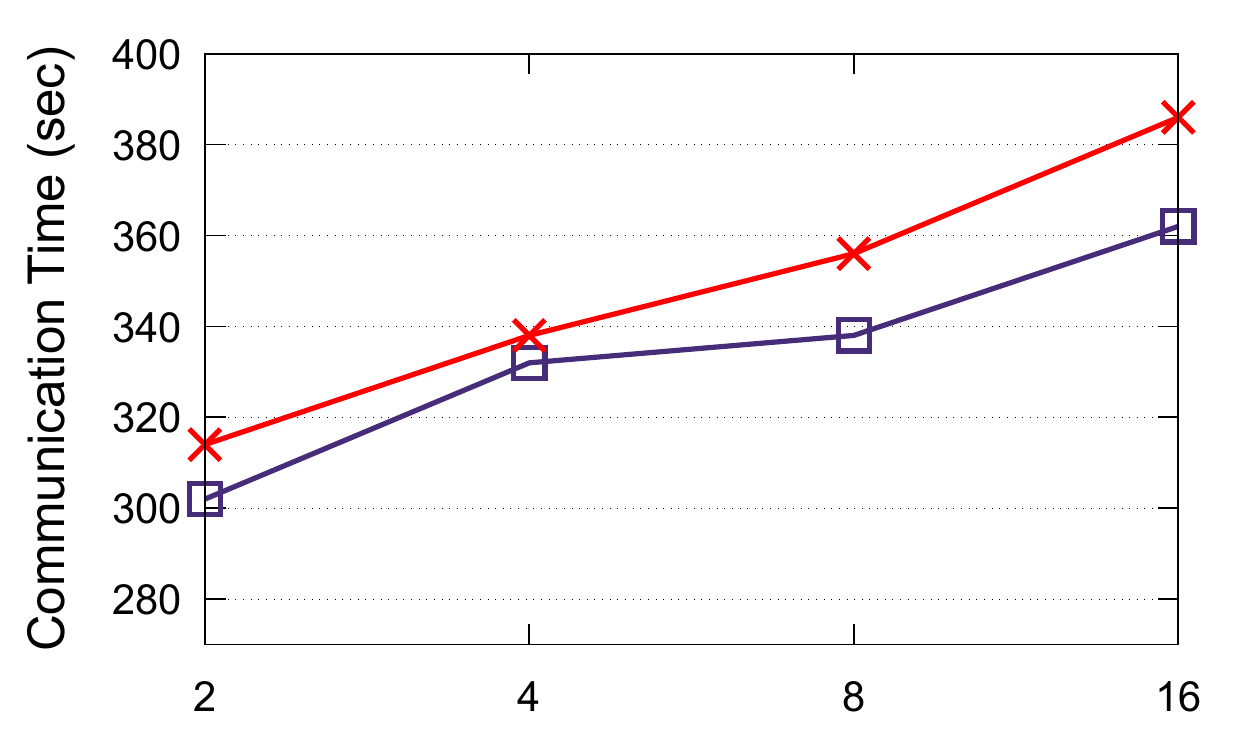}}}		
		\hfill\subfloat[\small{Vary $n$ (\imdb)}]{\label{fig:ccost(imdb)}
			{\includegraphics[width=4cm,height=3cm]{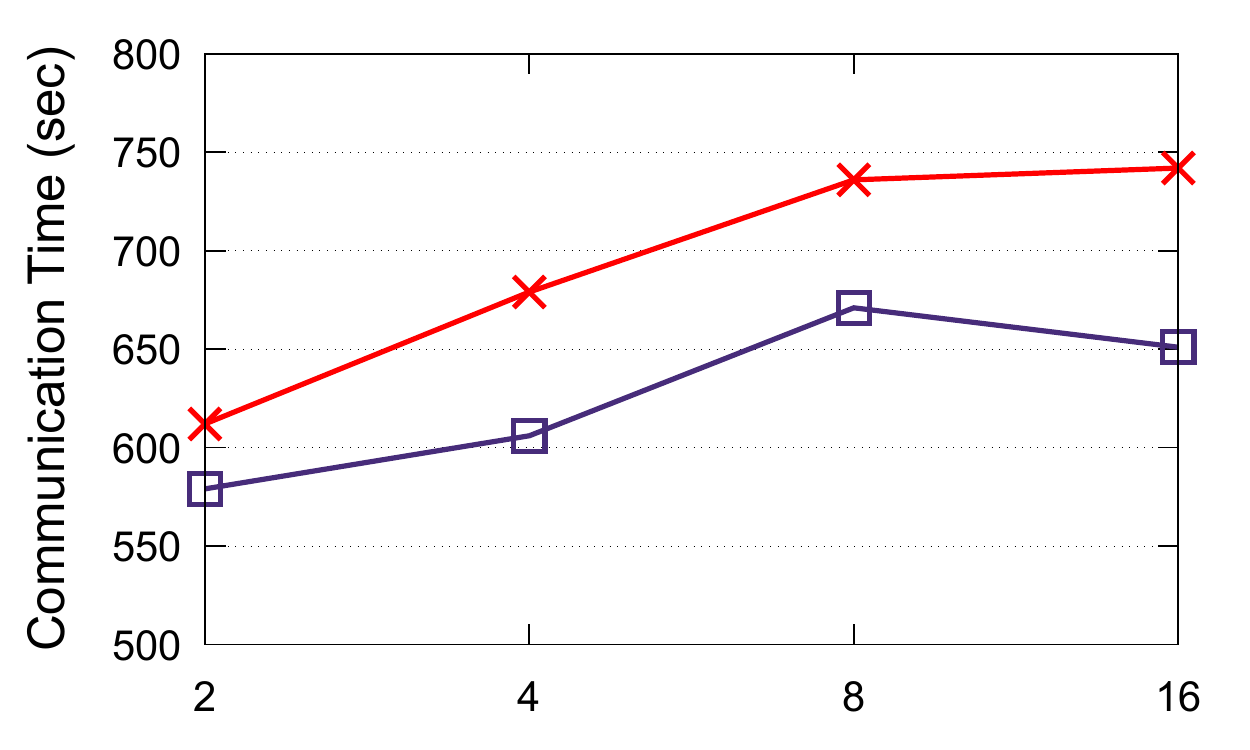}}}	
		\hfill\subfloat[\small{Vary Rate of Change} ]
		{\label{fig:burstiness(changeRate)}
			{\includegraphics[width=4cm,height=3cm]{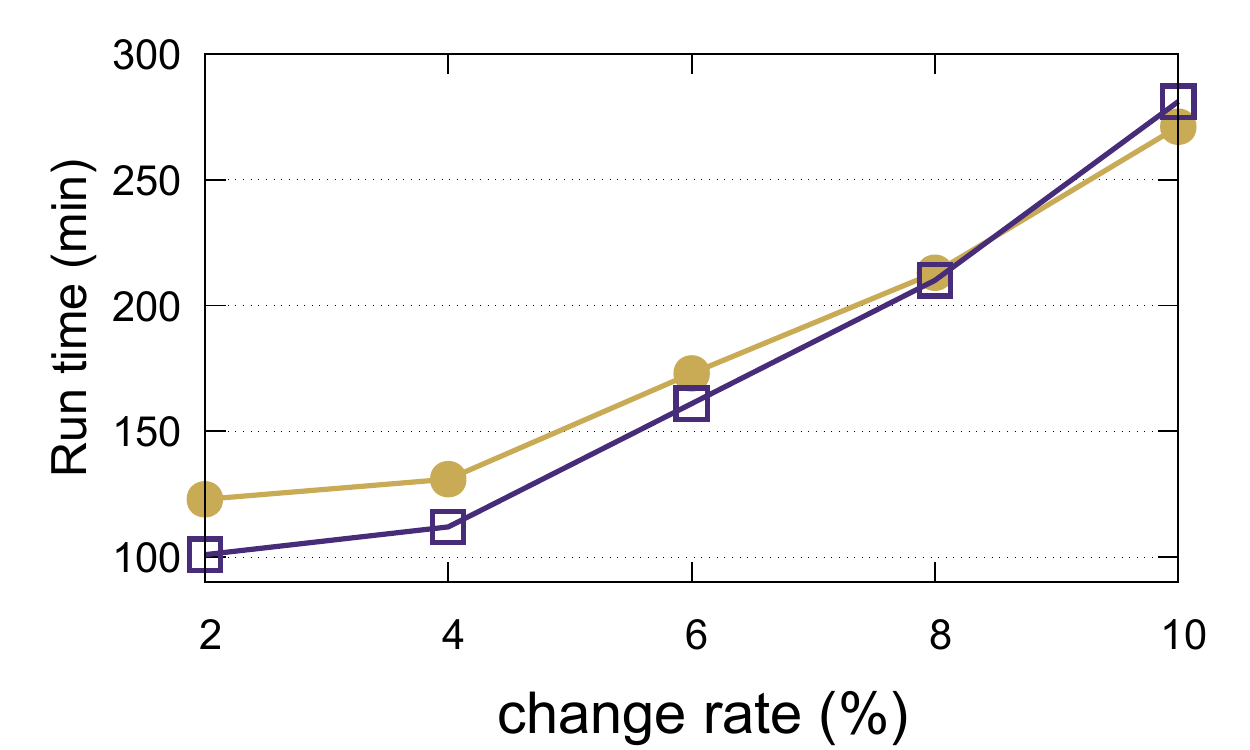}}}
		\hfill\subfloat[\small{Vary Types of Changes} ]
		{\label{fig:burstiness(typesOfChanges)}
			{\includegraphics[width=4cm,height=3cm]{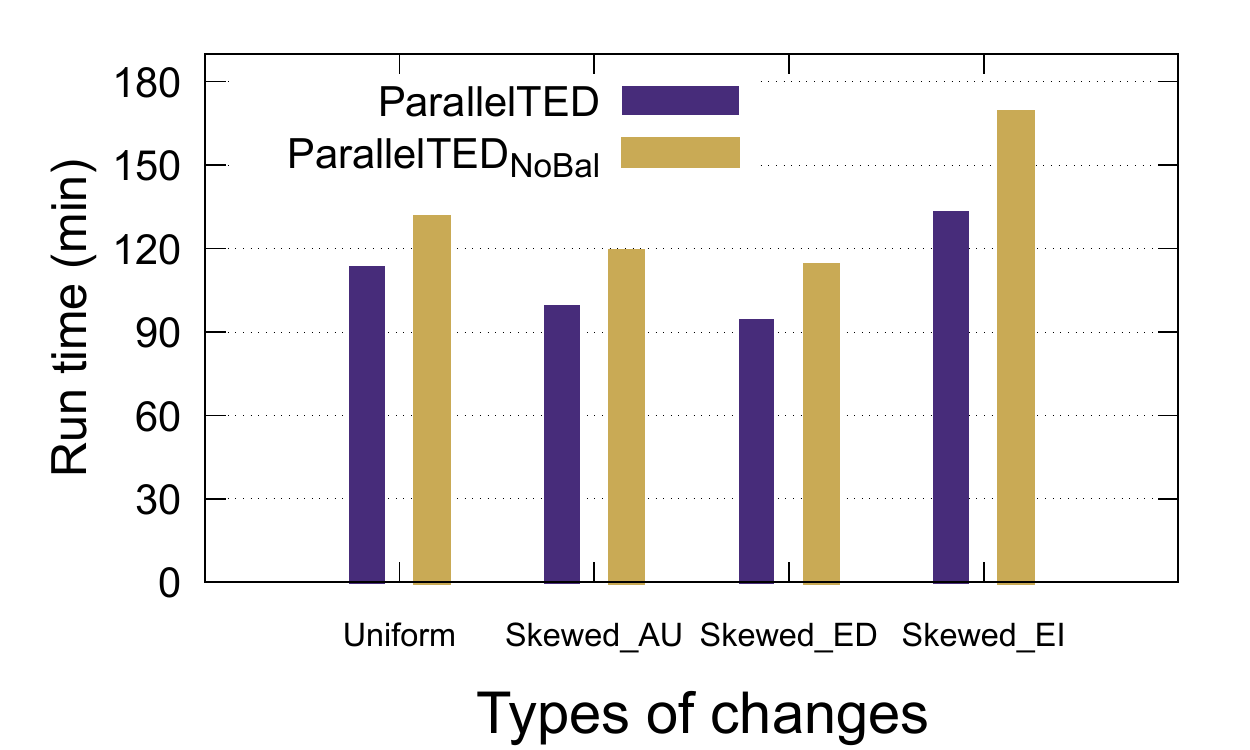}}}
\tr{	 \subfloat[\small{Vary Distribution of Changes} ]
		{\label{fig:burstiness(distribution)}
			{\includegraphics[width=4cm,height=3cm]{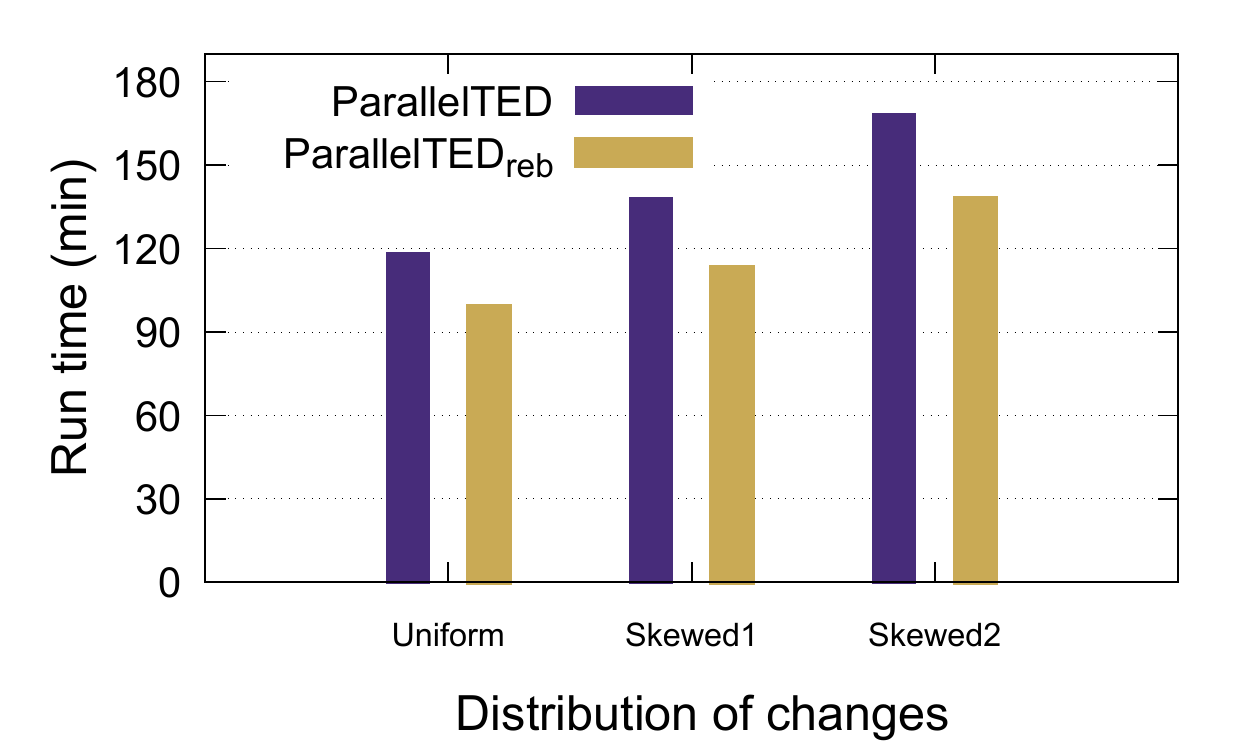}}}
}			
	\caption{\tgfd error detection efficiency and effectiveness.} \label{fig-efficiency} 
\end{figure*}

\stitle{
Vary $|\Sigma|$:}  
Figure~\ref{fig:sigma(dbpedia)} and Figure~\ref{fig:sigma(imdb)} show \tedparallel outperforms \seqted by an average 170\%, with larger gains in \imdb (more graph snapshots). We stopped executions of \tedbatchnaive over \imdb after 22hrs.  \tedparallel runs 22\% and 29\% faster than \gfdbasic over \dbpedia and \imdb, respectively, despite having to evaluate more matches.  This highlights the effectiveness of \tedincremental, and maintaining query path matches in the presence of changes. 

\stitle{
Vary $|Q|$.}  Figure~\ref{fig:patternSize(dbpedia)} and Figure~\ref{fig:patternSize(imdb)} show that larger graph patterns incur higher runtimes as the cost of local pattern matching increases. This is especially evident for sequential algorithms \tedbatchnaive and \seqted. \tedparallel is 80\% and 18\% faster than \seqted and \gfdbasic, respectively.

\stitle{
Vary $\Delta$.}  We found the performance of \tedbatchnaive is sensitive due to the increased number of pairwise matches that need to be compared for larger $\Delta$.  In contrast, \seqted and \tedparallel 
are largely insensitive due to their incremental checking strategies. \conf{We report figures in~\cite{AMCW21}.} \tr{Figure~\ref{fig:delta(imdb)} shows the results.}
\eat{\conf{Due to limited space, we show the figures in~\cite{AMCW21}.}  
}

\thesis{In this experiment, we vary the length of the time interval $\Delta$ for the \imdb dataset from 5 to 25 months. Shown in Figure~\ref{fig:delta(imdb)}, as expected, \tedbatchnaive grows in run time as we increase $\Delta$, since we have more pairs that their time difference lies within $\Delta$. On the other hand, all three algorithms \tedbatchopt, \tedincremental and \tedparallel are less sensitive to the length of $\Delta$. Using interval model of Section~\ref{sec:incdetect}, we avoid pairwise comparison in these algorithms and therefore, they run faster. However, on average, \tedincremental performs 1.7 and 2 times better than \tedbatchopt and \tedbatchnaive algorithms. \tedparallel runs 2.2 times faster than \tedincremental on a cluster of 4 workers.
}

\stitle{
Vary \changerate.} \eat{We vary the number of changes between snapshots, from 2\% to 10\%.  Figure~\ref{fig:change(dbpedia)} shows \blue{that} with more changes, all techniques incur longer runtimes.} \eat{\tedbatchnaive runtime increases an average of 10\%, while} Figure~\ref{fig:change(dbpedia)} shows \seqted and \tedparallel runtimes increase by 27\% and 26\%. respectively, for increased changes.  \seqted uses \isounit to compute matches which become more expensive as the number of changes localized to a subgraph increases.  \tedparallel uses boolean vectors to track changes to existing matches to avoid isomorphic computations  (Section~\ref{sec:querymatch}), and achieves the fastest runtime.  

\stitle{
Vary $|G|$.}  Figure~\ref{fig:graphSize(synthetic)} shows \blue{that} for increasingly large graph sizes, sequential algorithms are not feasible, i.e., we stopped \seqted \ after 20hrs due to its exponential growth.  However, \tedparallel shows that error detection is feasible over large graphs, outperforming \gfdbasic by 18\%.
\thesis{Figure~use the \synthetic data to evaluate runtimes on graphs up to 20M nodes, and 40M edges. Figure~\ref{fig:graphSize(synthetic)} shows exponential growth of \tedincremental for increasingly large $|G|$, where we stopped execution after 20hrs.  However, \tedparallel shows that error detection and matching is feasible over large graphs. \newtext{On average, \tedparallel achieves 18\% faster runtime vs. \gfdbasic \ due to matching over (simpler) query paths, and maintaining matches in the presence of updates using our proposed boolean vectors}.
}

\thesis{Using \synthetic dataset, we vary the size of the graph $|G|$ from $(5M,10M)$ to $(20M,40M)$ and run \tedincremental, \tedparallel and \gfdbasic algorithms. The results are shown in Figure~\ref{fig:graphSize(synthetic)} as the run time for \tedincremental grows exponentially and we stopped it on graph of size $(20M,40M)$ after 1200 minutes. However, using \tedparallel over the cluster of 4 nodes shows that the \tgfd error detection is feasible over large graphs. It also shows that \tedparallel runs 2.1 times faster that \tedincremental in those experiments that we have results for the incremental algorithm. Compare to \gfdbasic, we gain \tbf\% faster runtime on average in \tedparallel as use query paths to find and maintain the matches in the temporal graph.  
}

\stitle{
Vary $T$.} Figure~\ref{fig:snapshots(synthetic)} shows linear scaleup for increasing $T$.  For larger $T$, we expect more matches to occur (assuming changes are spread uniformly),  and validating a new match vs. existing matches in \hashmapXY \ can be done in constant time.

\thesis{
In this experiment, using the \synthetic graph of size $(5M,10M)$, we vary the number of timestamps from 5 to 20 using \tedincremental, \tedparallel and \gfdbasic algorithms. \tedincremental grows linearly as we increase the number of snapshots. For each snapshot, we set the change rate to be fixed at 4\% and since \tedincremental performs subgraph isomorphism for the same number of changes for two consecutive timestamps, it is less sensitive to the increase of $|T|$. Similar trend exists with the \tedparallel and \gfdbasic algorithms running on a cluster of 4 workers. However, \tedparallel runs 2 times faster than \tedincremental given the distributed nature of the algorithm and it runs \tbf\% faster than \gfdbasic with the same reasoning as the previous experiment. 
}

\stitle{
Vary $n$.}  Figure~\ref{fig:cores(dbpedia)} shows \eat{that both \tedparallel and \gfdbasic scale well with an increasing number of machine workers.} \tedparallel runs 29\% faster than \gfdbasic despite evaluating matches both \blue{within and between graph snapshots}\eat{(\gfdbasic only compares matches within a snapshot)}. \tedincremental's efficiency avoids  redundant pairwise comparisons, and adaptively tunes matches to avoid expensive subgraph isomorphism computations.  

\stitle{Communication Overhead.} 
\blue{
Figures~\ref{fig:ccost(dbpedia)} and \ref{fig:ccost(imdb)} show the communication overhead for an increasing number of  workers, $n$.  Using \dbpedia and \imdb, the communication cost comprises between 10-48\%, and 5-28\%, respectively, of the total error detection time.  As expected, for larger $n$, we see that both \tedparallel and \gfdbasic incur increased overhead as the need to exchange data increases.
}

\vspace{-2ex}
\subsection{Exp-2: Adapting to Workload Changes} \label{sec:exp-adaptiveLoadBalance}

\stitle{
Rate of Change.}
\blue{We study the impact of the burstiness time buffer $\zeta$ \eat{(see Workload Re-balancing in Section~\ref{sec:InitialWorkloadAssignment})}} by generating a graph of size (5M,10M), and implement a version of \tedparallel \ without $\zeta$ (no workload rebalancing) called \tednobalance.  Figure~\ref{fig:burstiness(changeRate)} shows \tedparallel runtimes with $\zeta = 0.1$ performs 16\% faster as change rates less than 8\% are accounted for by the burstiness buffer, requiring only at most two workload re-distributions.  Beyond this change point, larger $\zeta$ values are needed, which require more frequent workload re-distributions (up to six), and incur approximately 6\% overhead.  

\thesis{
\newtext{In this experiment, using the \synthetic graph, we varied the rate of the changes and run \tedparallel vs \tedparallelReb algorithms.
Figure~\ref{fig:burstiness(changeRate)} shows the runtime as we observe the followings: (1) for change rates 2\% to 6\%, \tedparallelReb runs 16\% faster than \tedparallel on average; (2) \tedparallelReb performed 1, 2 and 2 re-balancing during the execution for 2\%, 4\% and 6\% change rates respectively; (3) \tedparallelReb performed 3 re-balancing for the 8\% change rate, where the overhead of the re-balancing made the runtime to be almost equal as \tedparallel; (4) We performed re-balancing 6 times for the 10\% change rate, where \tedparallel performed around 10\% faster than \tedparallelReb due to the high cost of the re-balancing procedure.}
}

\stitle{
Vary Type and Distribution of Changes.}
Using the same generated graph, we distribute changes according to: (i) a \kw{Uniform} distribution that assigns attribute updates (AU) (40\%), edge deletions (ED) (30\%), and edge insertions (EI) (30\%); (ii) each of \kw{Skewed\_AU, Skewed\_ED, Skewed\_EI} assigns  85\% of changes to their respective type, and 7.5\% to the other two types. \eat{For example, \kw{Skewed\_ED} contains ED (85\%), EI (7.5\%), AU (7.5\%).}  Figure~\ref{fig:burstiness(typesOfChanges)} show that EI are slowest as pattern matching becomes more expensive to find new matches. AU cause matches to be added/removed, and are slower compared to ED, which only lead to fewer matches. Across all change types, Figure~\ref{fig:burstiness(typesOfChanges)} shows \tedparallel is 20\% faster, on average, than \tednobalance.  We also evaluated data partitioning schemes such that hot spots are: (i) uniform across four nodes; (ii) 80\% of hot spots across two nodes; and (iii) 70\% on a single node.  As expected, \tedparallel \ incurs 13\% more time, compared to \tednobalance at +19\%\conf{~\cite{AMCW21}}\tr{~(Figure\ref{fig:burstiness(distribution)})}.

\vspace{-2ex}
\subsection{Exp-3: Comparative Performance} \label{sec:accuracyExperiments}

\begin{figure*}[tb!]
	\captionsetup[subfloat]{justification=centering}
	\centering
		\subfloat[\small{F1 vs. \lit{Err} (\imdb)}] {\label{fig:f1(imdb)}
			{\includegraphics[width=4cm,height=3cm]{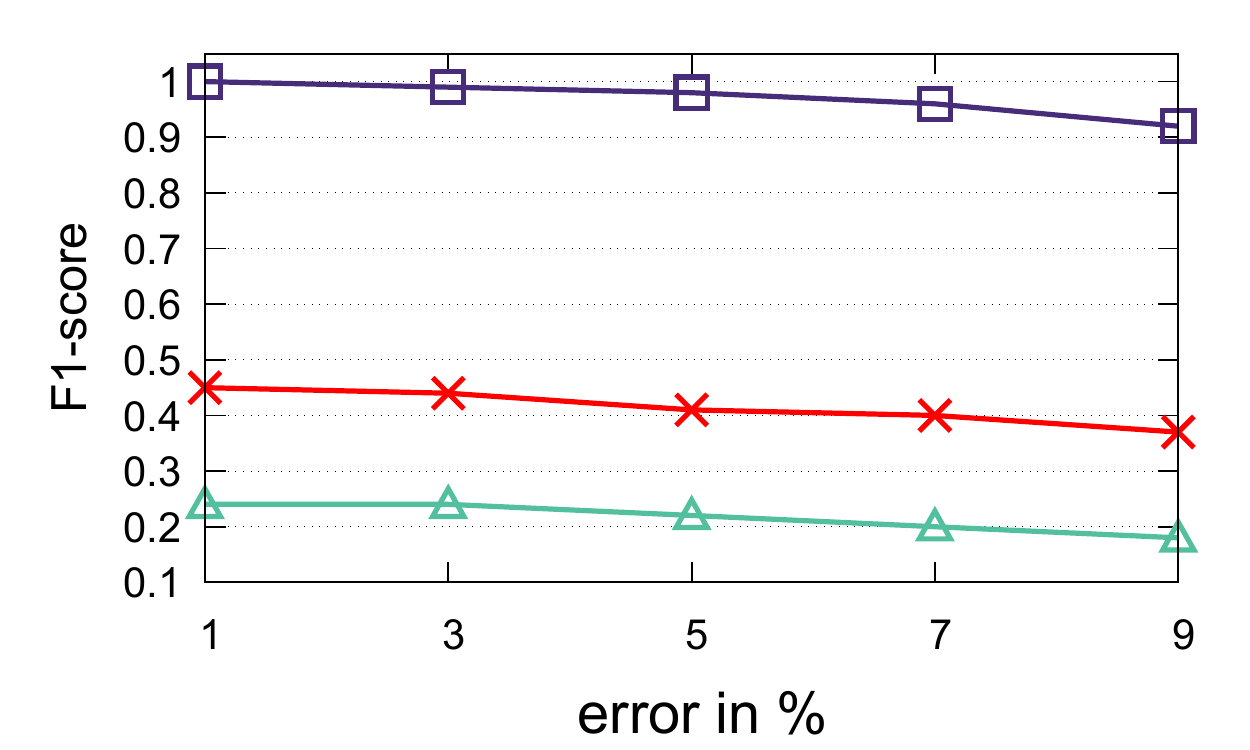}}}
		\hfill\subfloat[\small{F1 vs. \lit{Err} (\dbpedia)} ]	{\label{fig:f1(dbpedia)}
			{\includegraphics[width=4cm,height=3cm]{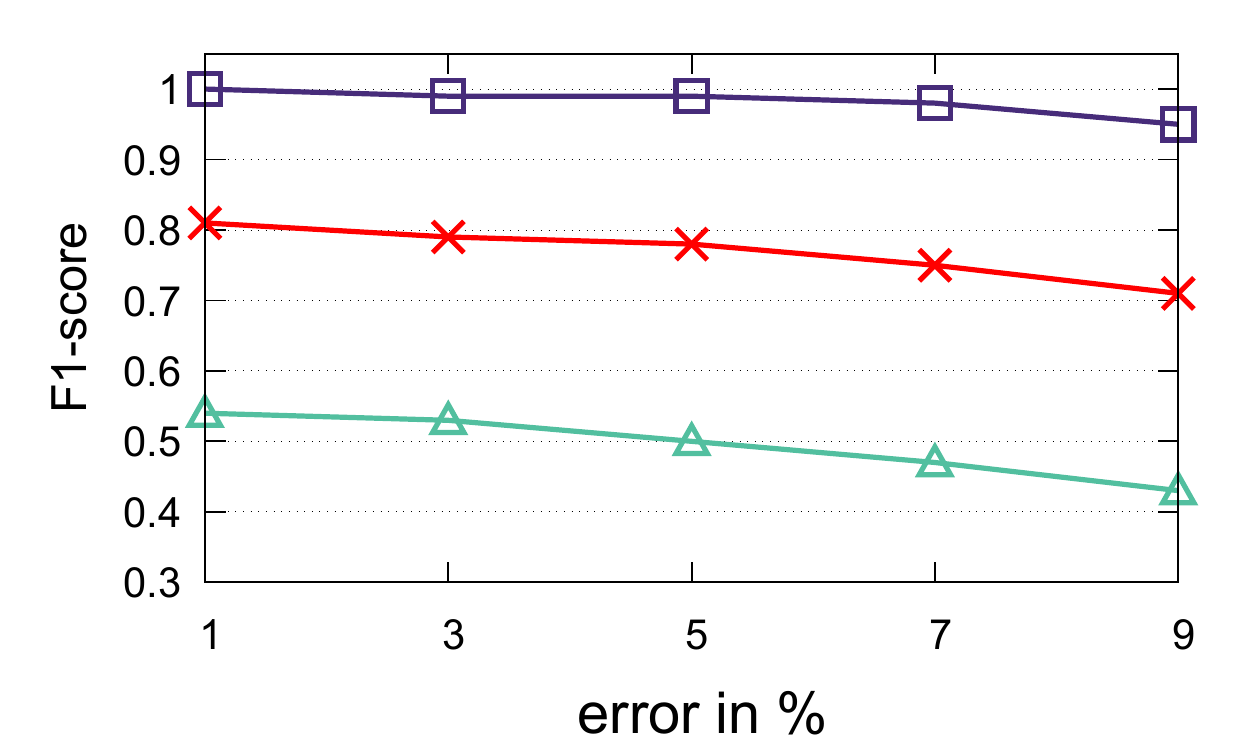}}}
		\hfill\subfloat[\small{F1 vs. \lit{Err} (\pdd)} ]	{\label{fig:f1(pdd)}
			{\includegraphics[width=4cm,height=3cm]{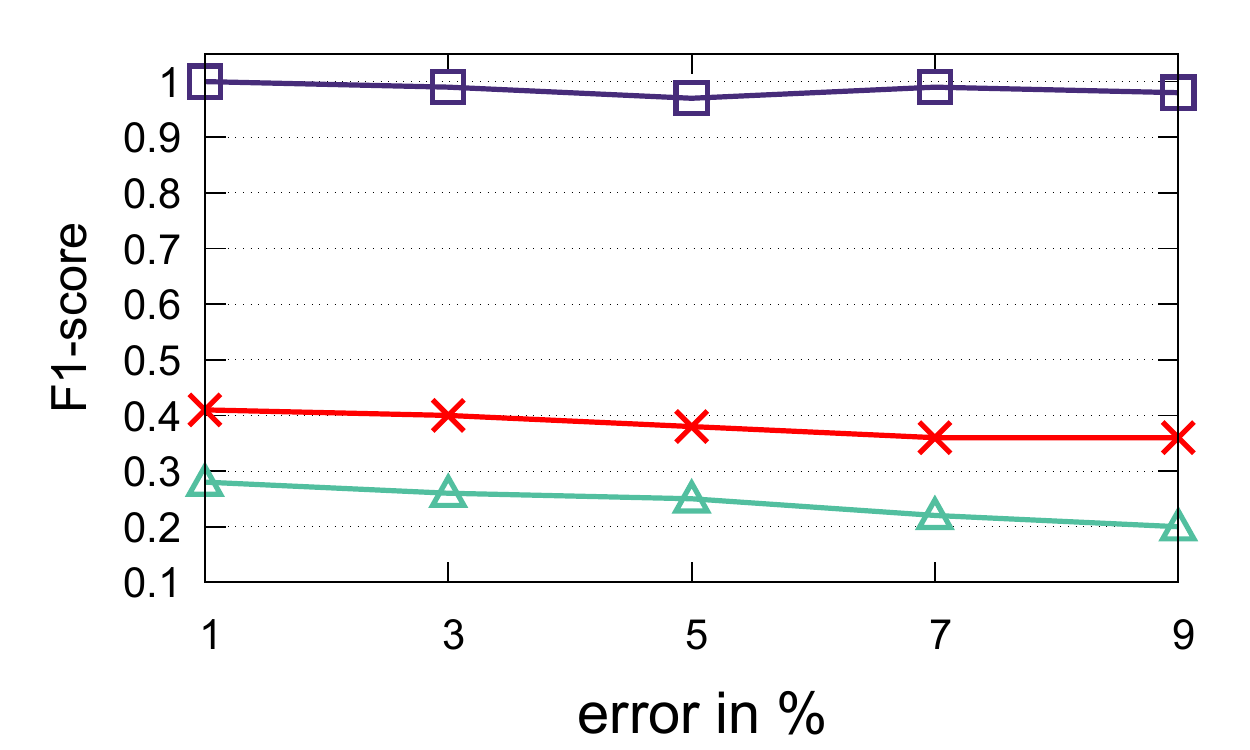}}}
		\hfill\subfloat[\small{Runtime vs. \lit{Err} (\imdb)} ]	{\label{fig:f1runtime(imdb)}
			{\includegraphics[width=4cm,height=3cm]{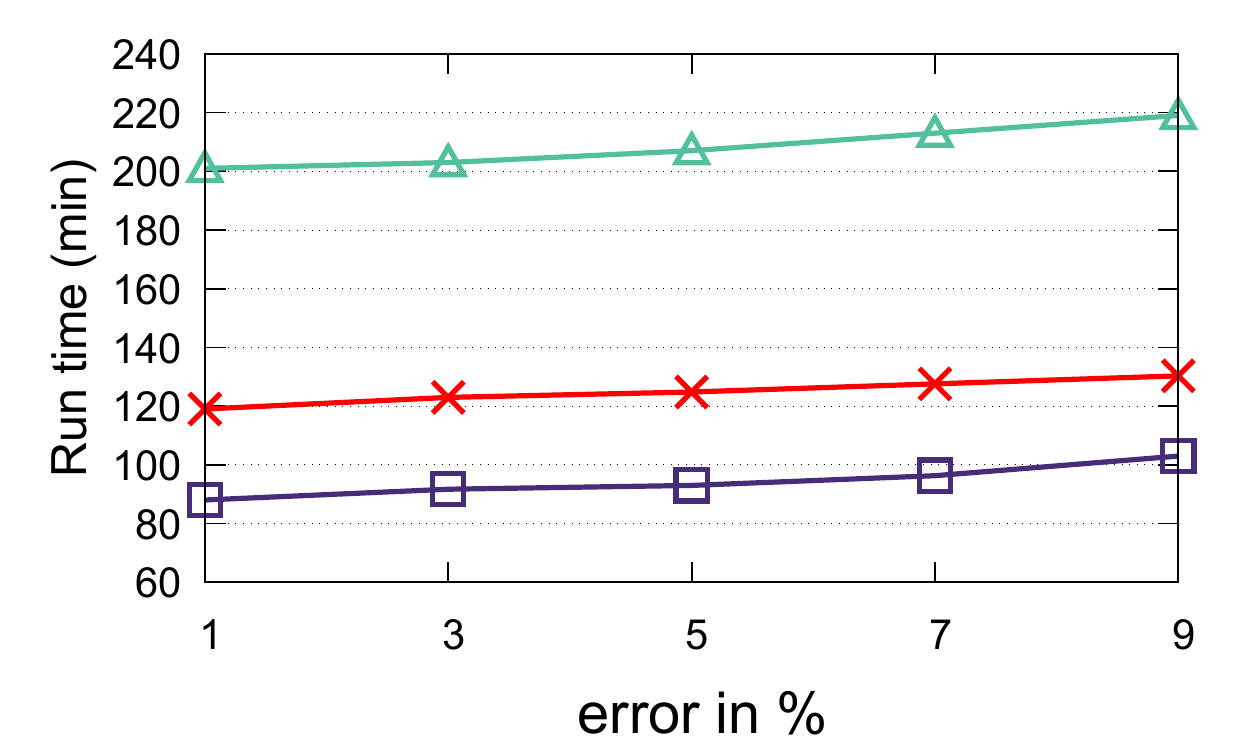}}}
	\tr{	\hfill\subfloat[\small{Vary \lit{err} (\dbpedia)} ]
		{\label{fig:f1runtime(dbpedia)}
			{\includegraphics[width=4cm,height=3cm]{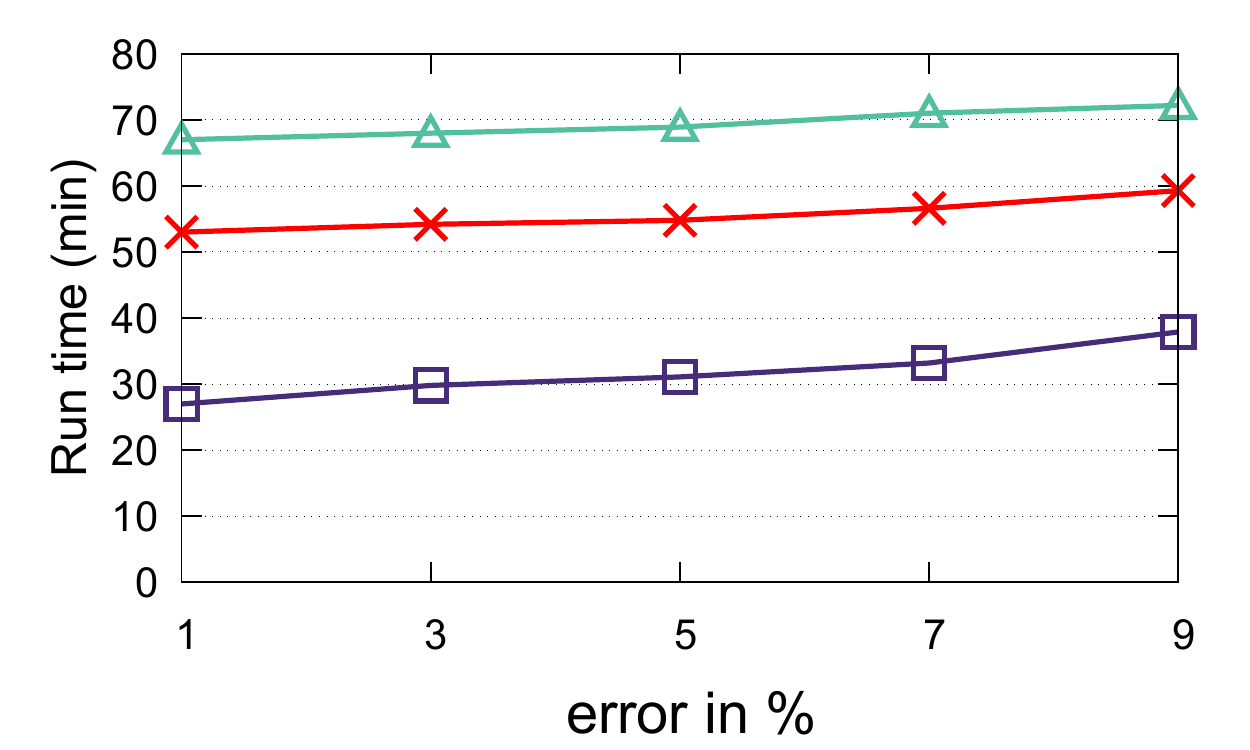}}}
		}
    \tr{	\subfloat[\small{Vary \lit{err} (\pdd)} ]
		{\label{fig:f1runtime(pdd)}
			{\includegraphics[width=4cm,height=3cm]{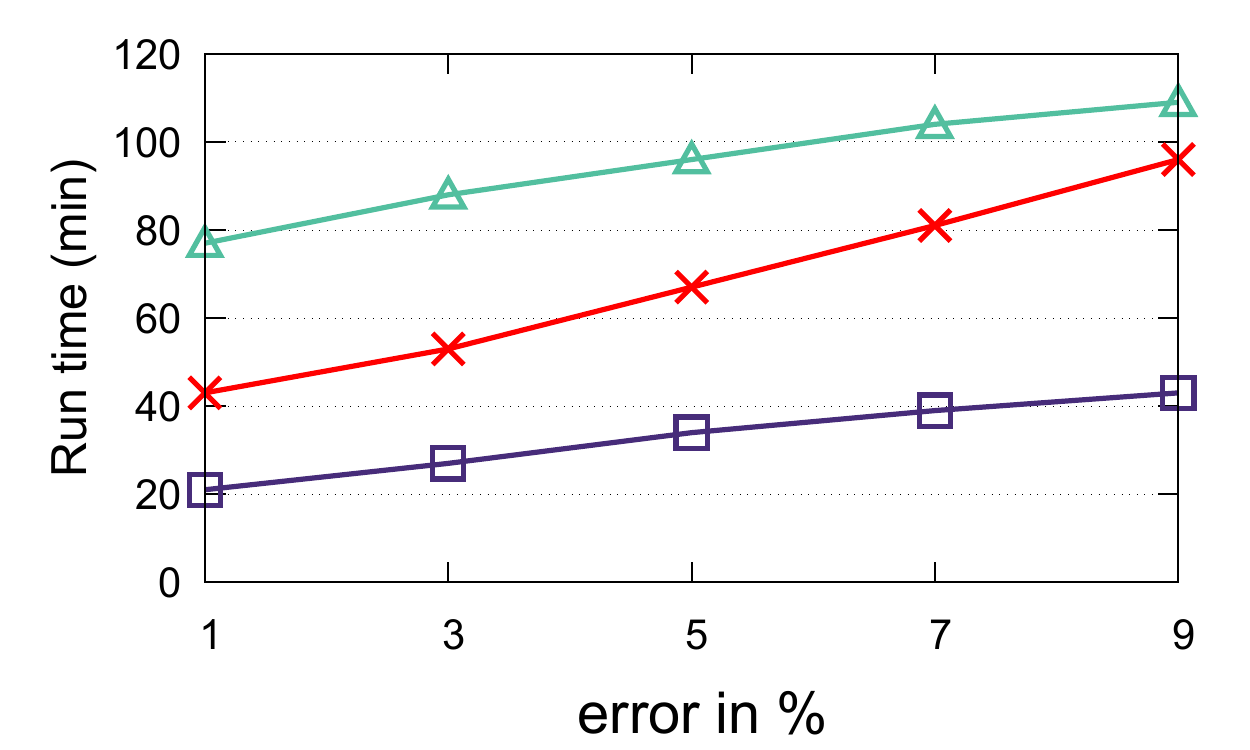}}}
        }	
	\caption{Baseline comparative performance.} \label{fig:BaselineComparison}
\end{figure*}

\vspace{-1ex}
\stitle{Comparing with \gfds and \gtars.} \rev{Figures~\ref{fig:f1(imdb)}-\ref{fig:f1(pdd)} show the \fscore-score of \tedparallel against \gfdbasic \ and \gtarSubIso \ for varying error rates over the \imdb, \dbpedia, and \pdd datasets.   \tedparallel outperforms \gfdbasic and \gtarSubIso on \imdb by an average of 55\% and 74\%, respectively.  Over \dbpedia (resp. \pdd),  \tedparallel achieves an average gain over \gfdbasic and \gtarSubIso of 20\% and 48\% (resp. 60\% and 73\%).}  \gfdbasic achieves 23\% recall using \imdb, while \tgfds capture more errors ``across'' graph snapshots than \gfds. \gfds exhibit greater sensitivity to negative errors and false positives for increasing error rates
as more match pairs are incorrectly detected as 
violations when temporal intervals are not considered.  In comparison to \gtarSubIso, with an average recall of 12\%, \tgfds have greater expressive power with variable and constant literals, to capture errors with non-zero lower bounds, and detect historical errors located in ``past'' matches.  \rev{We found approximately 70\% and 60\% of detected violations are missed if using \gtars, and \gfds, respectively.}  \rev{\tedparallel achieves lower runtimes than \gfdbasic and \gtarSubIso on \imdb by (32\%,  119\%)\conf{~(Figure\ref{fig:f1runtime(imdb)})}, \dbpedia (67\%, 140\%), and \pdd (73\%, 151\%), respectively\tr{~(Figure~\ref{fig:f1runtime(imdb)}-Figure~\ref{fig:f1runtime(pdd)})}. \conf{For brevity, runtime graphs for \dbpedia and \pdd can be found in~\cite{AMCW21}.}}

\begin{figure*}[tb!]
    \centering
	{\includegraphics[width=7cm,height=0.5cm]{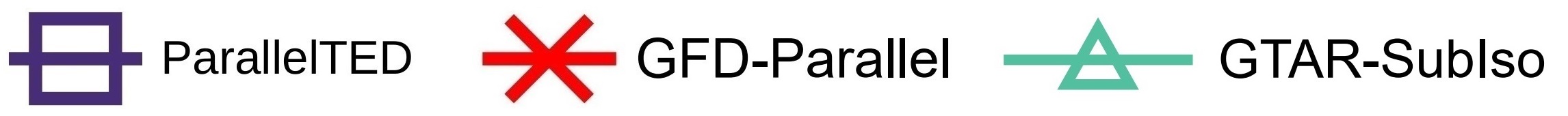}}
	\label{fig:legend}
	\vspace{-2ex}
\end{figure*}

\eat{In the same setting as Exp-1, we measured the total time of the communication by varying number of machines $n$ over \dbpedia and \imdb datasets, reported in Figures~\ref{fig:ccost(dbpedia)}, and \ref{fig:ccost(imdb)}, respectively. We find the following: (a) For the \dbpedia (resp. \imdb) dataset, the communication cost takes from 10\% to 48\% (resp. 5\% to 28\%) of the overall error detection
cost when n changes from 2 to 16; This confirms that adding new processors does not always reduce parallel running time as we need to ship data between the processors. (b) On average, \tedparallel have 5\% (resp. 10\%) less communication cost compared to \gfdbasic over \dbpedia (resp. \imdb) dataset that shows the effectiveness of our workload assignment and re-balancing techniques. (c)  Although we need to ship more data as we increase $n$, the communication time is not very sensitive to $n$ due to parallel shipment.}

\begin{figure}
    \centering
    \includegraphics[width=3.41in,keepaspectratio]{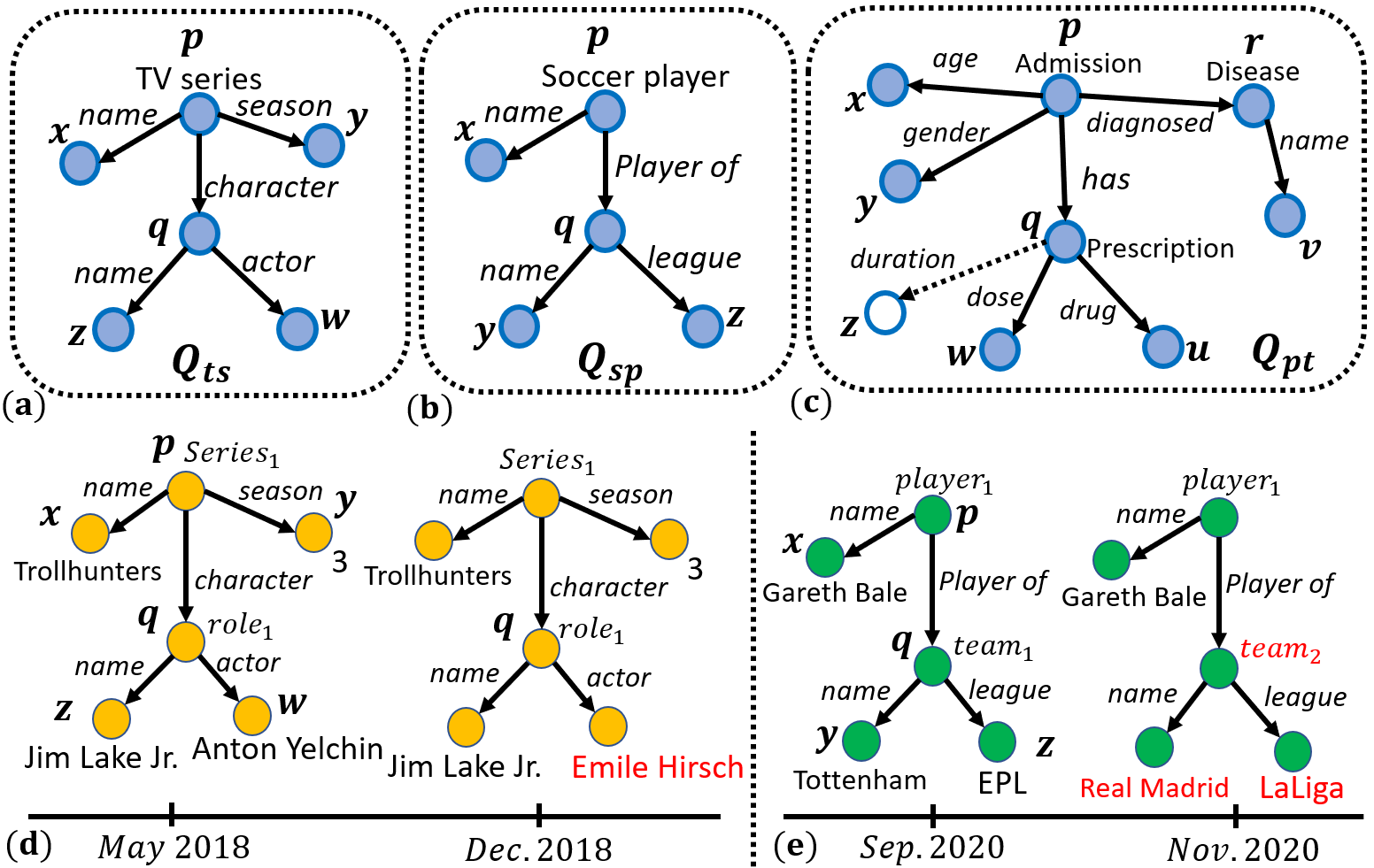}
    \vspace{-0.1cm}
    \caption{Case study: \tgfds in real-world graphs.  
    }
    \label{fig:caseStudy}
    \vspace{-1ex}
\end{figure}


\vspace{-2ex}
\subsection{Case Study: Example \tgfds} \label{sec:caseStudy}
\blue{We profiled \dbpedia \ \rev{and \pdd datasets} to identify \tgfds, and to validate their prevalence, and errors in real data~\cite{NoronhaC21}.}  \\
\underline{\tgfd 1:} \rev{Figure~\ref{fig:caseStudy}(a)} shows pattern $Q_{ts}$ for $\sigma_{ts}=(Q_{ts}[\bar{x}], (1,365), [x.name, y.season, z.name] \rightarrow [w.actor]$, \blue{which specifies \emph{``if a TV series in a given season is broadcast over a one-year period with a character role, then the actor playing this role must be unique.''}}  \rev{Figure~\ref{fig:caseStudy}(d)} shows a violation in \lit{Season 3}  of the series \lit{Trollhunters: Tales of Arcadia}, where character \lit{Jim Lake Jr.} was first played by \lit{Anton Yelchin} in \lit{May 2018}, and then by \lit{Emile Hirsch} in \lit{Dec. 2018}. \\
\underline{\tgfd 2:} 
\rev{Figure~\ref{fig:caseStudy}(b)} shows $Q_{sp}$ for $\sigma_{sp} = (Q_{sp}[\bar{x}]$, $(1,4)$, $[x.name] \rightarrow [y.league, z.team]$ that states \emph{``a football player must play for the same team and league between one and four months in a given year."} \eat{This requirement enforces player to team consistency for a given duration of time, and This only permits team changes during authorized transfer windows in a year.} \rev{Figure~\ref{fig:caseStudy}(e)} shows a violation for player \lit{Gareth Bale} who played for \lit{Tottenham Hotspur} on Sep. 2020, and then transferred to \lit{Real Madrid} on Nov. 2020.  \eat{The latter team change happened within four months, therefore leading to a violation. In both cases, existing graph constraints cannot  capture such temporal inconsistencies, and do not verify historical matches against current matches, particularly conditions involving variable literals~\cite{fan2016functional, namaki2017discovering}.} \\
\underline{\tgfd 3:} \rev{We profiled the \pdd dataset, and Figure~\ref{fig:caseStudy}(c) shows pattern $Q_{pt}$ (with only the blue vertices) for $\sigma_{pt} = (Q_{pt}[\bar{x}]$, $(1,7)$, $[x.age, y.gender, v.name, u.name] \rightarrow [w.dose]$. \tgfd $\sigma_{pt}$ specifies \emph{``for any two patients of the same age, gender, disease and prescribed drug, they must take the same daily dosage over seven days."}   We found that for 1323 patients with hypertension, 21\% were taking inconsistent dosages of the drug \lit{Metoprolol} \eat{(with drug bank id \lit{db00264})} ranging from  5mg to 100mg.} \\
\underline{\tgfd 4:} \rev{We define pattern $Q_{pt}'$ by augmenting $Q_{pt}$ with an edge from \lit{prescription} ($q$) to \lit{duration} ($z$) (shown as a dotted edge in Figure~\ref{fig:caseStudy}(c)). Consider \tgfd $\sigma_{pt}' = (Q_{pt}'[\bar{x}]$, $(1,31)$, $[x.age, y.gender, v.name, u.name,w.dose] \rightarrow [z.duration]$ that specifies \emph{``For patients with the same age, gender, disease, drug, and dosage, the drug must be taken over the same duration of days."}  However, we found that 41\% of patients were taking their drug dosages over inconsistent durations ranging from 0 to 14 days.}

%% file: relatedwork.tex
\vspace{-2.5ex}
\section{Related Work}
\label{sec:rw}

\stitle{Graph Dependencies.} \blue{
\eat{To preserve data quality over graphs, integrity constraints are defined using a graph pattern $Q$ and a data dependency $X \rightarrow Y$, where $Q$ determines the scope of the constraint.
\gfds impose the dependency on the matches of the graph pattern $Q$~\cite{fan2016functional}. Later,} \geds\cite{fan2019dependencies}  extend \gfds via literal equality of \emph{entity id} over two nodes in the graph pattern to subsume  \gfds and \gkeys. \gkeys and their ontological variants are defined to uniquely identify entities~\cite{fan2015keys,ma2019ontology}.
\NGDs~\cite{fan2018catching} are defined to extend \gfds with the support of linear arithmetic expressions and built-in comparison predicates. None of these dependencies are applicable to catch inconsistencies over temporal graphs. \tgfds use a time duration bound to \eat{are based on pairwise comparisons across timestamps satisfying the time duration bounds, which}induce the scope of historical and future matches relative to a current time. Association rules  \gpar~\cite{fan2015association} are defined over static graphs.  \gtars are soft rules for predictive analysis, and are unable to capture inconsistencies, particularly rare occurrences, across a given time duration.  
}

\stitle{Temporal Dependencies.}
\rev{FDs over temporal databases include \emph{Dynamic Functional Dependencies} (\dfds) that hold over consecutive snapshots (states) of a database, where attribute values from a current state determine values in the next state, e.g., a new salary is determined by the last salary and the last merit increase~\cite{JSS96}.  Wijsen et. al., define \emph{Temporal Functional Dependencies} (\tfds) that rely on object identity, include a valid-time on tuples,  and apply to the sequence of snapshots defining a temporal relation~\cite{WijsenVO93}. \dfds constrain pairs of adjacent states, while \tfds constrain a sequence of multiple valid-time states.  \tgfds impose topological constraints that are not captured in relational settings, and \tgfds model a different matching and temporal semantics.  \dfds and \tfds impose consistency over tuples from adjacent snapshots.  \tgfds are not restricted to compare consecutive matches, but constrain the time difference of  matches to lie within the $\Delta$ interval. Second, \tgfds model the time interval duration when topological and attribute consistency is expected, whereas \dfds and \tfds do not model such semantics.   
}



\stitle{Temporal Graph Mining}. 
Frequent dynamic subgraphs discover patterns beyond a given threshold over a time interval~\cite{borgwardt2006pattern}. Mining 
temporal motifs capture dynamic interactions with 
patterns~\cite{gurukar2015commit,paranjape2017motifs}, with extensions to time windows\eat{ known as $\delta$-temporal motifs}~\cite{paranjape2017motifs}.  
Mining algorithms identify dense temporal cliques with partitioning schemes over temporal graphs~\cite{sun2007graphscope}, MDL-based approaches to identify associations between changes~\cite{ferlez2008monitoring}, and mining for cross-graph quasi-cliques~\cite{pei2005mining}. \eat{TimeCrunch associates each subgraph across time to a template structure, and uses clustering to stitch these temporal graphs together to create an entity trajectory~\cite{shah2015timecrunch}. } While our work shares a similar spirit to identify consistent temporal patterns, \tgfds impose attribute dependency and temporal constraints over identified matches of a given pattern. 

\stitle{Constraint-Based Graph Cleaning}.
There has been extensive work to infer missing data in graphs~\cite{paulheim2017knowledge}. \eat{Graph Fact Checking rules~\cite{lin2018discovering} have been introduced to answer \emph{true/false} to generalized fact statements.} \cite{song2021explaining} proposed a constraint based approach to infer missing data in graphs. 
\eat{Graph identification~\cite{pujara2013knowledge} has been introduced to use probabilistic approaches to infer missing facts in  knowledge graphs.} Graph Quality Rules (\GQRs)~\cite{fan2019deducing} are defined to deduce \emph{certain} fixes over graphs by supporting conditional functional dependencies, graph keys and negative rules. These techniques are designed for static graphs to suggest changes or provide provenance information based on enforcing a set of rules. \tgfds serve a different purpose to model time-dependent data consistency requirements over temporal graphs that are conditioned by a given time interval.

%% file: conclusion.tex
\vspace{-2ex}
\section{Conclusion}
\label{sec:conclude}

We proposed \tgfds, a class of graph dependencies that characterize errors induced by graph patterns and time intervals for temporal graphs.  We established complexity results for fundamental problems, and 
introduced a sound and complete axiom system.  We introduced a parallel, and an incremental algorithm for 
\tgfd-based error detection. Our experimental results verified the effectiveness and efficiency of 
our error detection algorithms. \wu{ As next steps, we intend to explore non-parametrized methods to extend our workload rebalancing scheme to adapt to workload burstiness}.  Graph data cleaning with respect to \tgfds would also be an interesting next step.


%% file: main.bbl
\begin{thebibliography}{10}
\providecommand{\url}[1]{#1}
\csname url@samestyle\endcsname
\providecommand{\newblock}{\relax}
\providecommand{\bibinfo}[2]{#2}
\providecommand{\BIBentrySTDinterwordspacing}{\spaceskip=0pt\relax}
\providecommand{\BIBentryALTinterwordstretchfactor}{4}
\providecommand{\BIBentryALTinterwordspacing}{\spaceskip=\fontdimen2\font plus
\BIBentryALTinterwordstretchfactor\fontdimen3\font minus
  \fontdimen4\font\relax}
\providecommand{\BIBforeignlanguage}[2]{{%
\expandafter\ifx\csname l@#1\endcsname\relax
\typeout{** WARNING: IEEEtran.bst: No hyphenation pattern has been}%
\typeout{** loaded for the language `#1'. Using the pattern for}%
\typeout{** the default language instead.}%
\else
\language=\csname l@#1\endcsname
\fi
#2}}
\providecommand{\BIBdecl}{\relax}
\BIBdecl

\bibitem{fan2016functional}
W.~Fan, Y.~Wu, and J.~Xu, ``Functional dependencies for graphs,'' in
  \emph{SIGMOD International Conference on Management of Data}, 2016, pp.
  1843--1857.

\bibitem{fan2015keys}
W.~Fan, Z.~Fan, C.~Tian, and X.~L. Dong, ``Keys for graphs,'' \emph{Proceedings
  of the VLDB Endowment}, vol.~8, no.~12, pp. 1590--1601, 2015.

\bibitem{ma2019ontology}
H.~Ma, M.~Alipourlangouri, Y.~Wu, F.~Chiang, and J.~Pi, ``Ontology-based entity
  matching in attributed graphs,'' \emph{Proceedings of the VLDB Endowment},
  vol.~12, no.~10, pp. 1195--1207, 2019.

\bibitem{fan2015association}
W.~Fan, X.~Wang, Y.~Wu, and J.~Xu, ``Association rules with graph patterns,''
  \emph{Proceedings of the VLDB Endowment}, vol.~8, no.~12, pp. 1502--1513,
  2015.

\bibitem{wang2020extending}
X.~Wang, Y.~Xu, and H.~Zhan, ``Extending association rules with graph
  patterns,'' \emph{Expert Systems with Applications}, vol. 141, p. 112897,
  2020.

\bibitem{cdngov}
\BIBentryALTinterwordspacing
(2021) Canada’s {COVID}-19 economic response plan. [Online]. Available:
  \url{https://www.canada.ca/en/department-finance/economic-response-plan.html#businesses}
\BIBentrySTDinterwordspacing

\bibitem{fdacovid}
\BIBentryALTinterwordspacing
(2020) Fda approves first treatment for {COVID}-19. [Online]. Available:
  \url{https://www.fda.gov/news-events/press-announcements/fda-approves-first-treatment-covid-19}
\BIBentrySTDinterwordspacing

\bibitem{PG18}
P.~Augustyniak and G.~Slusarczyk, ``Graph-based representation of behavior in
  detection and prediction of daily living activities,'' \emph{Computers in
  Biology and Medicine}, vol.~95, pp. 261--270, 2018.

\bibitem{ismp22}
\BIBentryALTinterwordspacing
(2022) Institute for safe medication practices, fda advise-err: Reported
  medication errors with veklury (remdesivir) emergency use authorization.
  [Online]. Available:
  \url{https://www.ismp.org/resources/fda-advise-err-reported-medication-errors-veklury/remdesivir-emergency-use-authorization}
\BIBentrySTDinterwordspacing

\bibitem{fan2019dependencies}
W.~Fan and P.~Lu, ``Dependencies for graphs,'' \emph{ACM Transactions on
  Database Systems (TODS)}, vol.~44, no.~2, pp. 1--40, 2019.

\bibitem{namaki2017discovering}
M.~H. Namaki, Y.~Wu, Q.~Song, P.~Lin, and T.~Ge, ``Discovering graph temporal
  association rules,'' in \emph{CIKM}, 2017, pp. 1697--1706.

\bibitem{NoronhaC21}
L.~Noronha and F.~Chiang, ``Discovery of temporal graph functional dependencies
  (short paper),'' in \emph{{CIKM}, Virtual Event}, 2021, pp. 3348--3352.

\bibitem{Wijsen2009TemporalD}
J.~Wijsen, ``Temporal dependencies,'' in \emph{Encyclopedia of Database
  Systems}, 2009.

\bibitem{AMCW21}
\BIBentryALTinterwordspacing
M.~Alipourlangouri, A.~Mansfield, F.~Chiang, and Y.~Wu, ``Temporal graph
  functional dependencies, extended version,'' 2022. [Online]. Available:
  \url{https://arxiv.org/abs/2108.08719}
\BIBentrySTDinterwordspacing

\bibitem{GJ90}
M.~R. Garey and D.~S. Johnson, \emph{Computers and Intractability; A Guide to
  the Theory of NP-Completeness}.\hskip 1em plus 0.5em minus 0.4em\relax W. H.
  Freeman and Co., 1990.

\bibitem{FJJ11}
W.~Fan, J.~Li, J.~Luo, Z.~Tan, X.~Wang, and Y.~Wu, ``Incremental graph pattern
  matching,'' in \emph{Proceedings of the 2011 ACM SIGMOD International
  Conference on Management of Data}, 2011, p. 925–936.

\bibitem{BBT18}
T.~Bleifu\ss{}, L.~Bornemann, T.~Johnson, D.~V. Kalashnikov, F.~Naumann, and
  D.~Srivastava, ``Exploring change: A new dimension of data analytics,''
  \emph{Proc. VLDB Endow.}, vol.~12, no.~2, p. 85–98, Oct. 2018.

\bibitem{SRK07}
E.~P. Shironoshita, M.~T. Ryan, and M.~R. Kabuka, ``Cardinality estimation for
  the optimization of queries on ontologies,'' \emph{SIGMOD Rec.}, vol.~36,
  no.~2, p. 13–18, 2007.

\bibitem{shmoys1993approximation}
D.~B. Shmoys and {\'E}.~Tardos, ``An approximation algorithm for the
  generalized assignment problem,'' \emph{Mathematical programming}, vol.~62,
  no.~1, pp. 461--474, 1993.

\bibitem{datasite}
\BIBentryALTinterwordspacing
(2022) T{GFD} source code and data repository. [Online]. Available:
  \url{https://github.com/TGFD-Project/TGFD/}
\BIBentrySTDinterwordspacing

\bibitem{lehmann2015dbpedia}
J.~Lehmann, R.~Isele, M.~Jakob, A.~Jentzsch, D.~Kontokostas, P.~N. Mendes,
  S.~Hellmann, M.~Morsey, P.~Van~Kleef, S.~Auer \emph{et~al.}, ``D{B}pedia--a
  large-scale, multilingual knowledge base extracted from {W}ikipedia,''
  \emph{Semantic Web}, 2015.

\bibitem{imdb}
\BIBentryALTinterwordspacing
``I{MDB} dataset,'' 2021. [Online]. Available:
  \url{ftp://ftp.fu-berlin.de/pub/misc/movies/database/frozendata/}
\BIBentrySTDinterwordspacing

\bibitem{wang2017pdd}
M.~Wang, J.~Zhang, J.~Liu, W.~Hu, S.~Wang, X.~Li, and W.~Liu, ``Pdd graph:
  Bridging electronic medical records and biomedical knowledge graphs via
  entity linking,'' in \emph{International Semantic Web Conference}.\hskip 1em
  plus 0.5em minus 0.4em\relax Springer, 2017, pp. 219--227.

\bibitem{bagan2016generating}
G.~Bagan, A.~Bonifati, R.~Ciucanu, G.~H. Fletcher, A.~Lemay, and N.~Advokaat,
  ``Generating flexible workloads for graph databases,'' \emph{Proceedings of
  the VLDB Endowment}, vol.~9, no.~13, pp. 1457--1460, 2016.

\bibitem{foggia2001performance}
P.~Foggia, C.~Sansone, and M.~Vento, ``A performance comparison of five
  algorithms for graph isomorphism,'' in \emph{Proceedings of the 3rd IAPR
  TC-15 Workshop on Graph-based Representations in Pattern Recognition}, 2001,
  pp. 188--199.

\bibitem{fan2018catching}
W.~Fan, X.~Liu, P.~Lu, and C.~Tian, ``Catching numeric inconsistencies in
  graphs,'' in \emph{SIGMOD}, 2018, pp. 381--393.

\bibitem{JSS96}
C.~Jensen, R.~Snodgrass, and M.~Soo, ``Extending existing dependency theory to
  temporal databases,'' \emph{IEEE Transactions on Knowledge and Data
  Engineering}, vol.~8, no.~4, pp. 563--582, 1996.

\bibitem{WijsenVO93}
J.~Wijsen, J.~Vandenbulcke, and H.~Olivie, ``Functional dependencies
  generalized for temporal databases that include object-identity,'' in
  \emph{International Conference on the Entity-Relationship Approach}, ser.
  Lecture Notes in Computer Science, vol. 823, 1993, pp. 99--109.

\bibitem{borgwardt2006pattern}
K.~M. Borgwardt, H.-P. Kriegel, and P.~Wackersreuther, ``Pattern mining in
  frequent dynamic subgraphs,'' in \emph{Sixth International Conference on Data
  Mining (ICDM'06)}.\hskip 1em plus 0.5em minus 0.4em\relax IEEE, 2006, pp.
  818--822.

\bibitem{gurukar2015commit}
S.~Gurukar, S.~Ranu, and B.~Ravindran, ``Commit: A scalable approach to mining
  communication motifs from dynamic networks,'' in \emph{Proceedings of the
  2015 ACM SIGMOD International Conference on Management of Data}, 2015, pp.
  475--489.

\bibitem{paranjape2017motifs}
A.~Paranjape, A.~R. Benson, and J.~Leskovec, ``Motifs in temporal networks,''
  in \emph{Proceedings of the tenth ACM international conference on web search
  and data mining}, 2017, pp. 601--610.

\bibitem{sun2007graphscope}
J.~Sun, C.~Faloutsos, S.~Papadimitriou, and P.~S. Yu, ``Graphscope:
  parameter-free mining of large time-evolving graphs,'' in \emph{Proceedings
  of the 13th ACM SIGKDD international conference on Knowledge discovery and
  data mining}, 2007, pp. 687--696.

\bibitem{ferlez2008monitoring}
J.~Ferlez, C.~Faloutsos, J.~Leskovec, D.~Mladenic, and M.~Grobelnik,
  ``Monitoring network evolution using mdl,'' in \emph{2008 IEEE 24th
  International Conference on Data Engineering}.\hskip 1em plus 0.5em minus
  0.4em\relax IEEE, 2008, pp. 1328--1330.

\bibitem{pei2005mining}
J.~Pei, D.~Jiang, and A.~Zhang, ``On mining cross-graph quasi-cliques,'' in
  \emph{Proceedings of the eleventh ACM SIGKDD international conference on
  Knowledge discovery in data mining}, 2005, pp. 228--238.

\bibitem{paulheim2017knowledge}
H.~Paulheim, ``Knowledge graph refinement: A survey of approaches and
  evaluation methods,'' \emph{Semantic web}, vol.~8, no.~3, pp. 489--508, 2017.

\bibitem{song2021explaining}
Q.~Song, P.~Lin, H.~Ma, and Y.~Wu, ``Explaining missing data in graphs: A
  constraint-based approach,'' in \emph{2021 IEEE 37th International Conference
  on Data Engineering (ICDE)}.\hskip 1em plus 0.5em minus 0.4em\relax IEEE,
  2021, pp. 1476--1487.

\bibitem{fan2019deducing}
W.~Fan, P.~Lu, C.~Tian, and J.~Zhou, ``Deducing certain fixes to graphs,''
  \emph{Proceedings of the VLDB Endowment}, vol.~12, no.~7, pp. 752--765, 2019.

\end{thebibliography}
